\documentclass[aps,rmp,tightenlines,12pt,%showpacs,showkeys, %%ARXIV
groupedaddress,nofootinbib]{revtex4} %%ARXIV
\usepackage{hyperref}      % hypertext links %%ARXIV

\usepackage{graphicx} % needed for figures 
\usepackage{slashed}

\renewcommand{\citet}[1]{Ref.~\cite{#1}} %%ARXIV

\newcommand{\ttwo}[2]{\begin{minipage}{.3in}\noindent #1 \\
    \vspace*{-.05in} #2\end{minipage}} %%ARXIV
\newcommand{\mhbare}{m_{h\, 0}}
\newcommand{\sigmaSI}{\sigma_{\text{SI}}}
\newcommand{\sigmaSD}{\sigma_{\text{SD}}}
\newcommand{\sign}{\text{sign}}
\newcommand{\vrel}{v_{\text{rel}}}
\newcommand{\sigmath}{\sigma^{\text{th}}}

\newcommand{\zetaEM}{\zeta_{\text{EM}}}

\newcommand{\tBBN}{t_{\text{BBN}}}
\newcommand{\TBBN}{T_{\text{BBN}}}
\newcommand{\ThBBN}{T^h_{\text{BBN}}}

\newcommand{\TRH}{T_{\text{RH}}}

\newcommand{\ThRH}{T^h_{\text{RH}}}

\newcommand{\nequ}{n_{\text{eq}}}
\newcommand{\neff}{n_{\text{eff}}}
\newcommand{\mgmsb}{m_{\text{GMSB}}}
\newcommand{\mgrav}{m_{\text{grav}}}

\newcommand{\gweak}{g_{\text{weak}}}
\newcommand{\mweak}{m_{\text{weak}}}
\newcommand{\mgut}{m_{\text{GUT}}}
\newcommand{\mplanck}{M_{\text{Pl}}}
\newcommand{\mstar}{M_{*}}

\newcommand{\OmegaDM}{\Omega_{\text{DM}}}

\newcommand{\ev}{\text{eV}}
\newcommand{\kev}{\text{keV}}
\newcommand{\mev}{\text{MeV}}
\newcommand{\gev}{\text{GeV}}
\newcommand{\tev}{\text{TeV}}
\newcommand{\pb}{\text{pb}}

\newcommand{\cm}{\text{cm}}
\newcommand{\m}{\text{m}}
\newcommand{\km}{\text{km}}
\newcommand{\g}{\text{g}}
\newcommand{\kg}{\text{kg}}
\newcommand{\s}{\text{s}}

\newcommand{\eg}{{\em e.g.}}

\newcommand{\Eqref}[1]{Equation~(\ref{#1})}
\newcommand{\eqref}[1]{Eq.~(\ref{#1})}
\newcommand{\eqsref}[2]{Eqs.~(\ref{#1}) and (\ref{#2})}
\newcommand{\secref}[1]{Sec.~\ref{sec:#1}}
\newcommand{\secsref}[2]{Secs.~\ref{sec:#1} and \ref{sec:#2}}
\newcommand{\figref}[1]{Fig.~\ref{fig:#1}}
\newcommand{\figsref}[2]{Figs.~\ref{fig:#1} and \ref{fig:#2}}
\newcommand{\tableref}[1]{Table~\ref{table:#1}}

\newcommand{\WIMP}{\text{WIMP}}
\newcommand{\SWIMP}{\text{SWIMP}}

\newcommand{\mWIMP}{m_{\WIMP}}
\newcommand{\mSWIMP}{m_{\SWIMP}}

\newcommand{\mchi}{m_{\chi}}
\newcommand{\mgravitino}{m_{\gravitino}}

\newcommand{\gravitino}{\tilde{G}}
\newcommand{\Bino}{\tilde{B}}

\newcommand{\stau}{\tilde{\tau}}

\newcommand{\mgaugino}{M_{1/2}}
\newcommand{\epsEM}{\varepsilon_{\text{EM}}}
\newcommand{\mmess}{M_{\text{m}}}

\newcommand{\Omegachi}{\Omega_{\chi}}

\begin{document}

%\input psfig.sty
%\input epsf.tex

%\jname{Annu. Rev. Astron. Astrophys.} %%ARAA
%\jyear{2010} %%ARAA
%\jvol{48} %%ARAA
%\ARinfo{AR Info} %%ARAA

\preprint{UCI-TR-2009-13} %%ARXIV

\title{Dark Matter Candidates from Particle Physics and Methods of
Detection}

\markboth{Feng}{Dark Matter Candidates}

\author{Jonathan L.~Feng} %%ARXIV
\affiliation{Department of Physics and Astronomy, University of %%ARXIV
California, Irvine, CA 92697, USA}  %%ARXIV

%\author{Jonathan L.~Feng %%ARAA
%\affiliation{Department of Physics and Astronomy, University of %%ARAA
%California, Irvine, CA 92697, USA}}  %%ARAA


\begin{abstract}
The identity of dark matter is a question of central importance in
both astrophysics and particle physics.  In the past, the leading
particle candidates were cold and collisionless, and typically
predicted missing energy signals at particle colliders.  However,
recent progress has greatly expanded the list of well-motivated
candidates and the possible signatures of dark matter.  This review
begins with a brief summary of the standard model of particle physics
and its outstanding problems. We then discuss several dark matter
candidates motivated by these problems, including WIMPs, superWIMPs,
light gravitinos, hidden dark matter, sterile neutrinos, and axions.
For each of these, we critically examine the particle physics
motivations and present their expected production mechanisms, basic
properties, and implications for direct and indirect detection,
particle colliders, and astrophysical observations.  Upcoming
experiments will discover or exclude many of these candidates, and
progress may open up an era of unprecedented synergy between studies
of the largest and smallest observable length scales.
\end{abstract}

\maketitle

\tableofcontents %%ARXIV

\section{INTRODUCTION}
\label{sec:introduction}

The evidence that dark matter is required to make sense of our
Universe has been building for some time.  In 1933 Fritz Zwicky found
that the velocity dispersion of galaxies in the Coma cluster of
galaxies was far too large to be supported by the luminous
matter~\cite{Zwicky:1933gu}.  In the 1970s, Vera Rubin and
collaborators~\cite{Rubin:1970zz,Rubin:1980zd} and Albert
Bosma~\cite{1978PhDT.......195B} measured the rotation curves of
individual galaxies and also found evidence for non-luminous matter.
This and other ``classic'' evidence for non-luminous matter (see, \eg,
\citet{Trimble:1987ee}) has now been supplemented by data from
weak~\cite{Refregier:2003ct} and strong~\cite{Tyson:1998vp} lensing,
hot gas in clusters~\cite{Lewis:2002mfa}, the Bullet
Cluster~\cite{Clowe:2006eq}, Big Bang nucleosynthesis
(BBN)~\cite{Fields:2008}, further constraints from large scale
structure~\cite{Allen:2002eu}, distant
supernovae~\cite{Riess:1998cb,Perlmutter:1998np}, and the cosmic
microwave background (CMB)~\cite{Komatsu:2010fb}.

Together, these data now provide overwhelming evidence for the
remarkable fact that not only is there non-luminous matter in our
Universe, but most of it is not composed of baryons or any of the
other known particles.  Current data imply that dark matter is five
times more prevalent than normal matter and accounts for about a
quarter of the Universe.  More precisely, these data constrain the
energy densities of the Universe in baryons, non-baryonic dark matter
(DM), and dark energy $\Lambda$ to be~\cite{Komatsu:2010fb}
\begin{eqnarray}
\Omega_{\text{B}}  &\simeq& 0.0456 \pm 0.0016 \\
\OmegaDM           &\simeq& 0.227 \pm 0.014 \\
\Omega_{\Lambda}   &\simeq& 0.728 \pm 0.015 \ .
\label{Lambda}
\end{eqnarray}

Despite this progress, all of the evidence for dark matter noted above
is based on its gravitational interactions.  Given the universality of
gravity, this evidence does little to pinpoint what dark matter is.
At the same time, the identity of dark matter has far-reaching
implications: in astrophysics, the properties of dark matter determine
how structure forms and impact the past and future evolution of the
Universe; and in particle physics, dark matter is the leading
empirical evidence for new particles, and there are striking hints
that it may be linked to attempts to understand electroweak symmetry
breaking, the leading puzzle in the field today.  The identity of dark
matter is therefore of central importance in both fields and ties
together studies of the Universe at both the largest and smallest
observable length scales.

In this review, we discuss some of the leading dark matter candidates
and their implications for experiments and observatories. The wealth
of recent cosmological data does constrain some dark matter
properties, such as its self-interactions and its temperature at the
time of matter-radiation equality.  Nevertheless, it is still not at
all difficult to invent new particles that satisfy all the
constraints, and there are candidates motivated by minimality,
particles motivated by possible experimental anomalies, and exotic
possibilities motivated primarily by the desire of clever iconoclasts
to highlight how truly ignorant we are about the nature of dark
matter.

Here we will focus on dark matter candidates that are motivated not
only by cosmology, but also by robust problems in particle physics.
For this reason, this review begins with a brief summary of the
standard model of particle physics, highlighting its basic features
and some of its problems.  As we will see, particle physics provides
strong motivation for new particles, and in many cases, these
particles have just the right properties to be dark matter.  We will
find that many of them predict signals that are within reach of
current and near future experiments. We will also find that unusual
predictions for astrophysics emerge, and that cold and collisionless
dark matter is far from a universal prediction, even for candidates
with impeccable particle physics credentials.  At the same time, it
will become clear that even in favorable cases, a compelling solution
to the dark matter problem will not be easy to achieve and will likely
rely on synergistic progress along many lines of inquiry.

An outline of this review is provided by \tableref{summary}, which
summarizes the dark matter candidates discussed here, along with their
basic properties and opportunities for detection.  Some of the
acronyms and symbols commonly used in this review are defined in
\tableref{definitions}.

\begin{table}[tbp]
\begin{minipage}{\columnwidth}
\begin{tabular}{lcccccc} \hline\hline
\rule[1mm]{0mm}{5mm}
& WIMPs & SuperWIMPs & Light $\tilde{G}$ 
& Hidden DM & Sterile $\nu$ & Axions \vspace*{.1in} \\ \hline
Motivation \rule[3mm]{0mm}{5mm}
& GHP & GHP  & \ttwo{GHP}{NPFP} & \ttwo{GHP}{NPFP} & $\nu$ Mass &
Strong CP \vspace*{.2in} \\
\ttwo{Naturally}{Correct~$\Omega$} & Yes & Yes & No
& Possible & No & No \vspace*{.12in} \\
\ttwo{Production}{Mechanism} & Freeze Out & Decay & Thermal
& Various & Various & Various \vspace*{.12in} \\
Mass Range & GeV$-$TeV & GeV$-$TeV & eV$-$keV 
& GeV$-$TeV & keV & $\mu\ev - \text{meV}$ \vspace*{.12in} \\
Temperature & Cold & Cold/Warm & Cold/Warm
& Cold/Warm & Warm & Cold \vspace*{.12in} \\
Collisional & & & 
& $\surd$ & & \vspace*{.12in} \\
\ttwo{Early}{Universe} & & $\surd\surd$ & 
& $\surd$ & & \vspace*{.12in} \\
\ttwo{Direct}{Detection} & $\surd\surd$ & & 
& $\surd$ & & $\surd\surd$ \vspace*{.12in} \\
\ttwo{Indirect}{Detection} & $\surd\surd$ & $\surd$ & 
& $\surd$ & $\surd\surd$ & \vspace*{.12in} \\
\ttwo{Particle}{Colliders} & $\surd\surd$ & $\surd\surd$ & $\surd\surd$ 
& $\surd$ & & \vspace*{.12in} \\
\hline \hline
\end{tabular}
\end{minipage}
\caption{Summary of dark matter particle candidates, their properties,
and their potential methods of detection.  The particle physics
motivations are discussed in \secref{problems}; GHP and NPFP are
abbreviations for the gauge hierarchy problem and new physics flavor
problem, respectively.  In the last five rows, $\surd\surd$ denotes
detection signals that are generic for this class of dark matter
candidate and $\surd$ denotes signals that are possible, but not
generic.  ``Early Universe'' includes phenomena such as BBN and the
CMB; ``Direct Detection'' implies signatures from dark matter
scattering off normal matter in the laboratory; ``Indirect Detection''
implies signatures of late time dark matter annihilation or decay; and
``Particle Colliders'' implies signatures of dark matter or its
progenitors produced at colliders, such as the Large Hadron Collider
(LHC).  See the text for details.}
\vspace{0.2cm}
\label{table:summary}
\end{table}

\begin{table}[tbp]
\begin{minipage}{\columnwidth}
\begin{tabular}{ll} \hline\hline
$\chi \qquad \qquad \qquad $ & 
   lightest neutralino, a supersymmetric dark matter candidate \\
$\gravitino$ & gravitino, a supersymmetric dark matter candidate \\
$G_N$ & Newton's gravitational constant \\
GMSB & gauge-mediated supersymmetry breaking \\
LKP & lightest Kaluza-Klein particle \\
LSP & lightest supersymmetric particle \\
NLSP & next-to-lightest supersymmetric particle \\
$\mplanck$ & Planck mass $\simeq 1.2\times 10^{19}~\gev$ \\
$\mstar$ & reduced Planck mass $\simeq 2.4\times 10^{18}~\gev$ \\
\ttwo{minimal}{supergravity} & a simple version of the MSSM
specified by 5 parameters \\
MSSM & supersymmetric standard model with minimal number of extra particles \\
SM & standard model of particle physics \\
stau & scalar superpartner of the tau lepton \\
superWIMP & superweakly-interacting massive particle \\
UED & universal extra dimensions \\
WIMP & weakly-interacting massive particle \\
$X$ & general dark matter candidate \\
\hline \hline
    \end{tabular}
  \end{minipage}
\caption{Definitions of acronyms and symbols commonly used in this
  review.}
\vspace{0.2cm}
\label{table:definitions}
\end{table}

\section{THE STANDARD MODEL OF PARTICLE PHYSICS}

The standard model (SM) of particle physics is a spectacularly
successful theory of elementary particles and their interactions.  For
a brief, pedagogical introduction, see, \eg, \citet{Herrero:1998eq}.
At the same time, it has deficiencies, and the open questions raised
by the SM motivate many of the leading dark matter candidates and
provide guidance for dark matter searches.  We begin here with a brief
review of the SM's basic features and open problems, focusing on those
that are most relevant for dark matter.

\subsection{Particles}

\begin{figure}[tbp]
\includegraphics[width=0.80\columnwidth]{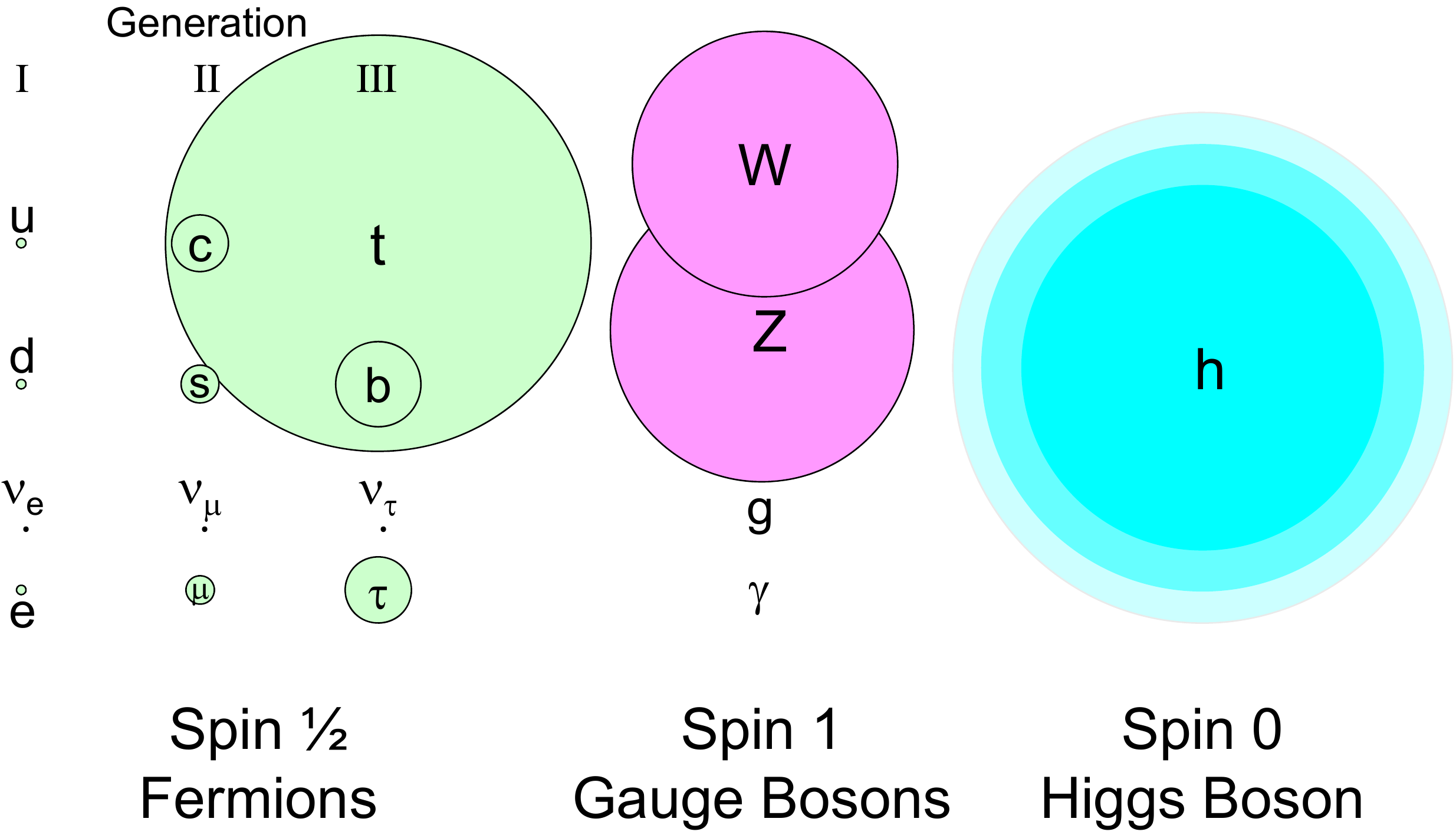}
\caption{The particles of the standard model, represented by circles
whose areas are proportional to their masses.  The photon and gluon
are massless. The Higgs boson has not yet been discovered --- its mass
has been taken in the allowed range $114.4~\gev < m_h < 186~\gev$ (see
text).
\label{fig:sm_particles}}
\end{figure}

The particles of the SM are shown in \figref{sm_particles}.  They may
be divided into three categories:
\begin{itemize}
\setlength{\itemsep}{1pt}\setlength{\parskip}{0pt}\setlength{\parsep}{0pt}
\item {\em Spin 1/2 Fermions.}  These matter particles include six
flavors of quarks (up, down, charm, strange, bottom, and top), three
flavors of charged leptons (electrons, muons, and taus), and three
flavors of neutral leptons (the electron, muon, and tau
neutrinos). These are grouped into three generations, as indicated in
\figref{sm_particles}.
\item {\em Spin 1 Gauge Bosons.} These force carrying particles
include the photon $\gamma$, which mediates electromagnetism; eight
gluons $g$, which mediate the strong force; and the $W$ and $Z$ gauge
bosons, which mediate the weak interactions.  The photon and gluons
are massless, but the $W$ and $Z$ have masses 80 GeV and 91 GeV,
respectively.
\item {\em Spin 0 Higgs Boson.} The SM Higgs particle is a spin 0
boson.  Although the Higgs boson has not yet been discovered, its mass
is constrained by a variety of collider results.  Assuming the SM,
null results from direct searches at the LEP $e^+e^-$ collider require
$m_h > 114.4~\gev$~\cite{Barate:2003sz}. Given this constraint,
precision measurements of electroweak observables at LEP require $m_h
< 186~\gev$~\cite{LEPEWWG:2009}.  In addition, the Tevatron $p\bar{p}$
collider, currently running, excludes the region $162~\gev < m_h <
166~\gev$~\cite{Aaltonen:2010yv}. These bounds may be relaxed in
extensions of the SM, but even in such theories, very general
arguments require $m_h \alt 1~\tev$ (see, \eg, \citet{Reina:2005ae}).
\end{itemize}

None of these SM particles is a good dark matter candidate.  Most of
the matter particles are unstable, with lifetimes far shorter than the
age of the Universe.  The remaining particles are the six lightest:
the electron, the up and down quarks, which may form stable protons
and neutrons in nuclei, and the three neutrinos.  Electrons may
contribute significantly to dark matter only if they are neutralized
through binding with protons, but protons (and neutrons) contribute to
the baryonic energy density $\Omega_B$, which is too small to be all
of dark matter.  In addition, current upper bounds on neutrino masses
from particle physics and cosmology imply that the neutrino relic
density $\Omega_{\nu} \simeq \sum_i m_{\nu_i} / 47~\ev \alt
0.012$~\cite{Komatsu:2010fb}.  The evidence for dark matter therefore
requires particles beyond the SM.

\subsection{Problems}
\label{sec:problems}

In addition to the need for dark matter, other problems motivate
physics beyond the SM.  These problems are of two types. The first and
most severe are experimental data that the SM cannot explain; at
present, aside from the existence of dark matter, these are confined
to the neutrino sector and are described in \secref{neutrinoproblem}.
The second are experimental data that can be explained, but only for
seemingly unnatural choices of parameters.  All of the remaining
problems are of this type.
  
\subsubsection{THE GAUGE HIERARCHY PROBLEM}
\label{sec:gaugehierarchyproblem}

The gauge hierarchy problem is the question of why the physical Higgs
boson mass $m_h$ is so small.  What is the natural value for $m_h$?
We know of three fundamental constants: the speed of light $c$,
Planck's constant $h$, and Newton's gravitational constant $G_N$.  One
combination of these has dimensions of mass, the Planck mass $\mplanck
\equiv \sqrt{hc/G_N} \simeq 1.2 \times 10^{19}~\gev$.  We therefore
expect dimensionful parameters to be either 0, if enforced by a
symmetry, or of the order of $\mplanck$.  In the SM, electroweak
symmetry is broken, and the Higgs boson mass is non-zero.  The gauge
hierarchy problem is the question of why $m_h \sim 100~\gev \ll
\mplanck$.

This problem is exacerbated in the SM by quantum corrections.  The
physical mass of the SM Higgs boson is $m_h^2 = \mhbare^2 + \Delta
m_h^2$, where $\mhbare^2$ is the tree-level mass, and
\begin{equation}
\Delta m_h^2 \sim \frac{\lambda^2}{16\pi^2} \int^{\Lambda} 
\frac{d^4 p}{p^2} \sim \frac{\lambda^2}{16\pi^2} \Lambda^2
\label{higgsquantumcorrections}
\end{equation}
is the quantum correction resulting from loop-level diagrams, where
the integral is over the momenta of particles in the loops.  The
parameter $\lambda$ is an ${\cal O}(1)$ dimensionless coupling, and
$\Lambda$ is the energy scale at which the SM is no longer a valid
description of nature.  Because $\Delta m_h^2$ is proportional to
$\Lambda^2$, it is natural to expect the Higgs mass to be pulled up to
within an order of magnitude of $\Lambda$ by quantum corrections.  In
the SM with $\Lambda \sim \mplanck$, this implies that $\mhbare^2$ and
$\Delta m_h^2$ must cancel to 1 part in $10^{36}$ to yield the correct
physical Higgs mass, which is hardly reasonable.

The gauge hierarchy problem may be eliminated if $\Lambda \alt
1~\tev$, implying new physics at the weak scale $\mweak \sim 10~\gev -
\tev$.  Alternatively, the Higgs boson may not be a fundamental
scalar, but in this case, too, its structure requires new physics at
the weak scale~\cite{Hill:2002ap}.  For these reasons, every attempt
to ameliorate the gauge hierarchy problem so far has implied new
particles with mass around $\mweak$.  The gauge hierarchy problem is
the leading motivation for dark matter candidates, such as WIMPs and
superWIMPs, weakly- and superweakly-interacting massive particles,
that are the topics of \secsref{wimps}{superwimps} below.

\subsubsection{THE NEW PHYSICS FLAVOR PROBLEM}
\label{sec:newphysicsflavorproblem}

The gauge hierarchy problem implies new particles with mass around
$\mweak$.  Such particles typically create many new problems, however,
because they may violate baryon number, lepton number, flavor, or CP,
where C and P are the discrete transformations of charge conjugation
and parity, respectively.  These symmetries are either beautifully
preserved or violated only slightly in the SM, but there is no
guarantee that new $\mweak$ particles will preserve them.  This set of
problems is collectively known as the new physics flavor problem.

The new physics flavor problem implies that not all solutions to the
gauge hierarchy problem are equally elegant.  For example, among
supersymmetric theories, it implies that those that naturally predict
very heavy squarks and sleptons, or those that predict squarks and
sleptons that are highly degenerate across different generations, are
favored, because these naturally suppress flavor-changing neutral
currents below current constraints.  This problem has implications for
the direct detection of WIMPs, as discussed in \secref{direct}, and
motivates light gravitino dark matter, reviewed in
\secref{lightgravitinos}, and some of the hidden sector dark matter
models described in \secref{hidden}.

\subsubsection{THE NEUTRINO MASS PROBLEM}
\label{sec:neutrinoproblem}

Fermion masses are described in quantum field theories by terms that
couple left- and right-handed fields together.  In the SM, however,
there are no right-handed neutrino fields, and so the SM predicts that
all neutrinos are massless. The observation of neutrino flavor
oscillations~\cite{Fukuda:1998mi,Ahmad:2002jz}, however, implies that
the three neutrinos are non-degenerate, and so at least two are
massive.  Neutrino masses and mixing provide the most direct and
compelling evidence that the SM of particle physics is incomplete, and
this problem motivates sterile neutrino dark matter, discussed in
\secref{sterileneutrinos}.

\subsubsection{THE STRONG CP PROBLEM}
\label{sec:strongcpproblem}

The SM Lagrangian includes the term $g_3^2 \theta_3/(32 \pi^2)
\epsilon^{\mu\nu\rho\sigma} G_{\mu\nu}^{\alpha}
G_{\rho\sigma}^{\alpha}$, where $g_3$ is the coupling of the strong
interactions, $\theta_3$ is an angle parameter,
$\epsilon^{\mu\nu\rho\sigma}$ is the totally anti-symmetric 4-index
tensor, and $G_{\mu\nu}$ is the gluon field strength.  This term
contributes to CP-violating, flavor-conserving observables, such as
the electric dipole moment (EDM) of the neutron $d_e$.  For $\theta_3
\sim 1$, one expects $d_e \sim 10^{-16}~e~\cm$.  The neutron EDM has
not yet been observed, but current constraints already imply $d_e <
2.9\times 10^{-26}~e~\cm$~\cite{Baker:2006ts}.  This is therefore a
fine-tuning problem of 1 part in $10^{10}$, and it motivates axions as
dark matter candidates, to be discussed in \secref{axions}.

\subsubsection{OTHER PROBLEMS}

In addition to the outstanding problems discussed above, there are
other open questions raised by the SM.  The {\em SM flavor problem},
distinct from the new physics flavor problem, is the puzzle of why the
fermion masses have such different values, as evident in
\figref{sm_particles}.  The {\em grand unification problem} is the
problem of trying to understand the strong, weak, and electromagnetic
interactions as different manifestations of a single underlying force.
In addition, at any given time there are always {\em experimental
anomalies}, measurements that do not agree with SM predictions.  At
present, the most compelling and persistent discrepancy is in the
anomalous magnetic moment of the muon, which disagrees with the SM
prediction at the level of
3.4$\sigma$~\cite{Hagiwara:2006jt}. Although the problems described in
this paragraph do not typically motivate dark matter candidates on
their own, they do sometimes play a supporting role, as discussed
below.

Finally, note that $\mhbare$ is only one of two dimensionful
parameters in the SM: there is also the term $\bar{\Lambda}^4$, which
contributes to dark energy or the cosmological constant.
\Eqref{Lambda} implies that the total energy density in dark energy is
$\Lambda \simeq (2.76~\text{meV})^4$. If the natural value of
$\bar{\Lambda}$ is $\mplanck^4$, it must cancel other contributions to
1 part in $10^{122}$, a fine-tuning that dwarfs even the gauge
hierarchy problem.  This is the {\em cosmological constant problem}.
Although one might hope for a unified solution to the cosmological
constant and dark matter problems, at present there is little
indication that they are related, and we will assume they are
decoupled in this review.

\section{WIMPS}
\label{sec:wimps}

WIMPs have mass in the range $\mweak \sim 10~\gev - \tev$ and
tree-level interactions with the $W$ and $Z$ gauge bosons, but not
with gluons or photons.  WIMPs are the most studied dark matter
candidates, as they are found in many particle physics theories,
naturally have the correct relic density, and may be detected in many
ways.  In this section, we discuss their production through thermal
freeze out, the examples of neutralino and Kaluza-Klein dark matter,
and their implications for direct detection, indirect detection, and
particle colliders.

\subsection{Thermal Freeze Out}

\subsubsection{THE WIMP MIRACLE}
\label{sec:wimpmiracle}

If a WIMP exists and is stable, it is naturally produced with a relic
density consistent with that required of dark matter.  This
tantalizing fact, sometimes referred to as the ``WIMP miracle,''
implies that particles that are motivated by the gauge hierarchy
problem, a purely microphysical puzzle, are excellent dark matter
candidates.

\begin{figure}[tbp]
\centering
\includegraphics[width=0.80\columnwidth]{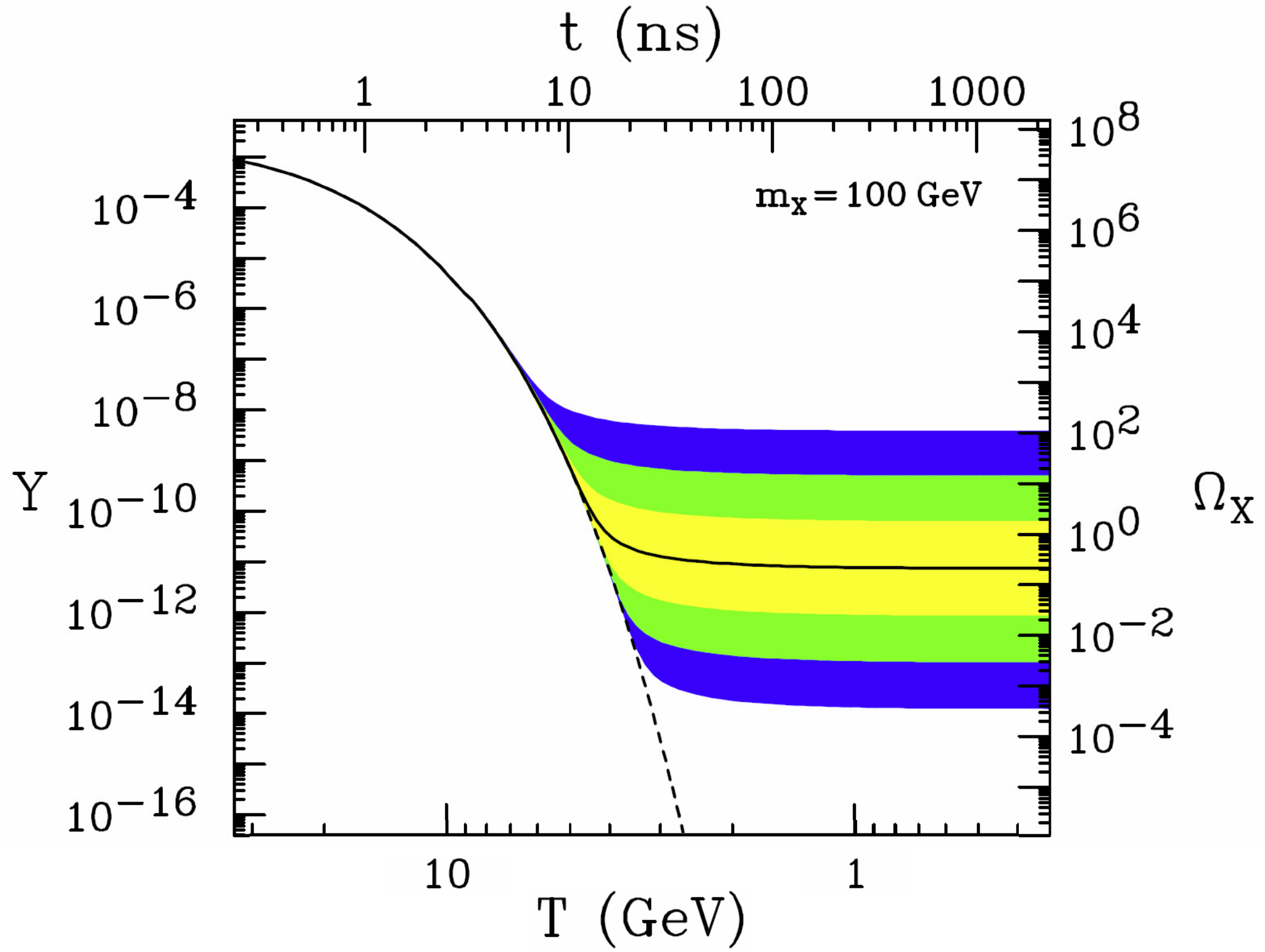}
\caption{The comoving number density $Y$ (left) and resulting thermal
relic density (right) of a 100 GeV, $P$-wave annihilating dark matter
particle as a function of temperature $T$ (bottom) and time $t$ (top).
The solid contour is for an annihilation cross section that yields the
correct relic density, and the shaded regions are for cross sections
that differ by 10, $10^2$, and $10^3$ from this value.  The dashed
contour is the number density of a particle that remains in thermal
equilibrium.
\label{fig:freezeout} }
\end{figure}

Dark matter may be produced in a simple and predictive manner as a
thermal relic of the Big Bang~\cite{Zeldovich:1965,%
Chiu:1966kg,Steigman:1979kw,Scherrer:1985zt}. The evolution of a
thermal relic's number density is shown in
\figref{freezeout}. Initially the early Universe is dense and hot, and
all particles are in thermal equilibrium.  The Universe then cools to
temperatures $T$ below the dark matter particle's mass $m_X$, and the
number of dark matter particles becomes Boltzmann suppressed, dropping
exponentially as $e^{-m_X/T}$.  The number of dark matter particles
would drop to zero, except that, in addition to cooling, the Universe
is also expanding.  In stage (3), the Universe becomes so large and
the gas of dark matter particles becomes so dilute that they cannot
find each other to annihilate.  The dark matter particles then
``freeze out,'' with their number asymptotically approaching a
constant, their thermal relic density.  Note that freeze out, also
known as chemical decoupling, is distinct from kinetic decoupling;
after thermal freeze out, interactions that change the number of dark
matter particles become negligible, but interactions that mediate
energy exchange between dark matter and other particles may remain
efficient.

This process is described quantitatively by the Boltzmann equation
\begin{equation}
\frac{dn}{dt} = -3 H n - \langle \sigma_A v \rangle
\left( n^2 - \nequ^2 \right) \ ,
\label{Boltzmann}
\end{equation}
where $n$ is the number density of the dark matter particle $X$, $H$
is the Hubble parameter, $\langle \sigma_A v \rangle$ is the thermally
averaged annihilation cross section, and $n_{\text{eq}}$ is the dark
matter number density in thermal equilibrium.  On the right-hand side
of \eqref{Boltzmann}, the first term accounts for dilution from
expansion.  The $n^2$ term arises from processes $X X \to \text{SM
SM}$ that destroy $X$ particles, where SM denotes SM particles, and
the $n_{\text{eq}}^2$ term arises from the reverse process $\text{SM
SM} \to X X$, which creates $X$ particles.

The thermal relic density is determined by solving the Boltzmann
equation numerically.  A rough analysis is highly instructive,
however.  Defining freeze out to be the time when $n \langle \sigma_A
v \rangle = H$, we have
\begin{equation}
n_f \sim (m_X T_f)^{3/2} e^{-m_X/T_f} \sim \frac{T_f^2}{\mplanck
  \langle \sigma_A v \rangle } \ ,
\end{equation}
where the subscripts $f$ denote quantities at freeze out.  The ratio
$x_f \equiv m_X/T_f$ appears in the exponential.  It is, therefore,
highly insensitive to the dark matter's properties and may be
considered a constant; a typical value is $x_f \sim 20$.  The thermal
relic density is, then,
\begin{equation}
\Omega_X = \frac{m_X n_0}{\rho_c} 
= \frac{m_X T_0^3}{\rho_c} \frac{n_0}{T_0^3} 
\sim \frac{m_X T_0^3}{\rho_c} \frac{n_f}{T_f^3} 
\sim \frac{x_f T_0^3}{\rho_c \mplanck} 
\langle \sigma_A v \rangle^{-1} \ ,
\end{equation}
where $\rho_c$ is the critical density and the subscripts $0$ denote
present day quantities.  We see that the thermal relic density is
insensitive to the dark matter mass $m_X$ and inversely proportional
to the annihilation cross section $\langle \sigma_A v
\rangle$.

Although $m_X$ does not enter $\Omega_X$ directly, in many theories it
is the only mass scale that determines the annihilation cross section.
On dimensional grounds, then, the cross section can be written
\begin{equation}
\sigma_A v = k \, \frac{\gweak^4}{16 \pi^2 m_X^2} \ 
(1\ \text{or} \ v^2)\ ,
\end{equation}
where the factor $v^2$ is absent or present for $S$- or $P$-wave
annihilation, respectively, and terms higher-order in $v$ have been
neglected.  The constant $\gweak \simeq 0.65$ is the weak interaction
gauge coupling, and $k$ parameterizes deviations from this estimate.

\begin{figure}[tbp]
\includegraphics[width=0.70\columnwidth]{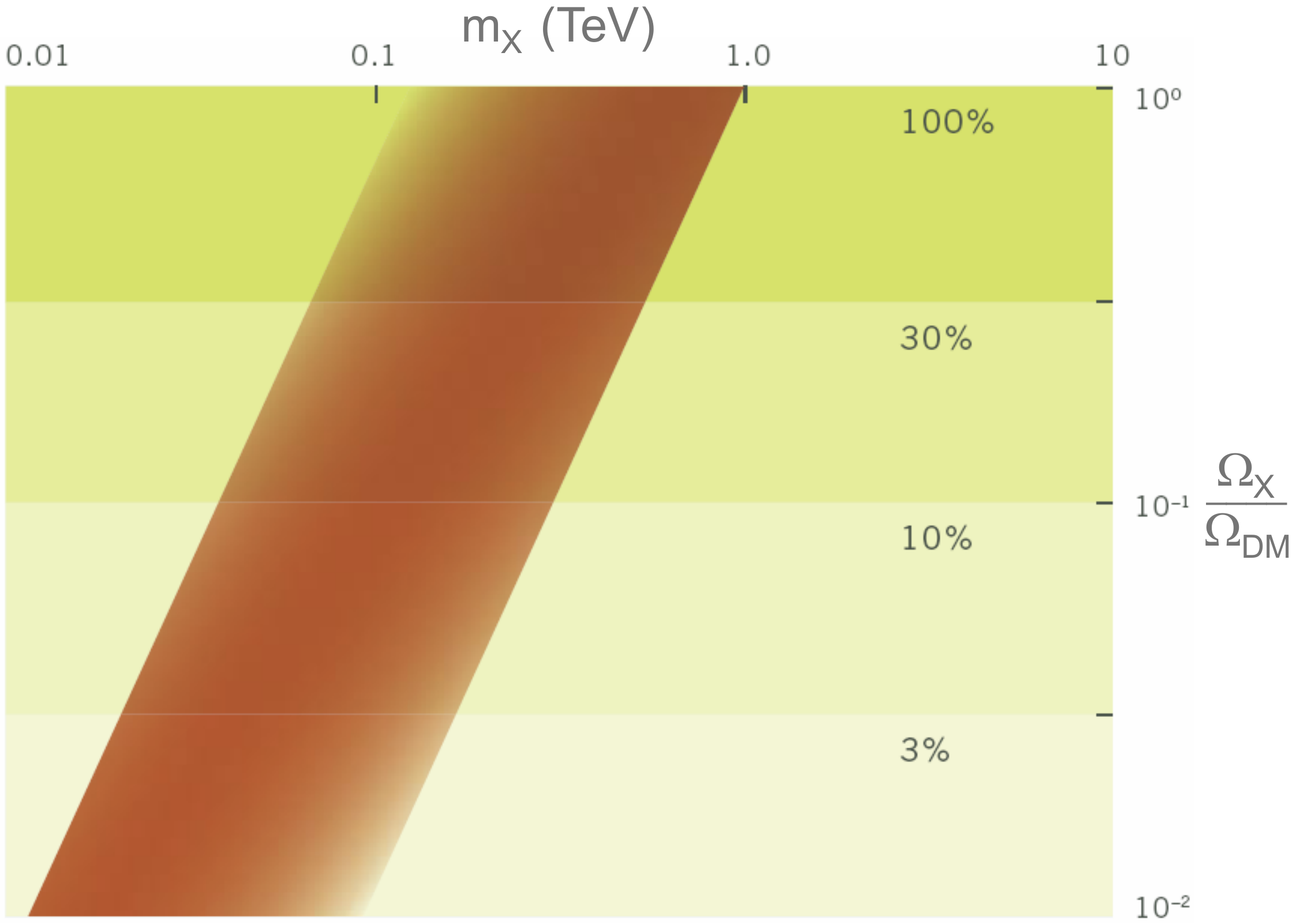}
\caption{A band of natural values in the $(m_X, \Omega_X/\OmegaDM)$
plane for a thermal relic $X$, where $\OmegaDM \simeq 0.23$ is the
required total dark matter density~\protect\cite{discovering}.}
\label{fig:mOmegaplane}
\end{figure}

With this parametrization, given a choice of $k$, the relic density is
determined as a function of $m_X$.  The results are shown in
\figref{mOmegaplane}.  The width of the band comes from considering
both $S$- and $P$-wave annihilation, and from letting $k$ vary from
$\frac{1}{2}$ to 2. We see that a particle that makes up all of dark
matter is predicted to have mass in the range $m_X \sim 100~\gev -
1~\tev$; a particle that makes up 10\% of dark matter has mass $m_X
\sim 30~\gev - 300~\gev$.  This is the WIMP miracle: weak-scale
particles make excellent dark matter candidates.  We have neglected
many details here, and there are models for which $k$ lies outside our
illustrative range, sometimes by as much as an order of magnitude or
two.  Nevertheless, the WIMP miracle implies that many models of
particle physics easily provide viable dark matter candidates, and it
is at present the strongest reason to expect that central problems in
particle physics and astrophysics may in fact be related.  Note also
that, for those who find the aesthetic nature of the gauge hierarchy
problem distasteful, the WIMP miracle independently provides a strong
motivation for new particles at the weak scale.

\subsubsection{STABILITY AND LEP'S COSMOLOGICAL LEGACY}
\label{sec:stability}

The entire discussion of \secref{wimpmiracle} assumes that the WIMP is
stable.  This might appear to be an unreasonable expectation; after
all, all particles heavier than a GeV in the SM decay on time scales
far shorter than the age of the Universe.

In fact, however, there are already indications that if new particles
exist at the weak scale, at least one of them should be stable.  This
is the cosmological legacy of LEP, the Large Electron-Positron
Collider that ran from 1989-2000.  Generically, new particles
introduced to solve the gauge hierarchy problem would be expected to
induce new interactions
\begin{equation}
\text{SM} \ \text{SM} \to \text{NP} \to \text{SM} \ \text{SM} \ , 
\label{fourpoint}
\end{equation}
where SM and NP denote standard model and new particles, respectively.
If the new particles are heavy, they cannot be produced directly, but
their effects may nevertheless be seen as perturbations on the
properties of SM particles.  LEP, along with the Stanford Linear
Collider, looked for the effects of these interactions and found none,
constraining the mass scale of new particles to be above $\sim 1 -
10~\tev$, depending on the SM particles involved (see, \eg,
\citet{Barbieri:1999tm}).  At the same time, to solve the gauge
hierarchy problem, the new particles cannot be decoupled completely.
At the very least, the new particles should modify the quantum
corrections to the Higgs boson mass given in
\eqref{higgsquantumcorrections}.  This implies that they must interact
with the Higgs boson through couplings
\begin{equation}
h \leftrightarrow \text{NP} \ \text{NP} \ ,
\label{higgsthreepoint}
\end{equation}
and their masses should not be significantly higher than $\mweak \sim
10~\gev - \tev$.

These apparently conflicting demands may be reconciled if there is a
conserved discrete parity that requires all interactions to involve an
even number of new particles~\cite{Wudka:2003se,Cheng:2003ju}.  Such a
conservation law would eliminate the problematic reactions of
\eqref{fourpoint}, while preserving the desired interactions of
\eqref{higgsthreepoint}.  As a side effect, the existence of a
discrete parity implies that the lightest new particle cannot decay.
The lightest new particle is therefore stable, as required for dark
matter.  Note that pair annihilation of dark matter particles is still
allowed.  The prototypical discrete parity is $R$-parity, proposed for
supersymmetry long before the existence of LEP
bounds~\cite{Farrar:1978xj}.  However, the existence of LEP
constraints implies that any new theory of the weak scale must
confront this difficulty.  The required discrete parity may be
realized in many ways, depending on the new physics at the weak scale;
an example in extra dimensions is given in \secref{KKWIMP}.

\subsubsection{IMPLICATIONS FOR DETECTION}
\label{sec:implications}

The WIMP miracle not only provides a model-independent motivation for
dark matter at the weak scale, but it also has strong implications for
how dark matter might be detected.  For WIMPs $X$ to have the observed
relic density, they must annihilate to other particles.  Assuming that
these other particles are SM particles, the necessity of $X X \to
\text{SM} \ \text{SM}$ interactions suggests three promising
strategies for dark matter detection:
\begin{itemize}
\setlength{\itemsep}{1pt}\setlength{\parskip}{0pt}\setlength{\parsep}{0pt}
\item Indirect detection: if dark matter annihilated in the early
Universe, it must also annihilate now through $X X \to \text{SM} \
\text{SM}$, and the annihilation products may be detected.
\item Direct detection: dark matter can scatter off normal matter
through $X \ \text{SM} \to X \ \text{SM}$ interactions, depositing
energy that may be observed in sensitive, low background detectors.
\item Particle colliders: dark matter may be produced at particle
colliders through $\text{SM} \ \text{SM} \to XX$.  Such events are
undetectable, but are typically accompanied by related production
mechanisms, such as $\text{SM} \ \text{SM} \to XX + \ \text{\{SM\}}$,
where ``\{SM\}'' denotes one or more standard model particles.  These
events are observable and provide signatures of dark matter at
colliders.
\end{itemize}

It is important to note that the WIMP miracle not only implies that
such dark matter interactions must exist, it also implies that the
dark matter-SM interactions must be efficient; although WIMPs may not
be a significant amount of the dark matter, they certainly cannot have
an energy density more than $\OmegaDM$. Cosmology therefore provides
{\em lower} bounds on interaction rates.  This fact provides highly
motivated targets for a diverse array of experimental searches that
may be able to detect WIMPs and constrain their properties.

To summarize, viable particle physics theories designed to address the
gauge hierarchy problem naturally (1) predict new particles with mass
$\sim \mweak$ that (2) are stable and (3) have the thermal relic
densities required to be dark matter.  The convergence of particle
physics and cosmological requirements for new states of matter has
motivated many new proposals for dark matter.  In the following
section, we discuss some prominent examples.

\subsection{Candidates}

\subsubsection{NEUTRALINOS}
\label{sec:neutralinos}

The gauge hierarchy problem is most elegantly solved by supersymmetry.
In supersymmetric extensions of the SM, every SM particle has a new,
as-yet-undiscovered partner particle, which has the same quantum
numbers and gauge interactions, but differs in spin by 1/2.  The
introduction of new particles with opposite spin-statistics from the
known ones supplements the SM quantum corrections to the Higgs boson
mass with opposite sign contributions, modifying
\eqref{higgsquantumcorrections} to
\begin{equation}
\Delta m_h^2 
\sim \left. \frac{\lambda^2}{16\pi^2} \int^{\Lambda} 
\frac{d^4 p}{p^2} \right|_{\text{SM}}
- \left. \frac{\lambda^2}{16\pi^2} \int^{\Lambda} 
\frac{d^4 p}{p^2} \right|_{\text{SUSY}}
\sim \frac{\lambda^2}{16\pi^2} 
\left(m_{\text{SUSY}}^2 - m_{\text{SM}}^2 \right)
\ln \frac{\Lambda}{m_{\text{SUSY}}} \ ,
\label{higgsquantumcorrectionsSUSY}
\end{equation}
where $m_{\text{SM}}$ and $m_{\text{SUSY}}$ are the masses of the SM
particles and their superpartners.  For $m_{\text{SUSY}} \sim \mweak$,
this is at most an ${\cal O}(1)$ correction, even for $\Lambda \sim
\mplanck$.  This by itself stabilizes, but does not solve, the gauge
hierarchy problem; one must also understand why $m_{\text{SUSY}} \sim
\mweak \ll \mplanck$.  There are, however, a number of ways to
generate such a hierarchy; for a review, see~\citet{Shadmi:1999jy}.
Given such a mechanism, the relation of
\eqref{higgsquantumcorrectionsSUSY} implies that quantum effects will
not destroy the hierarchy, and the gauge hierarchy problem may be
considered truly solved.

Not surprisingly, the doubling of the SM particle spectrum has many
implications for cosmology.  For dark matter, it is natural to begin
by listing all the new particles that are electrically neutral.  For
technical reasons, supersymmetric models require two Higgs bosons.
The neutral supersymmetric particles are then
\begin{eqnarray}
\text{Spin 3/2 Fermion:} && \text{Gravitino $\tilde{G}$} \\
\text{Spin 1/2 Fermions:} && \tilde{B}, \tilde{W}, \tilde{H}_u,
\tilde{H}_d \to \text{Neutralinos $\chi_1$, $\chi_2$, $\chi_3$, $\chi_4$} \\
\text{Spin 0 Scalars:} && \text{Sneutrinos $\tilde{\nu}_e$,
$\tilde{\nu}_\mu$, $\tilde{\nu}_\tau$} \ .
\end{eqnarray}
As indicated, the neutral spin 1/2 fermions mix to form four mass
eigenstates, the neutralinos.  The lightest of these, $\chi \equiv
\chi_1$, is a WIMP dark matter
candidate~\cite{Goldberg:1983nd,Ellis:1983ew}.  The sneutrinos are
{\em not} good dark matter candidates, as both their their
annihilation and scattering cross sections are large, and so they are
under-abundant or excluded by null results from direct detection
experiments for all masses near
$\mweak$~\cite{Falk:1994es,Arina:2007tm}.  The gravitino is not a
WIMP, but it is a viable and fascinating dark matter candidate, as
discussed in \secsref{superwimps}{hidden}.

A general supersymmetric extension of the SM contains many unknown
parameters.  To make progress, it is typical to consider specific
models in which simplifying assumptions unify many parameters, and
then study to what extent the conclusions may be generalized.  The
canonical model for supersymmetric studies is minimal supergravity,
which is minimal in the sense that it includes the minimum number of
particles and includes a large number of assumptions that drastically
reduces the number of independent model parameters.  Minimal
supergravity is defined by five parameters:
\begin{equation}
m_0, \mgaugino, A_0, \tan\beta, \sign(\mu) \ .
\end{equation}
The most important parameters are the universal scalar mass $m_0$ and
the universal gaugino mass $\mgaugino$, both defined at the scale of
grand unified theories $\mgut \simeq 2 \times 10^{16}~\gev$.  The
assumption of a universal gaugino mass and the choice of $\mgut$ are
supported by the fact that the three SM gauge couplings unify at
$\mgut$ in supersymmetric theories~\cite{Dimopoulos:1981yj}.  The
assumption of scalar mass unification is much more {\em ad hoc}, but
it does imply highly degenerate squarks and sleptons, which typically
satisfies the constraints of the new physics flavor problem.  Finally,
the parameter $A_0$ governs the strength of cubic scalar particle
interactions, and $\tan\beta$ and $\sign(\mu)$ are parameters that
enter the Higgs boson potential.  For all but their most extreme
values, these last three parameters have much less impact on collider
and dark matter phenomenology than $m_0$ and $\mgaugino$.

\begin{figure}[tbp]
\includegraphics[width=0.70\columnwidth]{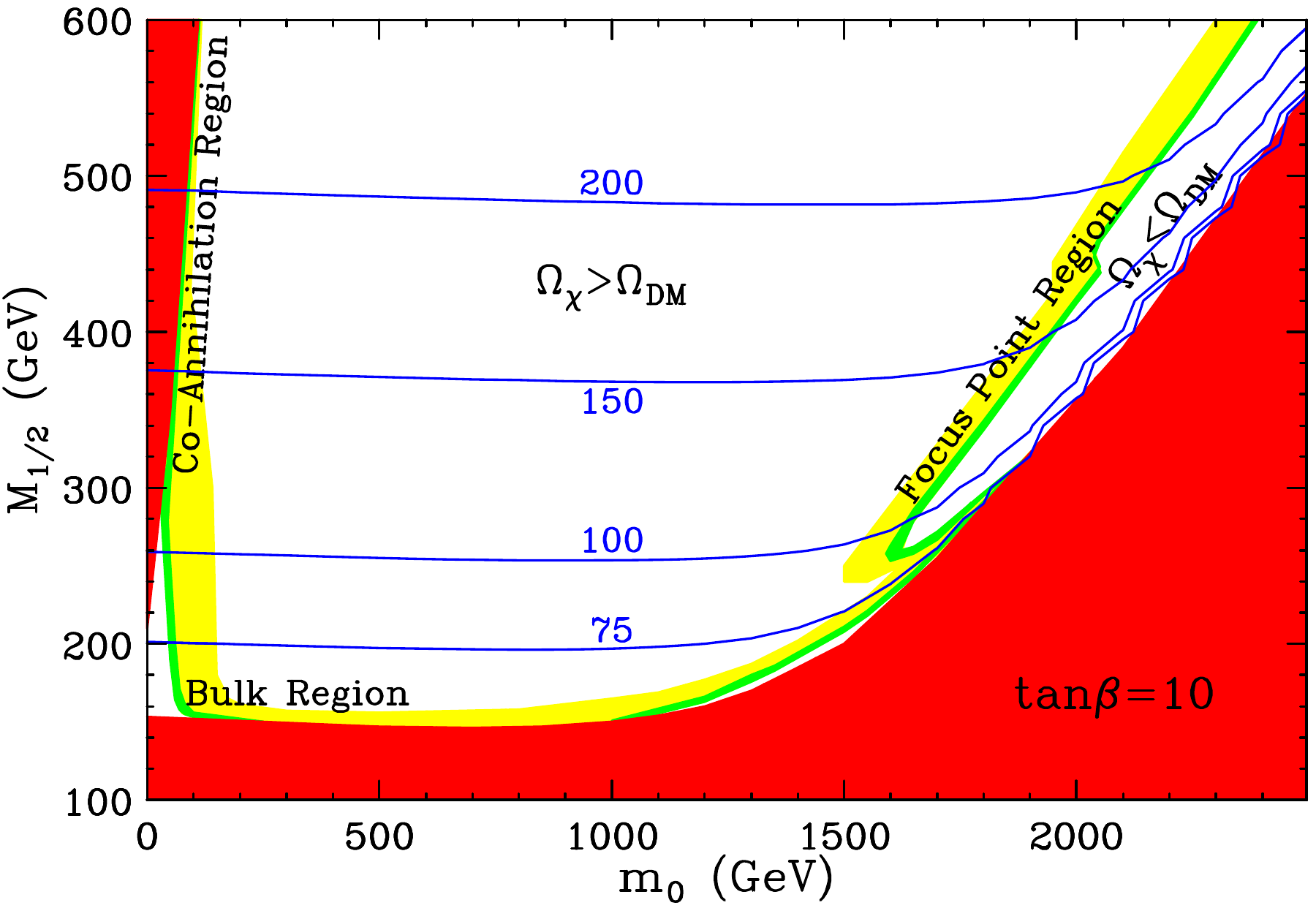}
\caption{Regions of minimal supergravity $(m_0, M_{1/2})$ parameter
space for fixed $A_0 = 0$, $\tan\beta = 10$, and $\mu > 0$. The green
(yellow) region is cosmologically favored with $0.20 < \Omega_{\chi} <
0.28$ ($0.2 < \Omega_{\chi} < 0.6$).  The names of
cosmologically-favored regions (focus point, bulk, and
co-annihilation) are indicated, along with regions with too much and
too little dark matter.  The lower right red shaded region is excluded
by collider bounds on chargino masses; the upper left red region is
excluded by the presence of a stable charged particle.  Contours are
for neutralino dark matter mass $m_{\chi}$ in GeV.  {}Adapted
from~\citet{Feng:2000zu}.
\label{fig:relicSUSY}}
\end{figure}

In the context of minimal supergravity, the thermal relic density is
given in the $(m_0, M_{1/2})$ plane for fixed values of $A_0$,
$\tan\beta$, and $\sign(\mu)$ in \figref{relicSUSY}.  We see that
current constraints on $\OmegaDM$ are highly constraining, essentially
reducing the cosmologically favored parameter space by one dimension.
The region of parameter space with the correct neutralino relic
density is further divided into three regions with distinct
properties: the bulk region, the focus point region, and the
co-annihilation region.  Of course, if one considers the full minimal
supergravity parameter space, other points in the $(m_0, M_{1/2})$
plane are possible (see, \eg, \citet{Trotta:2008bp}); notably, at
larger $\tan\beta$ there is another favored region, known as the
funnel region.

Note that for much of the region with $m_0, M_{1/2} \alt \tev$,
$\Omegachi$ is too large.  This is because neutralinos, although
widely studied, are very special: they are Majorana fermions, that is,
they are their own anti-particles.  If the initial state neutralinos
are in an $S$-wave state, the Pauli exclusion principle implies that
the initial state has total spin $S=0$ and total angular momentum
$J=0$.  Annihilation to fermion pairs with total spin $S=1$, such as
$e^-_R e^+_R$, is therefore $P$-wave suppressed, with an extra factor
of $v^2 \sim 0.1$ in the annihilation cross section.  As a result,
$\Omegachi$ is typically too large, and the correct $\Omegachi$ is
achieved for relatively light neutralinos, as evident in
\figref{relicSUSY}.

\subsubsection{KALUZA-KLEIN DARK MATTER}
\label{sec:KKWIMP}

An alternative possibility for new weak-scale physics is extra
dimensions.  The idea that there may be extra spatial dimensions is an
old one, going back at least as far as the work of Kaluza and Klein in
the 1920's~\cite{Klein:1926tv}. Their original idea is untenable, but
it has many modern descendants, of which the closest living relative
is universal extra dimensions (UED)~\cite{Appelquist:2000nn}.

In UED, all particles propagate in flat, compact extra dimensions of
size $10^{-18}~\m$ or smaller.  In the simplest UED model, minimal
UED, there is one extra dimension of size $R$ compactified on a
circle, with points with $y$ and $-y$ identified, where $y$ is the
coordinate of the extra dimension. Every SM particle has an infinite
number of partner particles, with one at every Kaluza-Klein (KK) level
$n$ with mass $\sim nR^{-1}$.  In contrast to supersymmetry, these
partner particles have the same spin.  As a result, UED models do not
solve the gauge hierarchy problem; in fact, their couplings become
large and non-perturbative at energies far below the Planck scale.
The motivation to consider UED models is that they provide an
interesting and qualitatively different alternative to supersymmetry,
but it assumes that UED are a low-energy approximation to a more
complete theory that resolves the gauge hierarchy problem and is
well-defined up to the Planck scale.

Minimal UED parameter space is extremely simple, as it is completely
determined by only two parameters: $m_h$, the mass of the SM Higgs
boson, and $R$, the compactification radius.  For the Higgs boson
mass, the direct search constraints on the SM also apply in UED and
require $m_h > 114.4~\gev$~\cite{Barate:2003sz}. However, the indirect
bounds are significantly weakened by the existence of many levels of
KK particles, and require only $m_h < 900~\gev \ (300~\gev)$ for
$R^{-1} = 250~\gev \ (1~\tev)$ at 90\% CL~\cite{Appelquist:2002wb}.

\begin{figure}[tbp]
\includegraphics[width=0.70\columnwidth]{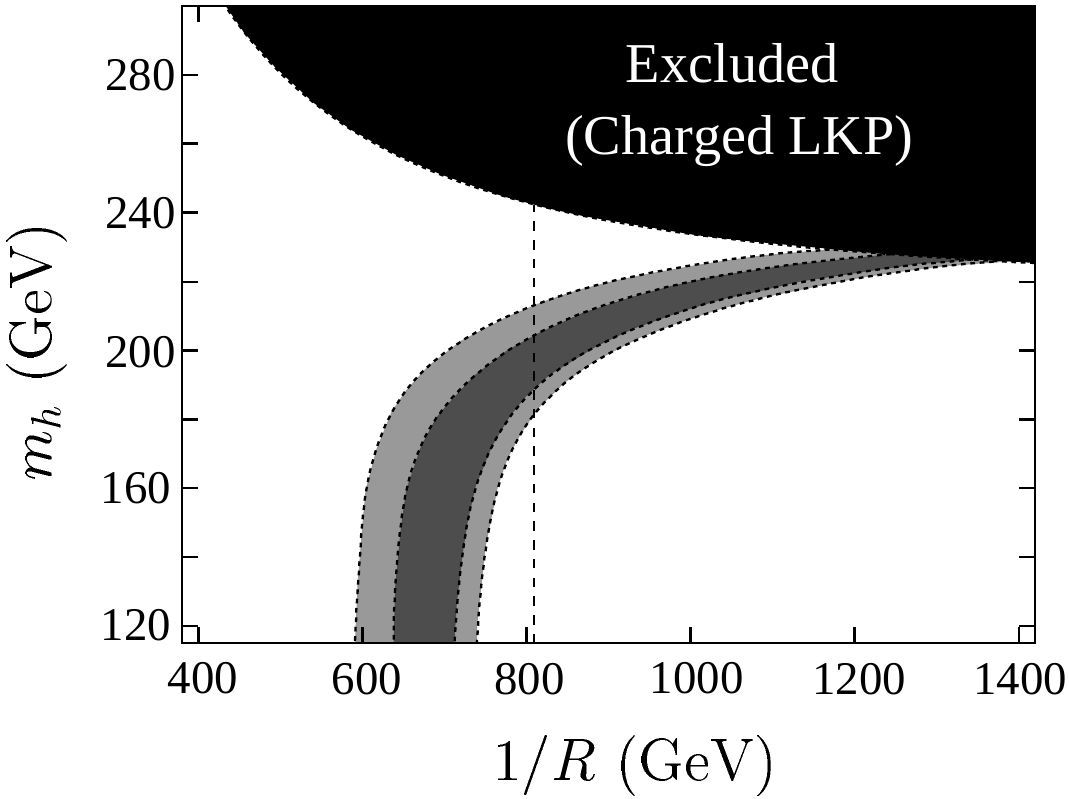}
\caption{Regions of minimal UED parameter space with the correct relic
density. The light (medium) shaded region has $\Omega_{B^1} h^2 =
0.099 \pm 0.020 \ (0.010)$.  The dark shaded region is excluded
because the LKP is charged.  {}From~\citet{Kakizaki:2006dz}.
\label{fig:relicKK}}
\end{figure}

The simplest UED models preserve a discrete parity known as KK-parity,
which implies that the lightest KK particle (LKP) is stable and a
possible dark matter candidate~\cite{Servant:2002aq,Cheng:2002ej}.
The lightest KK particle (LKP) is typically $B^1$, the level 1 partner
of the hypercharge gauge boson.  The regions of parameter space with
the correct $B^1$ thermal relic density have been investigated in a
series of increasingly sophisticated studies~\cite{Servant:2002aq,%
Burnell:2005hm,Kong:2005hn,Kakizaki:2006dz}; the end results are shown
in \figref{relicKK}.  As can be seen, the required LKP mass is
$600~\gev \alt m_{B^1} \alt 1.4~\tev$, a heavier range than for
neutralinos.  This is because LKPs annihilation is through $S$-wave
processes, and so is not $P$-wave suppressed, in contrast to
neutralinos.  Nevertheless, the required dark matter mass is still
$\sim 1~\tev$, as expected given the WIMP miracle.

\subsubsection{OTHERS}

Neutralinos are the prototypical WIMP, and KK dark matter provides an
instructive example of WIMPs that differ in important aspects from
neutralinos.  There are many other examples, however.  In the recent
years leading up to the start of the LHC, there has been a
proliferation of electroweak theories and an accompanying
proliferation of WIMP candidates.  These include branons in theories
with large extra dimensions~\cite{Cembranos:2003mr,Cembranos:2003fu},
$T$-odd particles in little Higgs
theories~\cite{Cheng:2003ju,Birkedal:2006fz}, and excited states in
theories with warped extra dimensions~\cite{Agashe:2004ci}.  As with
all WIMPs, these are astrophysically equivalent, in that they are
produced through thermal freeze out and are cold and collisionless,
but their implications for direct detection, indirect detection, and
particle colliders may differ significantly.

\subsection{Direct Detection}
\label{sec:direct}

As discussed in \secref{implications}, WIMP dark matter may be
detected by its scattering off normal matter through processes $X \
\text{SM} \to X \ \text{SM}$.  Given a typical WIMP mass of $m_X \sim
100~\gev$ and WIMP velocity $v \sim 10^{-3}$, the deposited recoil
energy is at most $\sim 100~\kev$, requiring highly sensitive, low
background detectors placed deep underground.  Such detectors are
insensitive to very strongly interacting dark matter, which would be
stopped in the atmosphere or earth and would be undetectable
underground.  However, such dark matter would be seen by rocket and
other space-borne experiments or would settle to the core of the
Earth, leading to other fascinating and bizarre implications.  Taken
together, a diverse quilt of constraints now excludes large scattering
cross sections for a wide range of dark matter
masses~\cite{Mack:2007xj,Albuquerque:2010bt}, and we may concentrate
on the weak cross section frontier probed by underground detectors.

\begin{figure}[tbp]
\includegraphics[width=0.70\columnwidth]{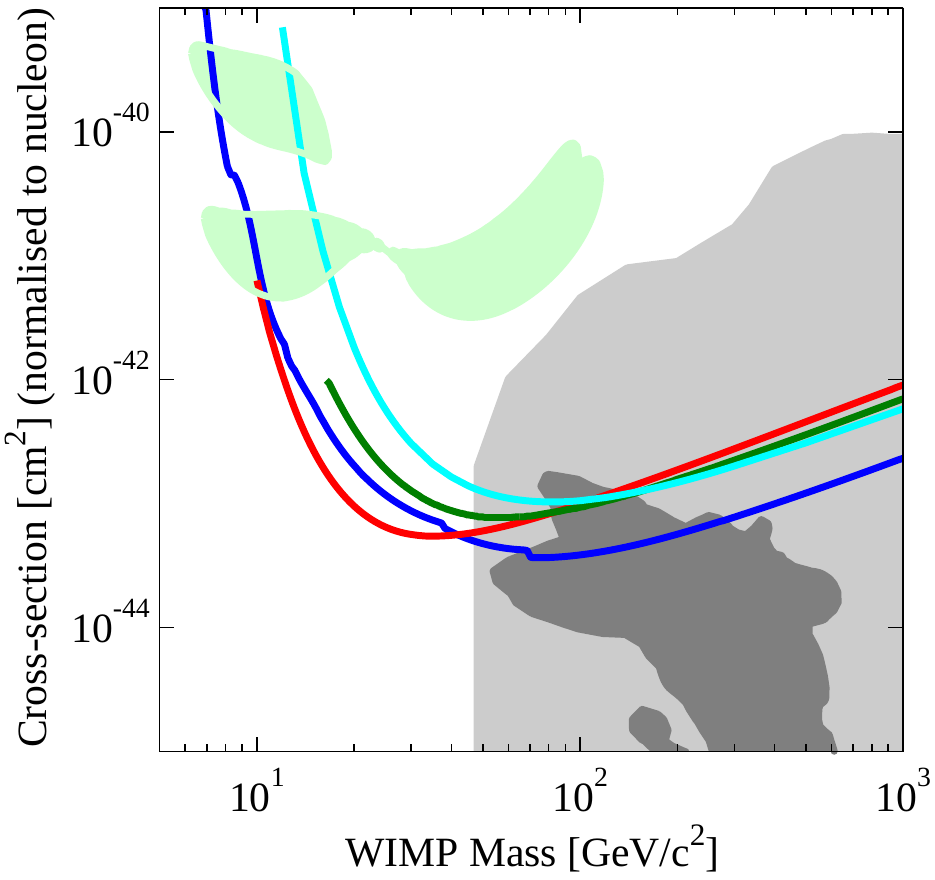}
\caption{Upper bounds on spin-independent WIMP-nucleon cross sections
$\sigmaSI$ from (top to bottom at 1 TeV WIMP mass)
XENON10~\cite{Angle:2007uj,Aprile:2008rc} (red), ZEPLIN
III~\cite{Lebedenko:2008gb} (green), EDELWEISS
II~\cite{Armengaud:2009hc} (light blue), and CDMS
II~\cite{Ahmed:2009zw} (blue), along with the combined $3\sigma$
favored regions (green shaded) from DAMA/LIBRA~\cite{Bernabei:2008yi}
with and without channeling~\cite{Savage:2008er}.  The lower left
shaded regions are predictions for neutralino dark matter in the
general minimal supersymmetric standard model~\cite{Kim:2002cy} (light
grey) and minimal supergravity~\cite{Trotta:2008bp} (dark grey).  Plot
produced by DM Tools~\cite{Gaitskell}.
\label{fig:directSUSY_SI}}
\end{figure}

\begin{figure}[tbp]
\includegraphics[width=0.70\columnwidth]{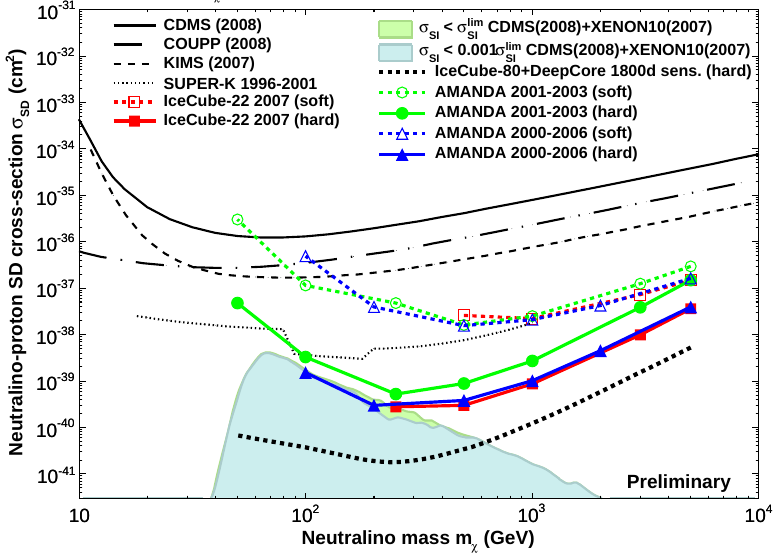}
\caption{Upper bounds on spin-dependent WIMP-proton cross sections
$\sigmaSD$ from CDMS~\cite{Ahmed:2008eu}, COUPP~\cite{Behnke:2008zza},
KIMS~\cite{Lee.:2007qn}, Super-K~\cite{Desai:2004pq}, and
IceCube~\cite{Abbasi:2009uz}, along with preliminary limits from
AMANDA~\cite{Braun:2009fr} and the projected 10 year sensitivity of
IceCube with DeepCore.  The shaded regions are predictions for
neutralino dark matter in the general minimal supersymmetric standard
model with $0.05 < \Omegachi h^2 < 0.20$.
{}From~\citet{Braun:2009fr}.
\label{fig:directSUSY_SD}}
\end{figure}

\begin{figure}[tbp]
\includegraphics[width=0.70\columnwidth]{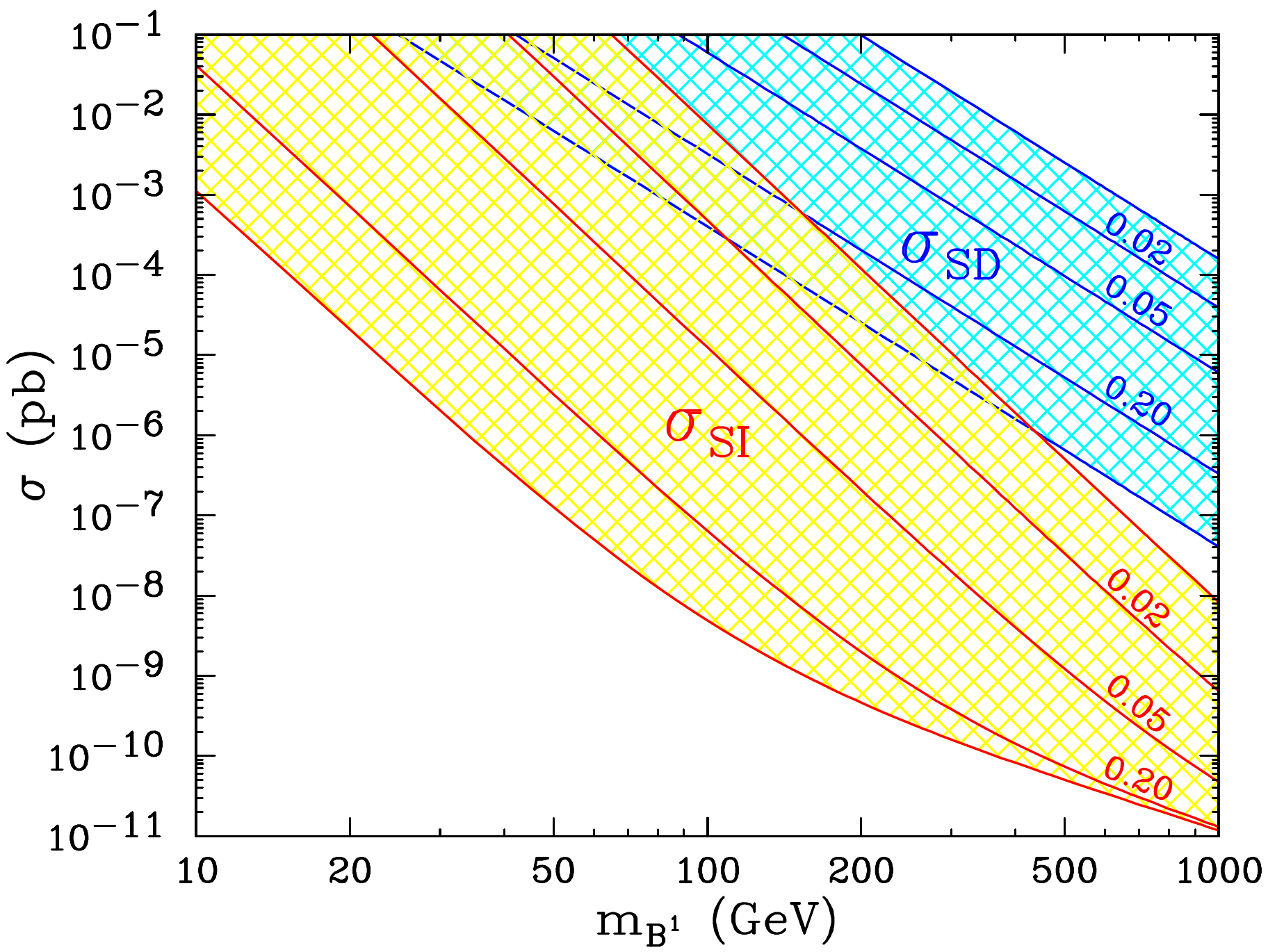}
\caption{Predicted spin-independent WIMP-nucleon cross sections (light
shaded, red) and spin-dependent WIMP-proton cross sections (dark
shaded, blue) in units of $\pb = 10^{-36}~\cm^2$ for $B^1$
Kaluza-Klein dark matter in universal extra dimensions with a
universal KK quark mass $m_{q^1}$.  The predictions are for $m_h =
120~\gev$ and $0.01 \le r = (m_{q^1} - m_{B^1} )/m_{B^1} \le 0.5$,
with contours for specific intermediate $r$ labeled.
{}From~\citet{Cheng:2002ej}.
\label{fig:directKK}}
\end{figure}

The field of direct detection is extremely active, with sensitivities
increasing by two orders of magnitude in the last decade and bright
prospects for continued rapid
improvement~\cite{DMSAG,Gaitskell:2004gd}.  WIMP scattering may be
through spin-independent couplings, such as interactions $\bar{X} X
\bar{q}q$, or spin-dependent couplings, such as interactions $\bar{X}
\gamma^{\mu} \gamma^5 X \bar{q} \gamma_{\mu} \gamma^5 q$, which reduce
to spin-spin couplings $S_X \cdot S_q$ in the non-relativistic
limit~\cite{Goodman:1984dc}.  The current state of affairs is
summarized in \figref{directSUSY_SI} for spin-independent searches and
\figref{directSUSY_SD} for spin-dependent searches.  These figures
also include scattering cross section predictions for neutralino dark
matter.  For comparison, the predictions for $B^1$ Kaluza-Klein dark
matter in UED for both spin-independent and spin-dependent cross
sections are given in \figref{directKK}.  These figures assume a
Maxwellian velocity distribution and local dark matter density of
$\rho = 0.3~\gev/\cm^3$; the impacts of halo modeling and Galactic
substructure on direct detection limits have been explored
by~\citet{Green:2002ht} and~\citet{Kamionkowski:2008vw}, respectively.

For spin-independent scattering, there are both an observed signal
from DAMA and null results from many other experiments. Putting aside
DAMA for the moment, as can be seen in
\figsref{directSUSY_SI}{directKK}, current bounds exclude some of the
parameter space of supersymmetry and UED, but do not test the bulk of
either parameter space. The experiments are improving rapidly,
however, and in the coming year, sensitivities to cross sections of
$\sigmaSI \sim 10^{-45} - 10^{-44}~\cm^2$ are possible.  

How significant will this progress be?  As evident in
\figref{directSUSY_SI}, supersymmetry predictions may be arbitrarily
small. However, many well-known supersymmetric theories predict
$\sigmaSI \sim 10^{-44}~\cm^2$.  In general, supersymmetric theories
suffer from the new physics flavor problem: the introduction of
squarks and sleptons with generic flavor mixing and weak scale masses
induces contributions to $K-\bar{K}$ mixing, $\mu \to e \gamma$, the
electric dipole moments of the neutron and electron, and a host of
other flavor- or CP-violating observables that badly violate known
constraints.  One generic solution to this problem is to assume heavy
squarks and sleptons, with masses above a TeV, so that they decouple
and do not affect low-energy observables.  This solution is realized
in the focus point region of minimal supergravity, and is also found
in many other models with greatly varying
motivations~\cite{Feng:1999mn,Feng:1999zg}.

These models have profound implications for dark matter searches.  In
general, the dominant contributions to neutralino annihilation are
$\chi \chi \to q \bar{q}, l \bar{l}$ through $t$-channel squarks and
sleptons, and $\chi \chi \to W^+ W^-, Z Z$ through $t$-channel
charginos and neutralinos.  In theories with decoupled squarks and
sleptons, the first class of processes are suppressed, and so
annihilation takes place through the second group, which depend
essentially only on the neutralino's mass and its Higgsino content.
The Higgsino content may be fixed by requiring the correct thermal
relic density.  In these models, then, the supersymmetry parameter
space is effectively reduced to one parameter, the $\chi$ mass.  More
detailed study shows that $\sigmaSI$ is almost independent of
$m_{\chi}$ and has a value near $10^{-44}~\cm^2$~\cite{Feng:2000gh}.

In the next year or so, then, direct detection will test all
supersymmetric scenarios with the correct relic density in which the
new physics flavor problem is solved by decoupled squarks and
sleptons. So far, direct detection experiments have trimmed a few
fingernails off the body of supersymmetry parameter space, but if
nothing is seen in the coming few years, it is arms and legs that will
have been lopped off.

In addition to the limits described above, the DAMA experiment
continues to find a signal in annual modulation~\cite{Drukier:1986tm}
with period and maximum at the expected values~\cite{Bernabei:2008yi}.
{}From a theorist's viewpoint, the DAMA/LIBRA result has been
puzzling, because the signal, if interpreted as spin-independent
elastic scattering, seemingly implied dark matter masses and
scattering cross sections that have been excluded by other
experiments.  Inelastic scattering, in which dark matter is assumed to
scatter through $X\ \text{SM} \to X' \ \text{SM}$, where $X'$ is
another new particle that is $\sim 100~\kev$ heavier than $X$, has
been put forward as one solution~\cite{TuckerSmith:2001hy}.  More
recently, astrophysics~\cite{Gondolo:2005hh} and
channeling~\cite{Drobyshevski:2007zj,Bernabei:2007hw}, a condensed
matter effect that effectively lowers the threshold for crystalline
detectors, have been proposed as possible remedies to allow elastic
scattering to explain DAMA without violating other constraints. If
these indications are correct, the favored parameters are $m_X \sim 1-
10~\gev$ and $\sigmaSI \sim 10^{-41} - 10^{-39}~\cm^2$.  This
interpretation is supported by unexplained events recently reported by
the CoGeNT direct detection search, which, if interpreted as a dark
matter signal, are best fit by dark matter with $m_X \sim 9~\gev$ and
$\sigmaSI \sim 6.7 \times 10^{-5}~\pb$~\cite{Aalseth:2010vx}.  This
mass is lower than typically expected for WIMPs, but even massless
neutralinos are allowed if one relaxes the constraint of gaugino mass
unification~\cite{Dreiner:2009ic}.  The cross section is, however,
very large; it may be achieved in corners of minimal supersymmetric
standard model (MSSM) parameter space~\cite{Bottino:2007qg}, but is
more easily explained in completely different frameworks, such as
those discussed in \secref{hidden}.

Spin-dependent scattering provides an independent method to search for
dark matter.  At the moment, the leading direct detection experiments,
such as CDMS, COUPP, and KIMS, are less promising in terms of probing
the heart of supersymmetric and UED WIMP parameter space, as seen in
\figsref{directSUSY_SD}{directKK}.  In addition, given some fairly
reasonable assumptions, indirect detection experiments looking for
dark matter annihilation to neutrinos in the Sun provide more
stringent constraints, as we discuss in the following section.

\subsection{Indirect Detection}
\label{sec:indirect}

After freeze out, dark matter pair annihilation becomes greatly
suppressed. However, even if its impact on the dark matter relic
density is negligible, dark matter annihilation continues and may be
observable.  Dark matter may therefore be detected indirectly: dark
matter pair-annihilates somewhere, producing something, which is
detected somehow.  There are many indirect detection methods being
pursued.  Their relative sensitivities are highly dependent on what
WIMP candidate is being considered, and the systematic uncertainties
and difficulties in determining backgrounds also vary greatly from one
method to another.

Searches for neutrinos are unique among indirect searches in that they
are, given certain assumptions, probes of scattering cross sections,
not annihilation cross sections, and so compete directly with the
direct detection searches described in \secref{direct}.  The idea
behind neutrino searches is the following: when WIMPs pass through the
Sun or the Earth, they may scatter and be slowed below escape
velocity.  Once captured, they then settle to the center, where their
densities and annihilation rates are greatly enhanced.  Although most
of their annihilation products are immediately absorbed, neutrinos are
not.  Some of the resulting neutrinos then travel to the surface of
the Earth, where they may convert to charged leptons through $\nu q
\to \ell q'$, and the charged leptons may be detected.

The neutrino flux depends on the WIMP density, which is determined by
the competing processes of capture and annihilation.  If $N$ is the
number of WIMPs captured in the Earth or Sun, $\dot{N} = C - A N^2$,
where $C$ is the capture rate and $A$ is the total annihilation cross
section times relative velocity per volume.  The present WIMP
annihilation rate is, then, $\Gamma_A \equiv A N^2 / 2 = C \tanh^2 (
\sqrt{CA} t_{\odot})/2$, where $t_{\odot} \simeq 4.5~\text{Gyr}$ is
the age of the solar system.  For most WIMP models, a very large
collecting body such as the Sun has reached equilibrium, and so
$\Gamma_A \approx C/2$.  The annihilation rate alone does not
completely determine the differential neutrino flux --- one must also
make assumptions about how the neutrinos are produced.  However, if
one assumes, say, that WIMPs annihilate to $b\bar{b}$ or $W^+W^-$,
which then decay to neutrinos, as is true in many neutralino models,
the neutrino signal is completely determined by the capture rate $C$,
that is, the scattering cross section.

Under fairly general conditions, then, neutrino searches are directly
comparable to direct detection.  The Super-Kamiokande, IceCube, and
AMANDA Collaborations have looked for excesses of neutrinos from the
Sun with energies in the range $10~\gev \alt E_{\nu} \alt
1~\tev$. Given the assumptions specified above, their null results
provide the leading bounds on spin-dependent scattering cross
sections, as seen in \figref{directSUSY_SD}.  These experiments are
just beginning to probe relevant regions of supersymmetric and UED
parameter space.

Neutrino searches are also sensitive to spin-independent cross
sections.  For typical WIMP masses, they are not competitive with
direct searches, but future neutrino searches at Super-Kamiokande may
have lower thresholds and so provide leading bounds on low mass WIMPs.
In this way, Super-Kamiokande may test the DAMA and CoGeNT signal
regions at high $\sigmaSI$ and $m_X \sim
1-10~\gev$~\cite{Hooper:2008cf,Feng:2008qn,Kumar:2009ws}.

\begin{figure}[tbp]
\includegraphics[width=0.48\columnwidth]{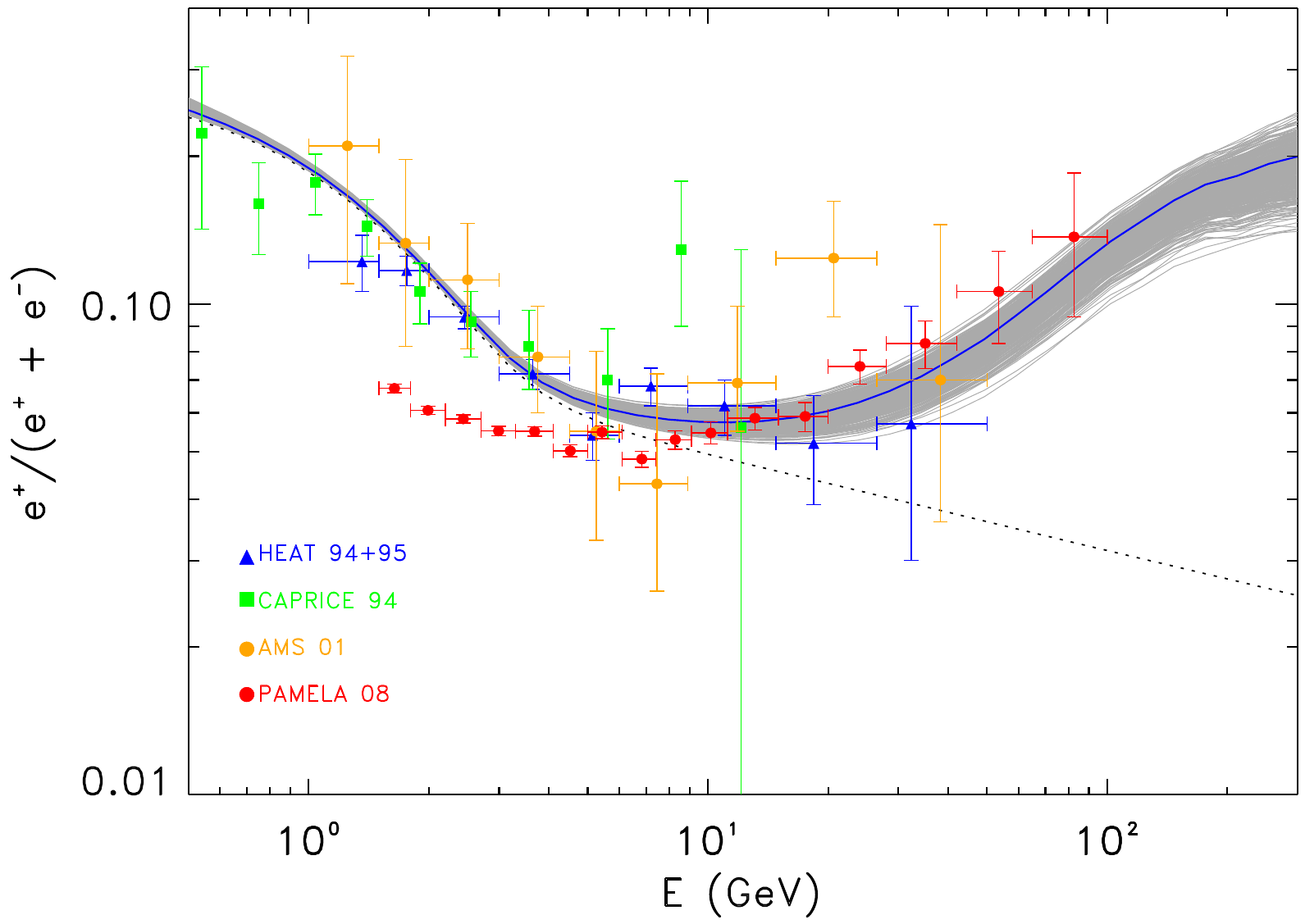}
\hfil
\includegraphics[width=0.48\columnwidth]{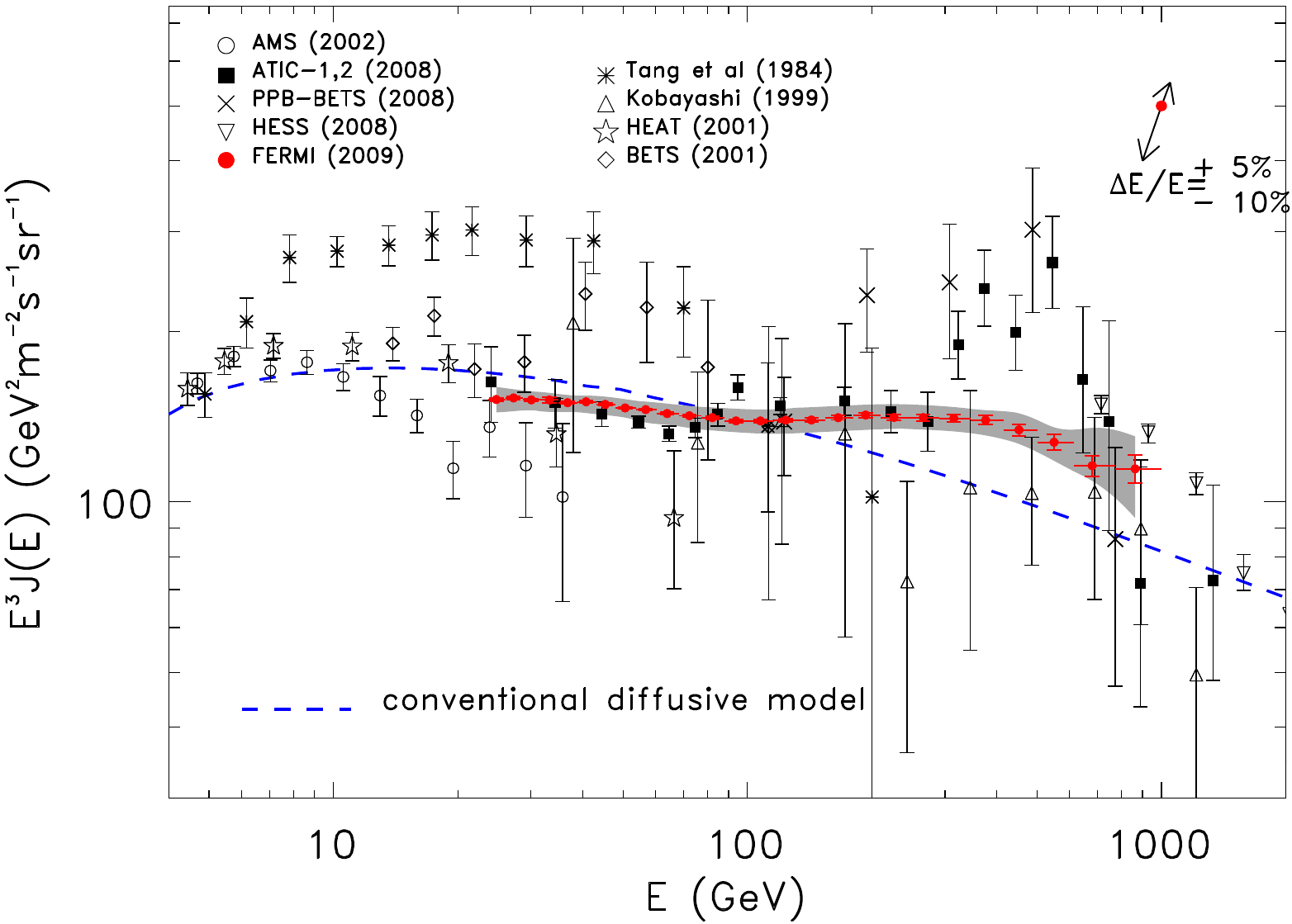}
\caption{Left: the cosmic positron fraction measured by PAMELA and
other experiments and the predictions of pulsars with various
parameters (grey contours)~\cite{Grasso:2009ma}. Discrepancies at
energies below 10 GeV are claimed to arise from solar modulation.
Right: the total $e^+ +e^-$ flux measured by ATIC, Fermi, and other
experiments~\cite{Abdo:2009zk}.  In both cases, the dashed contours
represent the predicted backgrounds from GALPROP~\cite{Strong:2009xj}.
\label{fig:fermi}}
\end{figure}

In addition to neutrinos, there are many other particles that may be
signals of dark matter annihilation.  In contrast to direct detection,
there have been many reported anomalies in indirect detection, and
some of these have been interpreted as possible evidence for dark
matter.  The most prominent recent example is the detection of
positrons and electrons with energies between 10 GeV and 1 TeV by the
PAMELA~\cite{Adriani:2008zr}, ATIC~\cite{:2008zzr}, and Fermi
LAT~\cite{Abdo:2009zk} Collaborations.  These data are shown in
\figref{fermi}, and reveal an excess above an estimate of the expected
background, as modeled by GALPROP~\cite{Strong:2009xj}.

These data have several plausible astrophysical explanations.  The
ATIC and Fermi experiments are unable to distinguish positrons from
electrons, and so constrain the total $e^+ + e^-$ flux.  As seen in
\figref{fermi}, the ATIC ``bump'' is not confirmed by the Fermi LAT
data, which has much higher statistics.  The Fermi data may be
explained by modifying the spectral index of the cosmic ray
background~\cite{Grasso:2009ma}.  This exacerbates the PAMELA
discrepancy, but the PAMELA data, with or without the modified
spectral index, is consistent with expectations from pulsars derived
both before~\cite{1989ApJ...342..807B,2001A&A...368.1063Z} and
after~\cite{Hooper:2008kg,Yuksel:2008rf,Profumo:2008ms} the PAMELA
data (see \figref{fermi}), and may also have other astrophysical
explanations~\cite{Dado:2009ux,Biermann:2009qi,Katz:2009yd}.

\begin{figure}[tbp]
\includegraphics[width=0.70\columnwidth]{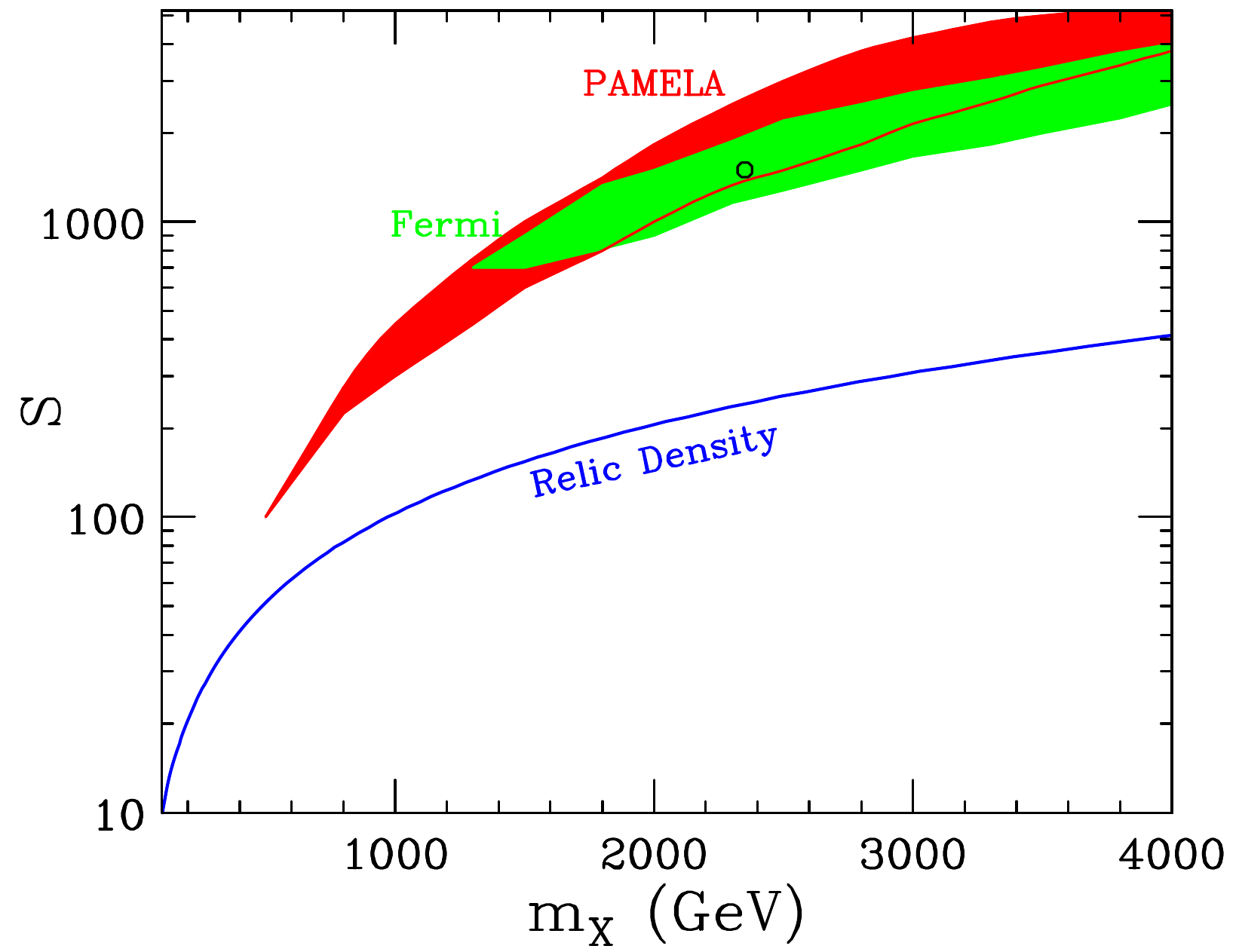}
\caption{The annihilation cross section enhancement factor $S$
required to explain the PAMELA and Fermi data, as a function of dark
matter mass $m_X$ (shaded regions)~\cite{Bergstrom:2009fa}, along with
upper bounds on $S$ from the requirement that the dark matter have the
right thermal relic density. {}From~\citet{Feng:2009hw}.
\label{fig:Sbounds}}
\end{figure}

Despite the astrophysical explanations, one may explore the
possibility that the positron excesses arise from dark matter
annihilation.  The energies of the excess, around $\mweak$, are as
expected for WIMPs.  Unfortunately, the observed fluxes are far larger
than expected for generic WIMPs.  For a WIMP annihilating through
$S$-wave processes to have the desired thermal relic density, its
annihilation cross section must be $\sigmath \equiv \langle \sigma_A v
\rangle \approx 3 \times 10^{-26}~\cm^3/\s$.  This must be enhanced by
two or three orders of magnitudes to explain the positron data, as
shown in \figref{Sbounds}.  Astrophysical boosts from substructure are
unable to accommodate such large enhancements, and so one must look to
particle physics.

A seemingly attractive solution is to postulate that dark matter
interacts with a light force carrier $\phi$ with fine structure
constant $\alpha_X \equiv \lambda^2/(4\pi)$.  For massless $\phi$, this
enhances the annihilation cross section by the Sommerfeld enhancement
factor
\begin{equation}
S = \frac{\pi \, \alpha_X / \vrel}{1 - e^{-\pi \alpha_X / \vrel}} \ ,
\label{Sbar}
\end{equation}
an effect first derived for the case of $e^+e^-$
annihilation~\cite{Sommerfeld:1931}. For massive $\phi$, $S$ is
typically cut off at a value $\propto \alpha_X m_X/
m_{\phi}$~\cite{Hisano:2002fk,Cirelli:2007xd,Cirelli:2008pk,%
ArkaniHamed:2008qn}.  The relative velocity of colliding dark matter
particles is $\vrel \sim 1/3$ at freeze out and $\vrel \sim 10^{-3}$
now.  The Sommerfeld enhancement therefore provides an elegant
mechanism for boosting annihilations now.  The case $m_\phi=0$ is
excluded by constraints from dark matter annihilation in protohalos
with $\vrel \sim 10^{-8}$~\cite{Kamionkowski:2008gj}.  However, taking
$m_X \sim \tev$ and $m_\phi \sim \mev - \gev$, and assuming $\langle
\sigma_A \vrel \rangle \approx \sigmath$, one may seemingly still
generate $S \sim 10^3$ to explain the positron excesses, while the
cutoff allows one to satisfy the protohalo constraint.

Unfortunately, for the annihilation cross section for $XX \to \phi
\phi$ to give the correct relic density, $\alpha_X$ cannot be too
large, which bounds $S$ from above.  Even ignoring the $S$ cutoff for
massive $\phi$, the resulting constraints exclude the possibility
that, with standard astrophysical assumptions, Sommerfeld enhancement
alone can explain the PAMELA and Fermi excesses, as shown in
\figref{Sbounds}. For particular choices of $\alpha_X$, $m_X$ and
$m_\phi$, $S$ may in fact be resonantly enhanced, but these
enhancements also reduce the thermal relic
density~\cite{Dent:2009bv,Zavala:2009mi}; including the effect on the
relic density in fact increases the discrepancy significantly.

There are other proposed dark matter explanations: for example, the
annihilation cross section may be boosted by resonances from states
with mass $\sim 2 m_X$~\cite{Feldman:2008xs,Ibe:2008ye,%
Guo:2009aj,Ibe:2009mk}, or the dark matter may be produced not by
thermal freeze out, but by decays~\cite{Arvanitaki:2008hq}.  At
present, however, the dark matter explanations are considerably more
exotic than the astrophysical ones.  Additional data from, for
example, Fermi and the Alpha Magnetic Spectrometer (AMS), an
anti-matter detector to be placed on the International Space Station,
may be able to distinguish the various proposed explanations for the
positron excesses, as well as be sensitive to canonical WIMP models,
but it remains to be seen whether the astrophysical backgrounds may be
sufficiently well understood for these experiments to realize their
dark matter search potential.

In addition to neutrinos from the Sun and positrons from the galactic
halo, there are several other promising indirect detection search
strategies.  Searches for anti-protons and anti-deuterons from WIMP
annihilation in the galactic halo provide complementary searches, as
they are sensitive to dark matter candidates that annihilate primarily
to quarks.  In addition, searches for gamma rays by space-based
experiments, such as Fermi and AMS, and by ground-based atmospheric
Cherenkov telescopes are also promising.  The most striking gamma ray
signal would be mono-energetic photons from $XX \to \gamma \gamma$,
but since WIMPs cannot be charged, these processes are typically
loop-induced or otherwise highly suppressed. More commonly, gamma rays
are produced when WIMPs annihilate to other particles, which then
radiate photons, leading to a smooth distribution of gamma ray
energies.  On the other hand, photons point back to their source,
providing a powerful diagnostic.  Possible targets for gamma ray
searches are the center of the Galaxy, where signal rates are high but
backgrounds are also high and potentially hard to estimate, and dwarf
galaxies, where signal rates are lower, but backgrounds are also
expected to be low.

\subsection{Particle Colliders}
\label{sec:collidersWIMPs}

If WIMPs are the dark matter, what can colliders tell us?  Given the
energy of the LHC and the requirement that WIMPs have mass $\sim
\mweak$ and interact through the weak force, WIMPs will almost
certainly be produced at the LHC.  Unfortunately, direct WIMP
production of $XX$ pairs is invisible.  The next best targets are
mono-jet or mono-photon signals from $XXj$ and $XX\gamma$ production,
respectively, where the jet $j$ or photon comes from initial state
radiation.  At the International Linear Collider, a proposed high
energy $e^+ e^-$ collider, such signals can been disentangled from
background, using the fact that the initial state particles have
definite energy and may be polarized, which provides a useful
diagnostic~\cite{Birkedal:2004xn}.  Unfortunately, these features are
missing at hadron colliders, where the initial state protons have
fixed energy but the quarks and gluons do not.  As a result, at the
Tevatron and LHC, the mono-jet and mono-photon signals are completely
obscured by backgrounds such as $Zj$ and $Z\gamma$ followed by $Z \to
\nu \bar{\nu}$~\cite{Feng:2005gj}.

Searches for dark matter at the LHC therefore rely on indirect
production.  For example, in supersymmetry, the LHC will typically
produce pairs of squarks and gluinos.  These will then decay through
some cascade chain, eventually ending up in neutralino WIMPs, which
escape the detector.  Their existence is registered through the
signature of missing energy and momentum, a signal that is a staple of
searches for physics beyond the SM.

Although the observation of missing particles is consistent with the
production of dark matter, it is far from compelling evidence.  The
observation of missing particles only implies that a particle was
produced that was stable enough to exit the detector, typically
implying a lifetime $\tau \agt 10^{-7}~\s$, a far cry from the
criterion $\tau \agt 10^{17}~\s$ required for dark matter.

Clearly more is needed.  In the last few years, there has been a great
deal of progress in this direction.  The main point of this progress
has been to show that colliders can perform detailed studies of new
physics, and this can constrain the dark matter candidate's properties
so strongly that the candidate's thermal relic density can be
precisely determined.  The consistency of this density with the
cosmologically observed density would then be strong evidence that the
particle produced at colliders is, in fact, the cosmological dark
matter.

This approach is analogous to the well-known case of BBN.  For BBN,
data from nuclear physics experiments stringently constrain cross
sections involving the light nuclei.  Along with the assumption of a
cooling and expanding Universe, this allows one to predict the light
element abundances left over from the Big Bang, and the consistency of
these predictions with observations gives us confidence that the light
elements were actually created in this way.  For dark matter, the idea
is that particle physics experiments at the LHC may stringently
constrain cross sections involving dark matter and related particles.
Along with the assumption of a cooling and expanding Universe, this
microscopic data allows one to predict the dark matter relic density,
basically by following the relic density curves of \figref{freezeout}.
This thermal relic density may be compared to the observed density of
dark matter, and their consistency would give us confidence that dark
matter is actually produced in this way and is made of the particles
produced at the collider.

\begin{figure}[tbp]
\centering
\includegraphics[width=0.70\columnwidth]{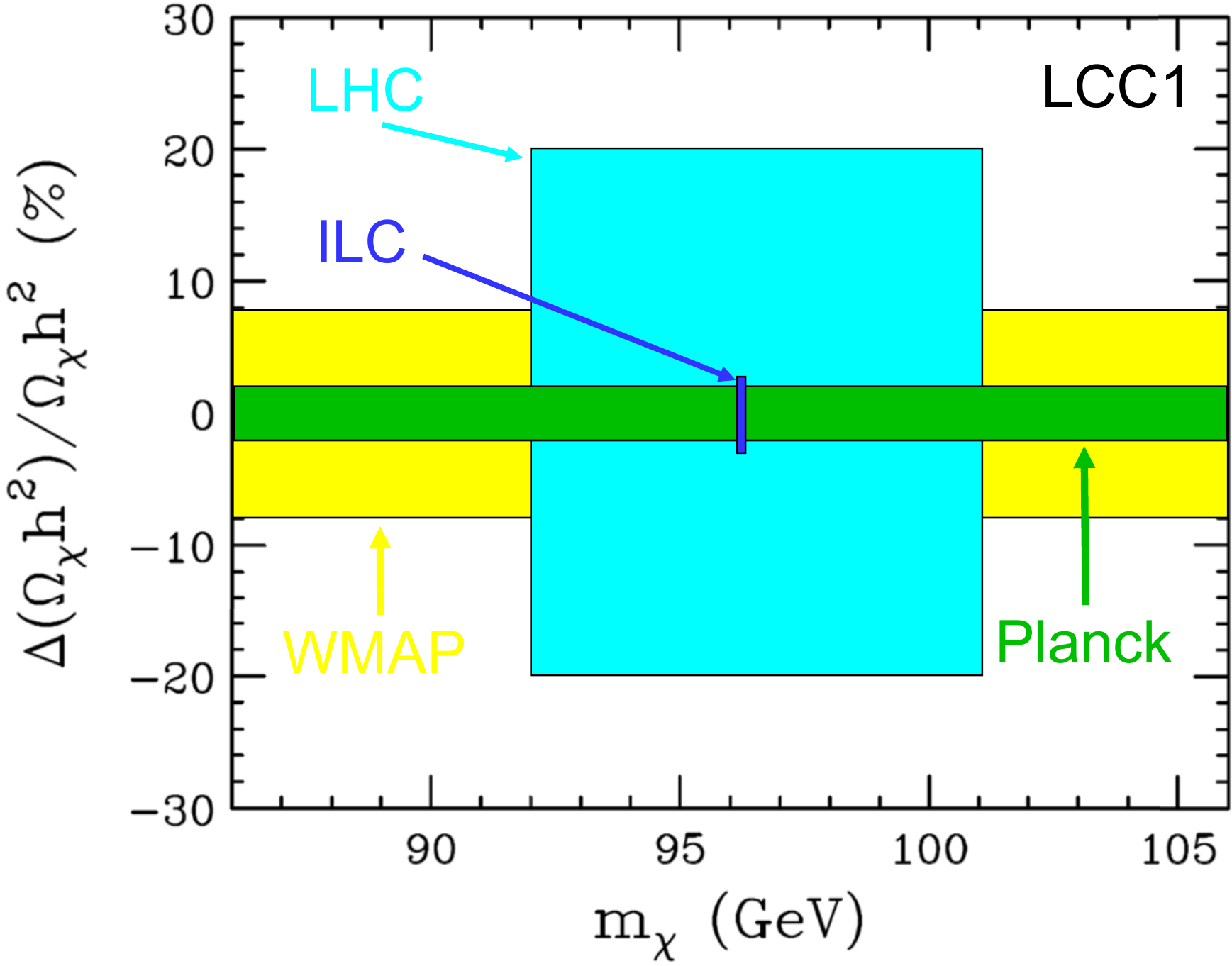}
\caption{Constraints in the $(\mchi, \Omegachi)$ plane from the LHC
and the International Linear Collider, and from the WMAP and Planck
satellites~\cite{Feng:2005nz,Baltz:2006fm}. WMAP and Planck measure
$\Omegachi$, but are insensitive to the dark matter mass $\mchi$; the
collider experiments bound both.  These results are for LCC1, a
supersymmetric model with minimal supergravity parameters $(m_0,
M_{1/2}, \tan\beta, A_0, \sign(\mu)) = (100~\gev, 250~\gev, 10,
-100~\gev, +)$. }
\label{fig:feng_omegaconstraints}
\end{figure}

How well can the LHC do?  The answer depends sensitively on the
underlying dark matter scenario, but several qualitatively different
cases have now been studied~\cite{Allanach:2004xn,Moroi:2005nc,%
Birkedal:2005jq,Baltz:2006fm}.  The results of one (exemplary) case
study are given in \figref{feng_omegaconstraints}.  In conjunction
with other cosmological observations, the WMAP satellite constrains
the dark matter relic density $\Omegachi$ to a fractional uncertainty
of $\pm 8\%$.  Its successor, Planck, is expected to sharpen this to
$\pm 2\%$.  Of course, CMB experiments do not constrain the dark
matter mass.  At the same time, precision studies at the LHC can
determine so many of the supersymmetric model parameters that the
neutralino mass can be determined to $\pm 5~\gev$ and the thermal
relic density can be predicted to $\pm 20\%$.  Measurements at the
International Linear Collider could improve these constraints on mass
and relic density to $\pm 50~\mev$ and $\pm 3\%$, respectively.

Consistency between the particle physics predictions and the
cosmological observations would provide compelling evidence that the
particle produced at the LHC is in fact dark matter.  Along the way,
the colliders will also determine the dark matter's mass, spin, and
many other properties.  In this way, colliders may finally help solve
the question of the microscopic identity of dark matter.  Note also
that, just as BBN gives us confidence that we understand the Universe
back to 1 second after the Big Bang and temperatures of $1~\mev$, such
dark matter studies will provide a window on the era of dark matter
freeze out at 1 nanosecond after the Big Bang and temperatures of
$\sim 10~\gev$.  Of course, the thermal relic density prediction from
colliders and the cosmological observations need not be consistent.
In this case, there are many possible lines of inquiry, depending on
which is larger.

\section{SUPERWIMPS}
\label{sec:superwimps}

The WIMP miracle might appear to require that dark matter have weak
interactions if its relic density is naturally to be in the right
range.  This is not true, however --- in recent years, two other
mechanisms have been shown to be viable and lead to dark matter
particles that also exploit the WIMP miracle to have the correct relic
density, but have vastly different interactions and implications for
detection.  These two possibilities are the topics of this section and
\secref{hidden}.

In this section, we discuss superWIMPs, superweakly-interacting
massive particles, which have the desired relic density, but have
interactions that are much weaker than weak.  The extremely weak
interactions of SuperWIMPs might naively be thought to be a nightmare
for searches for dark matter.  In fact, superWIMP scenarios predict
signals from cosmic rays, at colliders, and in astrophysics that can
be far more striking than in WIMP scenarios, making superWIMPs highly
amenable to experimental investigation.

\subsection{Production Mechanisms}

\subsubsection{DECAYS}

\begin{figure}[tbp]
\includegraphics[width=0.98\columnwidth]{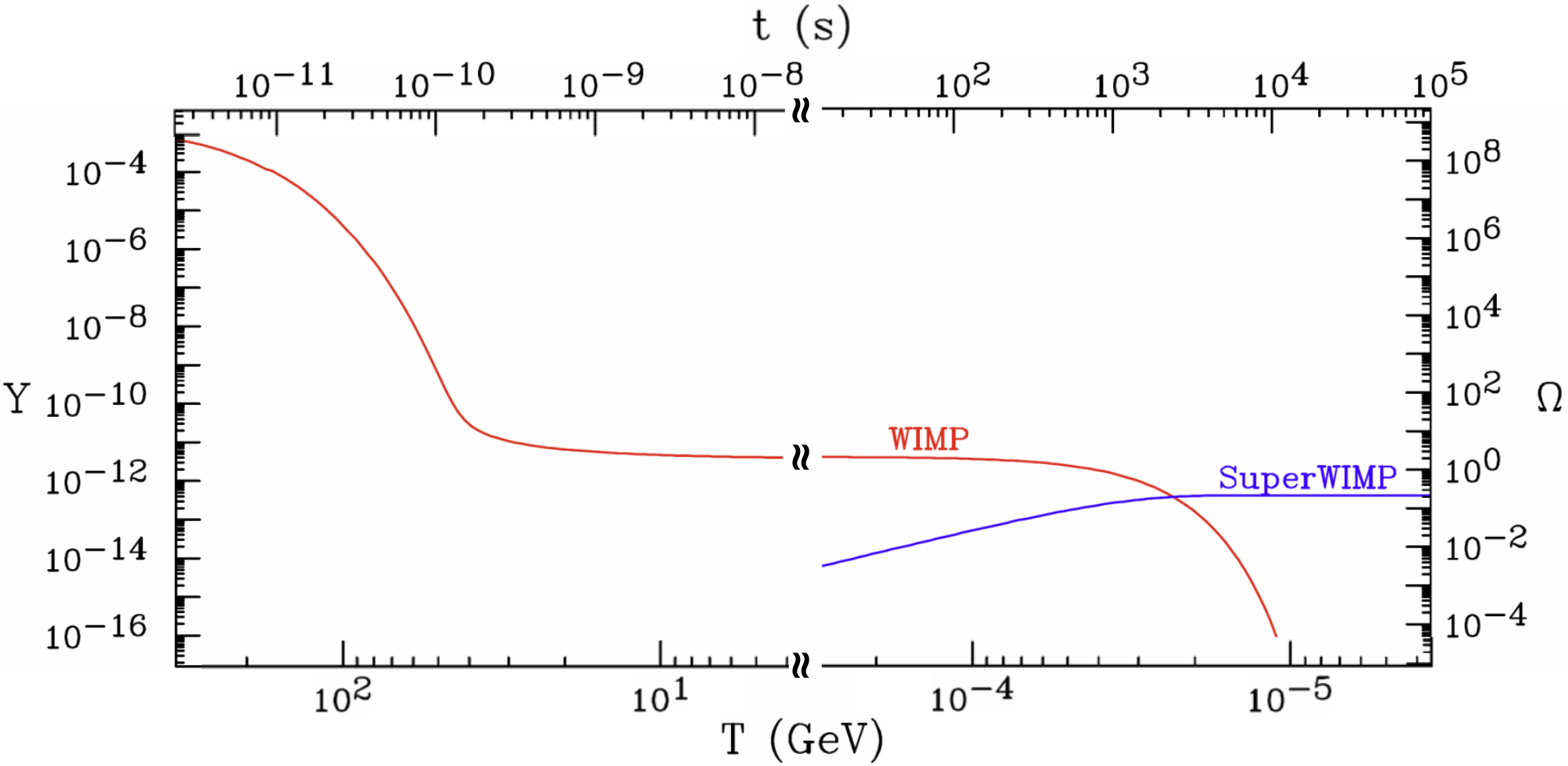}
\caption{In superWIMP scenarios, WIMPs freeze out as usual, but then
decay to superWIMPs, superweakly-interacting massive particles that
are the dark matter. This figure shows the WIMP comoving number
density $Y$ (left) and the superWIMP relic density (right) as
functions of temperature $T$ (bottom) and time $t$ (top). The WIMP is
a 1 TeV, $S$-wave annihilating particle with lifetime $10^3~\s$, and
the superWIMP has mass 100 GeV.  }
\label{fig:freezeout_swimp}
\end{figure}

In the superWIMP framework, dark matter is produced in late decays:
WIMPs freeze out as usual in the early Universe, but later decay to
superWIMPs, which form the dark matter that exists
today~\cite{Feng:2003xh,Feng:2003uy}.  Because superWIMPs are very
weakly interacting, they have no impact on WIMP freeze out in the
early Universe, and the WIMPs decouple, as usual, with a thermal relic
density $\Omega_{\text{WIMP}} \sim \OmegaDM$.  Assuming that each WIMP
decay produces one superWIMP, the relic density of superWIMPs is
\begin{equation}
\Omega_{\text{SWIMP}} = \frac{m_{\text{SWIMP}}}{m_{\text{WIMP}}}
\Omega_{\text{WIMP}} \ .
\label{superWIMPomega}
\end{equation}
SuperWIMPs therefore inherit their relic density from WIMPs, and for
$m_{\text{SWIMP}} \sim m_{\text{WIMP}}$, the WIMP miracle also implies
that superWIMPs are produced in the desired amount to be much or all
of dark matter.  The evolution of number densities is shown in
\figref{freezeout_swimp}.  The WIMP decay may be very late by particle
physics standards.  For example, if the superWIMP interacts only
gravitationally, as is true of many well-known candidates, the natural
timescale for WIMPs decaying to superWIMP is $1/(G_N \mweak^3) \sim
10^3 - 10^7~\s$.

Because the WIMP is unstable and not the dark matter, it need not be
neutral in this context --- to preserve the naturalness of the
superWIMP relic density, all that is required is $\Omega_{\text{WIMP}}
\sim \OmegaDM$.  In the case of supersymmetry, for example, the WIMP
may be a neutralino, but it may also be a charged slepton.  Even
though charged sleptons interact with photons, on dimensional grounds,
their annihilation cross sections are also necessarily governed by the
weak scale, and so are $\sim \gweak^4/\mweak^2$, implying roughly the
same relic densities as their neutral counterparts.  Of course,
whether the WIMP is charged or not determines the properties of the
other particles produced in WIMP decay, which has strong consequences
for observations, as we will see below.

\subsubsection{REHEATING}

SuperWIMPs may also be produced after reheating, when the energy of
the inflaton potential is transferred to SM and other particles.  This
creates a hot thermal bath, and, if the temperature is high enough,
may be a significant source of superWIMPs~\cite{Krauss:1983ik,%
Nanopoulos:1983up,Khlopov:1984pf,Ellis:1984eq,Juszkiewicz:1985gg}.

After reheating, the Universe is characterized by three rates: the
interaction rate of particles that have SM interactions,
$\sigma_{\text{SM}} n$; the expansion rate, $H$; and the rate of
interactions involving a superWIMP, $\sigma_{\text{SWIMP}} n$.  Here
$n$ is the number density of SM particles.  After reheating, the
Universe is expected to have a temperature well below the Planck
scale, but still well above SM masses.  Assuming the superWIMP is
gravitationally interacting, dimensional analysis implies that these
rates are hierarchically separated:
\begin{equation}
\sigma_{\text{SM}} n \sim T \gg H \sim \frac{T^2}{\mplanck} \gg 
\sigma_{\text{SWIMP}} n \sim \frac{T^3}{\mplanck^2} \ .
\end{equation}
This implies a very simple picture: after reheating, particles with SM
interactions are in thermal equilibrium with each other.  Occasionally
these interact to produce a superWIMP.  The produced superWIMPs then
propagate through the Universe essentially without interacting and
without annihilating, surviving to the present day.

To determine the superWIMP abundance, we turn once again to the
Boltzmann equation of \eqref{Boltzmann}, where now $n$ is the number
density of superWIMPs. To be concrete, consider the case of gravitino
superWIMPs.  In this case, the source term $\nequ^2$ arises from
interactions that produce superWIMPs, such as $g g \to \tilde{g}
\gravitino$.  In contrast to our previous application of the Boltzmann
equation in \secref{wimpmiracle}, however, here the sink term $n^2$ is
negligible.  Changing variables with $t \to T$ and $n \to Y \equiv
n/s$, where $s$ is the entropy density, we find
\begin{equation}
\frac{dY}{dT} = - \frac{ \langle \sigma_{\gravitino} v \rangle }
{H T s} n^2 \ .
\end{equation}
The right-hand side is independent of $T$, since $n \propto T^3$, $H
\propto T^2$ and $s \propto T^3$.  We thus find an extremely simple
relation --- the superWIMP relic number density is linearly
proportional to the reheating temperature $T_R$.

The constant of proportionality is the gravitino production cross
section.  Including all such production mechanisms, the gravitino
relic density can be determined as a function of reheating temperature
$T_R$ and gravitino mass.  For gravitino mass $m_{\gravitino} \sim
100~\gev$, the constraint on $\OmegaDM$ requires $T_R \alt
10^{10}~\gev$~\cite{Bolz:2000fu}.  Of course, if this bound is nearly
saturated, gravitinos produced after reheating may be a significant
component of dark matter, adding to the relic density from decays.
  
\subsection{Candidates}

\subsubsection{WEAK-SCALE GRAVITINOS}

The superWIMP scenario is realized in many particle physics models.
The prototypical superWIMP is the gravitino
$\gravitino$~\cite{Feng:2003xh,Feng:2003uy,Ellis:2003dn,%
Buchmuller:2004rq,Wang:2004ib,Roszkowski:2004jd}.  Gravitinos are the
spin 3/2 superpartners of gravitons, and they exist in all
supersymmetric theories.  The gravitino's mass is
\begin{equation}
m_{\gravitino} = \frac{F}{\sqrt{3} \mstar} \ ,
\label{gravitinomass}
\end{equation}
where $F$ is the supersymmetry-breaking scale squared and $\mstar = (8
\pi G_N)^{-1/2} \simeq 2.4 \times 10^{18}~\gev$ is the reduced Planck
mass.  In the simplest supersymmetric models, where supersymmetry is
transmitted to SM superpartners through gravitational
interactions, the masses of SM superpartners are
\begin{equation}
\tilde{m} \sim \frac{F}{\mstar} \ .
\label{tildem}
\end{equation}
A solution to the gauge hierarchy problem requires $F \sim
(10^{11}~\gev)^2$, and so all superpartners and the gravitino have
weak-scale masses.  The precise ordering of masses depends on unknown,
presumably ${\cal O}(1)$, constants in \eqref{tildem}. There is no
theoretical reason to expect the gravitino to be heavier or lighter
than the lightest SM superpartner, and so in roughly ``half'' of the
parameter space, the gravitino is the lightest supersymmetric particle
(LSP).  Its stability is guaranteed by $R$-parity conservation, and
since $m_{\gravitino} \sim \tilde{m}$, the gravitino relic density is
naturally $\Omega_{\text{SWIMP}} \sim \OmegaDM$.

In gravitino superWIMP scenarios, the role of the decaying WIMP is
played by the next-to-lightest supersymmetric particle (NLSP), a
charged slepton, sneutrino, chargino, or neutralino.  The gravitino
couples SM particles to their superpartners through
gravitino-sfermion-fermion interactions
\begin{equation}
L = - \frac{1}{\sqrt{2} \mstar} \partial_{\nu} \tilde{f} \, \bar{f}
\, \gamma^{\mu} \gamma^{\nu} \tilde{G}_{\mu} 
\label{gravfermion}
\end{equation}
and gravitino-gaugino-gauge boson interactions
\begin{equation}
L = - \frac{i}{8 \mstar}
\bar{\tilde{G}}_{\mu} \left[ \gamma^{\nu}, \gamma^{\rho} \right]
\gamma^{\mu}\tilde{V} F_{\nu\rho} \ .
\label{gravgauge}
\end{equation}
The presence of $\mstar$ in \eqsref{gravfermion}{gravgauge} implies
that gravitinos interact only gravitationally, a property dictated by
the fact that they are the superpartners of gravitons.  These
interactions determine the NLSP decay lifetime.  As an example, if the
NLSP is a stau, a superpartner of the tau lepton, its lifetime is
\begin{equation}
 \tau(\tilde{\tau} \to \tau \tilde{G}) = \frac{6}{G_N}
 \frac{m_{\tilde{G}}^2}{m_{\tilde{\tau}}^5}
 \left[1 -\frac{m_{\tilde{G}}^2}{m_{\tilde{\tau}}^2} \right]^{-4} 
 \approx 3.6\times 10^7~\s
\left[\frac{100~\gev}{m_{\tilde{\tau}} - m_{\gravitino}}\right]^4
\left[\frac{m_{\tilde{G}}}{100~\gev}\right]\ ,
\label{sfermionwidth}
\end{equation}
where the approximate expression holds for $\mgravitino /
m_{\tilde{\tau}} \approx 1$.  We see that decay lifetimes of the order
of hours to months are perfectly natural.  At the same time, the
lifetime is quite sensitive to the underlying parameters and may be
much longer for degenerate $\tilde{\tau}-\gravitino$ pairs, or much
shorter for light gravitinos.

\subsubsection{OTHERS}

In addition to gravitinos, other well-motivated examples of superWIMPs
include axinos~\cite{Rajagopal:1990yx,Covi:1999ty,Covi:2001nw}, the
supersymmetric partners of axions, particles introduced to resolve the
strong CP problem described in \secref{strongcpproblem}.  Axions may
also be cold dark matter (see \secref{axions}), and the possibility
that both axions and axinos contribute to dark matter is one of the
better motivated multi-component dark matter
scenarios~\cite{Baer:2009vr}.  UED models also have superWIMP
candidates in the form of Kaluza-Klein graviton and axion
states~\cite{Feng:2003nr}. One of these may be the lightest KK state;
in fact, in minimal UED, the KK graviton is the lightest KK state for
all $R^{-1} < 800~\gev$, where $R$ is the compactification
radius~\cite{Cembranos:2006gt}. If one of these is the lightest KK
state, it is stabilized by KK-parity conservation and has properties
very similar to its supersymmetric analogue.  Other superWIMPs
candidates include quintessinos in supersymmetric
theories~\cite{Bi:2003qa}, stable particles in models that
simultaneously address the problem of baryon
asymmetry~\cite{Kitano:2005ge}, and other particles produced in
decays, where the decay time is greatly lengthened by mass
degeneracies~\cite{Strigari:2006jf,Cembranos:2008bw}.

In summary, there are many superWIMP candidates that possess all of
the key virtues of WIMPs: they exist in the same well-motivated
frameworks and, since they inherit their relic density from WIMPs, are
also naturally produced with the desired relic density.  As we will
see, however, they have completely different implications for
detection.

\subsection{Indirect Detection}

SuperWIMPs are so weakly interacting that they cannot be detected by
direct searches, and their annihilation cross sections are so
suppressed that their annihilation signal rates are completely
negligible.  However, if the decaying WIMP is charged, the superWIMP
scenario implies long-lived charged particles, with striking
implications for indirect detection.

One interesting possibility is that long-lived charged particles may
be produced by ultra-high energy cosmic rays, resulting in exotic
signals in cosmic ray and cosmic neutrino
experiments~\cite{Albuquerque:2003mi,Bi:2004ys}.  As an example, in
the gravitino superWIMP scenario with a stau NLSP, ultra-high energy
neutrinos may produce events with two long-lived staus through $\nu q
\to \tilde{\tau} \tilde{q}'$ followed by the decay $\tilde{q}' \to
\tilde{\tau}$.  The metastable staus then propagate to neutrino
telescopes~\cite{Huang:2006ie}, where they have a typical transverse
separation of hundreds of meters.  They may therefore be detected
above background as events with two upward-going, extremely high
energy charged tracks in experiments such as IceCube.

\subsection{Particle Colliders}
\label{sec:colliderssuperwimps}

Particle colliders may also find evidence for superWIMP scenarios.
This evidence may come in one of two forms.  Collider experiments may
see long-lived, charged particles.  Given the stringent bounds on
charged dark matter, such particles presumably decay, and their decay
products may be superWIMPs.  Alternatively, colliders may find
seemingly stable particles, but the precision studies described in
\secref{collidersWIMPs} may imply that these particles have a thermal
relic density that is too large.  These two possibilities are not
mutually exclusive.  In fact, the discovery of long-lived, charged
particles with too-large predicted relic density is a distinct
possibility and would strongly motivate superWIMP dark matter.

As an example, again consider gravitino superWIMPs with a charged
slepton NLSP.  (The possibility of a neutralino NLSP is essentially
excluded by considerations of BBN, as discussed in \secref{BBN}.)  In
the slepton NLSP scenario, supersymmetric events at colliders are not
characterized by missing energy and momentum as predicted in WIMP dark
matter scenarios, but rather by pairs of heavy, long-lived, charged
particles. Such particles lead to spectacular signals and require far
less luminosity for discovery than missing energy
signals~\cite{Drees:1990yw,Rajaraman:2006mr}.

To determine the superWIMP relic density of \eqref{superWIMPomega}, we
must determine the superWIMP's mass.  This is not easy, since the WIMP
lifetime may be very large, implying that superWIMPs are typically
produced long after the WIMPs have escaped collider detectors.  At the
same time, because some sleptons will be slowly moving and
highly-ionizing, they may be trapped and studied.  As an example,
sleptons may be trapped in water tanks placed outside collider
detectors.  These water tanks may then be drained periodically to
underground reservoirs where slepton decays can be observed in quiet
environments.  The number of sleptons that may be trapped is
model-dependent, but a 1 meter thick tank of water may trap as many as
a thousand sleptons per year~\cite{Feng:2004yi}.  Other possibilities
for capturing sleptons include using the LHC detectors themselves as
the slepton traps~\cite{Hamaguchi:2004df}, or carefully tracking
sleptons as they exit the detector and digging them out of the walls
of the detector halls, giving new meaning to the phrase ``data
mining''~\cite{DeRoeck:2005bw}.

If thousands of sleptons are trapped, the slepton lifetime may be
determined to the few percent level simply by counting the number of
slepton decays as a function of time.  The slepton mass will be
constrained by analysis of the collider event kinematics. Furthermore,
the outgoing lepton energy can be measured, and this provides a high
precision measurement of the gravitino mass, and therefore a
determination of the gravitino relic density through
\eqref{superWIMPomega}.  Consistency at the percent level with the
observed dark matter relic density will provide strong evidence that
dark matter is indeed composed of gravitino superWIMPs.

Perhaps as interesting, the determination of $\tau$, $\mgravitino$,
and $m_{\tilde{l}}$ in \eqref{sfermionwidth} implies that one can
differentiate the various superWIMP
candidates~\cite{Brandenburg:2005he} and determine Newton's constant
on the scale of fundamental
particles~\cite{Buchmuller:2004rq,Feng:2004gn}.  According to
conventional wisdom, particle colliders are insensitive to gravity,
since it is such a weak force.  We see that this is not true --- if
$G_N$ enters in a decay time, one can achieve the desired sensitivity
simply by waiting a long time.  In this case, one can measure the
force of gravity between two test particles with masses $\sim
10^{-27}~\kg$, a regime that has never before been probed.  If this
force is consistent with gravity, these studies will show that the
newly discovered particle is indeed interacting gravitationally, as is
required for the gravitino to be the graviton's superpartner, and
demonstrate that gravity is in fact extended to supergravity in
nature.

\subsection{Astrophysical Signals}

Because superWIMPs are very weakly interacting, the decays of WIMPs to
superWIMPs may be very late and have an observable effects on BBN and
the CMB. In addition, in contrast to WIMPs, superWIMP dark matter may
behave as warm dark matter.

\subsubsection{COSMIC MICROWAVE BACKGROUND}
\label{sec:CMB}

\begin{figure}[tbp]
\includegraphics[width=0.70\columnwidth]{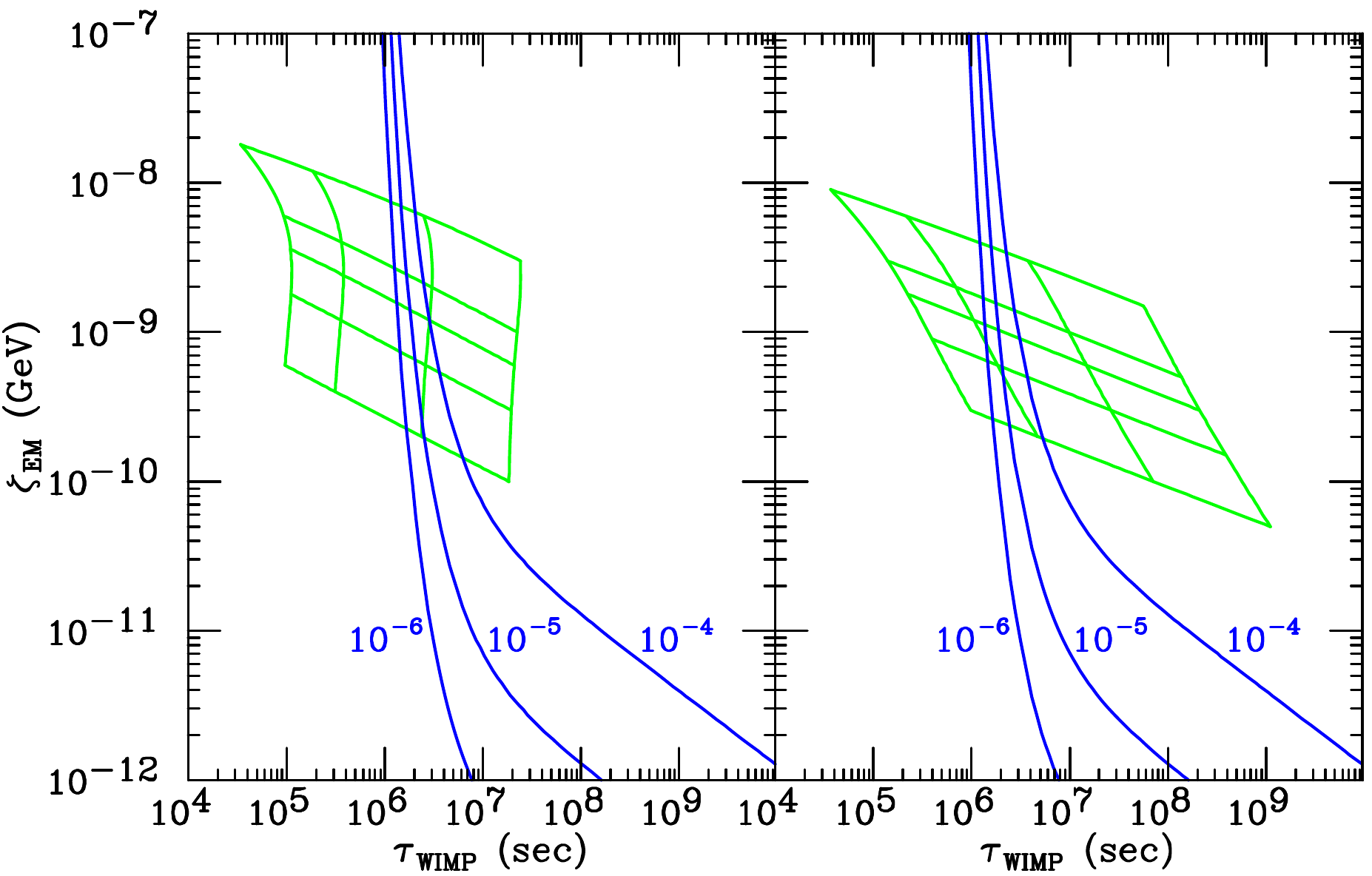}
\caption{Predicted values of WIMP lifetime $\tau$ and electromagnetic
energy release $\zetaEM \equiv \epsEM Y_{\text{NLSP}}$ in the $\Bino$
(left) and $\stau$ (right) superWIMP scenarios for $\mSWIMP =
100~\gev$, $300~\gev$, $500~\gev$, $1~\tev$, and $3~\tev$ (top to
bottom) and $\Delta m \equiv \mWIMP - \mSWIMP = 600~\gev$, $400~\gev$,
$200~\gev$, and $100~\gev$ (left to right).  The contours are for
$\mu$, which parameterizes the distortion of the CMB from a Planckian
spectrum.  {}From~\citet{Feng:2003uy}.
\label{fig:mu} }
\end{figure}

When WIMPs decay to superWIMPs, the accompanying particles may distort
the frequency dependence of the CMB away from its ideal black body
spectrum~\cite{Hu:1993gc,Feng:2003uy,Lamon:2005jc}.  The impact on the
CMB is determined by the NLSP lifetime and the energy released in
visible decay products when the WIMP decays.  The energy release is
conveniently expressed in terms of
\begin{equation}
\zeta_{\text{EM}} \equiv \epsilon_{\text{EM}} Y_{\text{NLSP}}
\label{eq:xi_EM}
\end{equation}
for electromagnetic energy, with a similar expression for hadronic
energy.  Here $\epsilon_{\text{EM}}$ is the initial EM energy released
in NLSP decay and $Y_{\text{NLSP}} \equiv n_{\text{NLSP}}/n_{\gamma}$
is the NLSP number density just before it decays, normalized to the
background photon number density $n_{\gamma} = 2 \zeta(3) T^3 /
\pi^2$.  Once the NLSP is specified, and assuming superWIMPs from late
decays make up all of the dark matter, with $\Omega_{\gravitino} =
\Omega_{\text{DM}} = 0.23$, both the lifetime and energy release are
determined by only two parameters: $m_{\gravitino}$ and
$m_{\text{NLSP}}$.  The results for neutralino and slepton NLSPs are
given in \figref{mu}.

For the decay times of interest with redshifts $z \sim 10^5$ to
$10^7$, the resulting photons interact efficiently through $\gamma e^-
\to \gamma e^-$ and $e X \to e X \gamma$, where $X$ is an ion, but
photon number is conserved, since double Compton scattering $\gamma
e^- \to \gamma \gamma e^-$ is inefficient.  The spectrum therefore
relaxes to statistical but not thermodynamic equilibrium, resulting in
a Bose-Einstein distribution function
\begin{equation}
f_{\gamma}(E) = \frac{1}{e^{E/(kT) + \mu} - 1} \ ,
\end{equation}
with chemical potential $\mu \ne 0$.  Figure \ref{fig:mu} also
includes contours of $\mu$.  The current bound is $\mu < 9\times
10^{-5}$~\cite{Fixsen:1996nj}. Although there are at present no
indications of deviations from black body, current limits are already
sensitive to the superWIMP scenario, and future improvements may
further probe superWIMP parameter space.

\subsubsection{BIG BANG NUCLEOSYNTHESIS}
\label{sec:BBN}

Late time energy release after $t \sim 1~\s$ also destroys and creates
light elements, potentially distorting the predictions of standard
BBN.  The impact depends sensitively on what the decaying NLSP is.
For example, in the neutralino NLSP case, the neutralino decays
generically through $\chi \to Z \gravitino, h \gravitino, \gamma
\gravitino$.  The first two modes lead to hadrons, which are very
dangerous.  Constraints from BBN on hadronic energy release
essentially exclude the neutralino WIMP scenario, allowing only the
fine-tuned case in which the neutralino is photino-like, $\chi \approx
\tilde{\gamma}$, in which case the decays to $Z \gravitino$ and $h
\gravitino$ are suppressed~\cite{Feng:2004zu,Feng:2004mt}.  In the
charged slepton NLSP case, the decaying WIMP may first bind with
nuclei, which may enhance the effect of its decays on
BBN~\cite{Pospelov:2006sc,Kohri:2006cn,Kaplinghat:2006qr,%
Kawasaki:2007xb,Takayama:2007du}.

\begin{figure}[tbp]
\includegraphics[width=0.70\columnwidth]{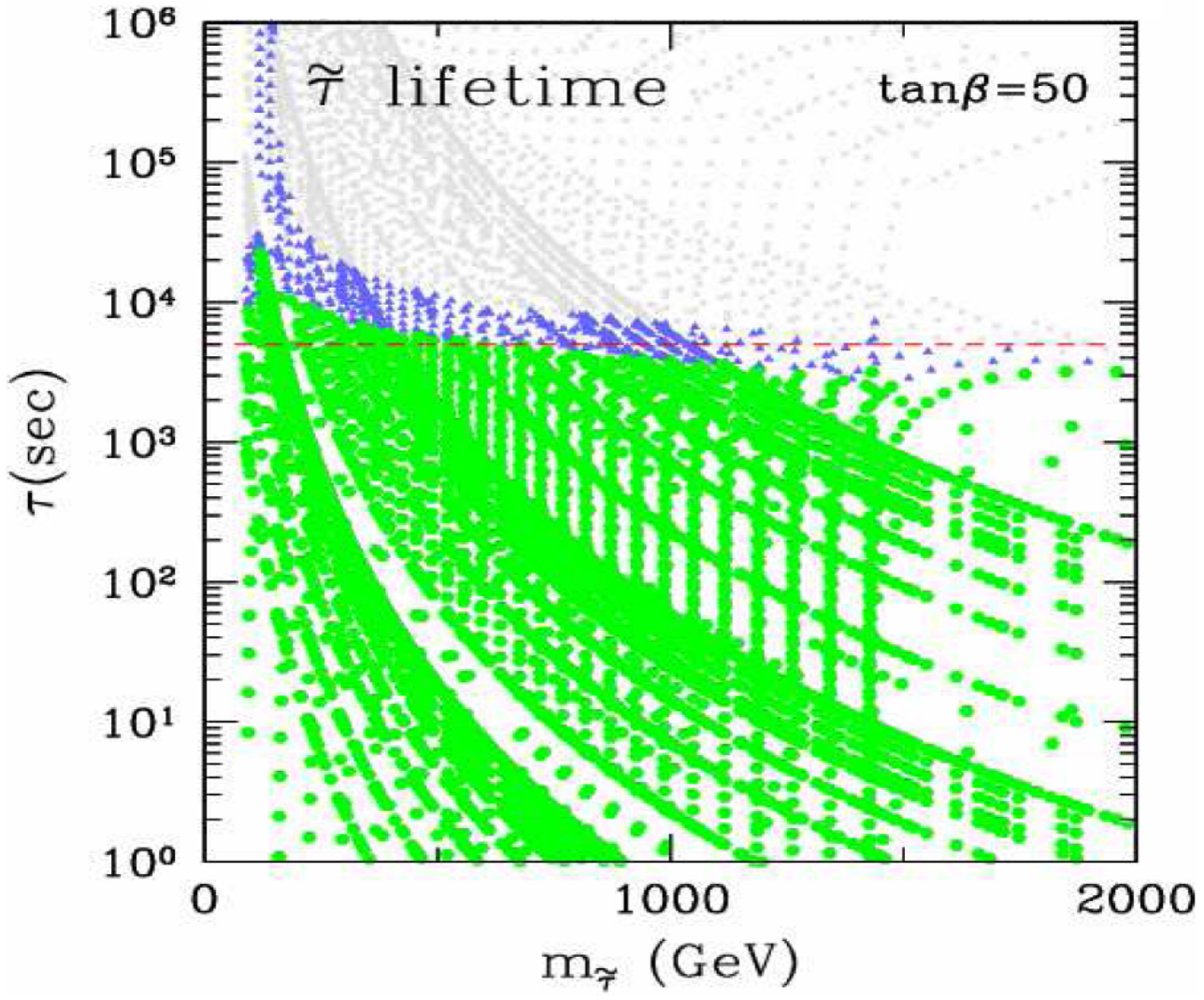}
\caption{Allowed values for the stau mass and lifetime in gravitino
superWIMP scenarios scanning over minimal supergravity parameters with
$\tan\beta = 50$, and assuming gravitinos are produced both after
reheating and in late decays.  The grey dots represent models that
satisfy collider constraints; the green dots represent models that
also satisfy all BBN constraints; the blue dots represent models that
are allowed for slightly loosened BBN constraints on
$^6$Li/$^7$Li. {}From~\protect\citet{Bailly:2009pe}.}
\label{fig:staulifetime}
\end{figure}

BBN constraints on the gravitino superWIMP possibility have been
considered in a number of studies.  The results of one study, which
considered minimal supergravity models with a stau NLSP decaying to a
gravitino superWIMP, are given in \figref{staulifetime}. Without BBN
constraints, we find that extremely large stau lifetimes are possible.
The BBN constraints typically exclude the largest lifetimes (although
there are interesting exceptions~\cite{Ratz:2008qh}).  Nevertheless,
lifetimes as large as $10^4~\s$ are still allowed for all stau masses,
and even larger lifetimes are possible for light staus with mass $\sim
100~\gev$.  Models with large lifetimes and light staus are the most
promising for the collider studies described in
\secref{colliderssuperwimps}, and we see that BBN constraints do not
exclude these scenarios.

Finally, late decays to superWIMPs may in fact improve the current
disagreement of standard BBN predictions with the observed $^7$Li and
$^6$Li abundances, although this typically requires that the decaying
slepton be heavy, with mass above a
TeV~\cite{Jittoh:2007fr,Cumberbatch:2007me,Bailly:2008yy}.  For a
review, see~\citet{Jedamzik:2009uy}.

\subsubsection{SMALL SCALE STRUCTURE}

\begin{figure}[tbp]
\includegraphics[width=0.60\columnwidth]{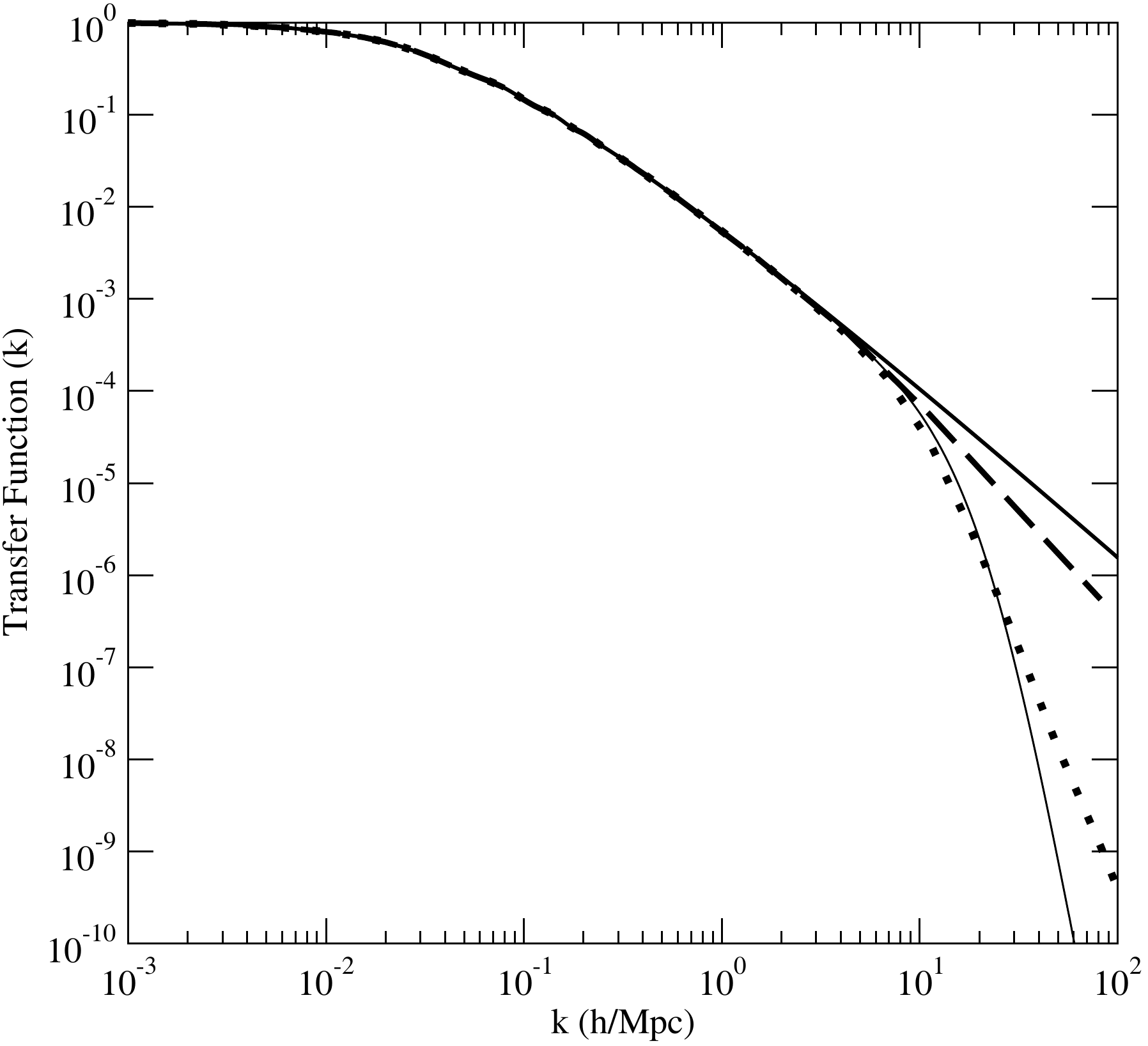}
\caption{The power spectrum for scenarios in which dark matter is
completely composed of WIMPs (solid), half WIMPs and half superWIMPs
(dashed), and completely composed of superWIMPs (dotted).  For
comparison, the lower solid curve is for 1 keV sterile neutrino warm
dark matter.  {}From~\citet{Kaplinghat:2005sy}.
\label{fig:small_scale}
}
\end{figure}

In contrast to WIMPs, superWIMPs are produced with large velocities at
late times.  This has two, {\em a priori} independent, effects.
First, the velocity dispersion reduces the phase space density,
smoothing out cusps in dark matter halos.  Second, such particles damp
the linear power spectrum, reducing power on small
scales~\cite{Lin:2000qq,Sigurdson:2003vy,Profumo:2004qt,%
Kaplinghat:2005sy,Cembranos:2005us,Jedamzik:2005sx,Borzumati:2008zz}.
As seen in \figref{small_scale}, superWIMPs may suppress small scale
structure as effectively as a 1 keV sterile neutrino, the prototypical
warm dark matter candidate (see \secref{sterileneutrinos}).  Some
superWIMP scenarios may therefore be differentiated from standard cold
dark matter scenarios by their impact on small scale structure; for a
review, see~\citet{Primack:2009jr}.

\section{LIGHT GRAVITINOS}
\label{sec:lightgravitinos}

The gravitino dark matter candidates discussed in \secref{superwimps}
have masses around $\mweak$.  Other well-motivated, if somewhat
problematic, candidates are light gravitinos, with masses in the range
$\ev - \kev$ and very different implications for experiments.

\subsection{Thermal Production}

\subsubsection{GAUGE-MEDIATED SUPERSYMMETRY BREAKING}
\label{sec:gmsb}

Low-energy supersymmetry elegantly addresses the gauge hierarchy
problem but does not by itself solve the new physics flavor problem.
In fact, the goal of solving the new physics flavor problem is the
prime driver in the field of supersymmetric model building, and
motivates a particularly elegant subset of supersymmetric theories
known as gauge-mediated supersymmetry breaking (GMSB)
models~\cite{Dine:1981za,Dimopoulos:1981au,%
Nappi:1982hm,AlvarezGaume:1981wy,Dine:1994vc,Dine:1995ag}.  In these
models, supersymmetry-breaking is transmitted from a
supersymmetry-breaking sector to the MSSM by so-called ``messenger''
particles through both gauge interactions and gravity.  The resulting
squark and slepton mass matrices, in $3\times 3$ generation space, are
of the form
\begin{equation}
m_{\tilde{f}}^2 = \left( \begin{array}{ccc} 
\mgmsb^2 & 0 & 0 \\ 
0 & \mgmsb^2 & 0 \\ 
0 & 0 & \mgmsb^2 \end{array} \right) 
+ \left( \begin{array}{ccc} 
\sim \mgrav^2 & \sim \mgrav^2 & \sim \mgrav^2 \\
\sim \mgrav^2 & \sim \mgrav^2 & \sim \mgrav^2 \\
\sim \mgrav^2 & \sim \mgrav^2 & \sim \mgrav^2 
\end{array} \right) \ ,
\end{equation}
where
\begin{equation}
\mgmsb \sim \frac{g^2}{16\pi^2} \frac{F}{\mmess} 
\quad \text{and} \quad
\mgrav \sim \frac{F}{\mstar} \ .
\end{equation}
The parameter $g$ denotes gauge coupling constants, $F$ is the
supersymmetry-breaking scale squared, $\mstar$ is the reduced Planck
mass, and $\mmess$ is another mass scale determined by the
supersymmetry-breaking sector and is related to the mass of the
messenger particles.

The essential feature is that the GMSB contributions are determined by
gauge coupling constants only and so are generation-blind.  They
therefore do not mediate flavor-changing effects, and if they are
dominant, such theories ameliorate the new physics flavor problem.
For them to be dominant, one assumes $\mmess \ll \mstar$ and $F \ll
(10^{11}~\gev)^2$ subject to the constraint $F/\mmess \sim 100~\tev$,
so that the superpartner masses are at the weak scale, but the
flavor-violating gravity contributions are negligible.

As seen in \eqref{gravitinomass}, however, $\mgrav \sim \mgravitino$,
that is, the gravity contributions to the squark and sleptons masses
are the same size as the gravitino mass.  In GMSB scenarios, then, the
LSP is the gravitino.  To sufficiently suppress flavor-violating
effects, one typically requires $\mgrav, \mgravitino \alt 1~\gev$. As
a result, WIMPs and superWIMPs are typically not viable dark matter
candidates in GMSB models: WIMPs decay through $R$-parity conserving
interactions such as $\chi \to \gamma \gravitino$, and superWIMPs are
under-abundant, since $\mgravitino \ll m_{\text{WIMP}}$ implies
$\Omega_{\text{SWIMP}} \ll \Omega_{\text{WIMP}}$.

\subsubsection{RELIC DENSITY}

Light gravitinos may be dark matter candidates in GMSB models,
however~\cite{Pagels:1981ke}.  As with neutrinos, light gravitinos may
be in thermal equilibrium in the early hot Universe.  They then
decouple, with relic density
\begin{equation}
\Omega_{\gravitino}^{\text{th}} \approx 
0.25 \, \frac{\mgravitino}{100~\ev} \ .
\end{equation}
When originally proposed, $\OmegaDM \sim 1$ was possible, and
constraints from structure formation allowed $\mgravitino \sim 100~\ev
- \kev$.  Such ``keV gravitinos'' were then viable dark matter
candidates.  There is no theoretical reason to favor the 100 eV$-$keV
mass range for gravitinos, and so this scenario does not naturally
explain the relic density in the way that the WIMP miracle does.  In
contrast to WIMPs and superWIMPs, however, light gravitinos are
well-motivated by their presence in models that solve not only the
gauge hierarchy problem, but also the new physics flavor problem.

The light gravitino scenario has become much less simple in recent
years, however.  The allowed value for the dark matter relic density
has been reduced to $\OmegaDM \simeq 0.23$.  In addition, there are
much more stringent limits on how hot dark matter can be.  Among the
strongest bounds are those from Lyman-$\alpha$ forest observations,
which constrain the distribution of gas between distant objects and
us.  Observations of density fluctuations on relatively small scales
implies that dark matter should not have erased power on these scales.
Lyman-$\alpha$ constraints therefore require that the bulk of dark
matter be sufficiently cold, implying $\mgravitino \agt
2~\kev$~\cite{Seljak:2006qw,Viel:2006kd}.  Together, these
developments have closed the window on the minimal light gravitino
dark matter scenario.

There are, however, at least two viable extensions.  In the
one-component gravitino scenario, sometimes abbreviated $\Lambda$WDM,
the gravitino has mass $\agt 2~\kev$ and is cold enough to agree with
Lyman-$\alpha$ constraints, but its thermal relic density is either
not realized, for example, because of a low reheating temperature, or
is significantly diluted, for example, through late entropy
production~\cite{Baltz:2001rq}.  Alternatively, in the two-component
gravitino scenario, typically denoted $\Lambda$CWDM, the gravitino has
mass $\alt 16~\ev$~\cite{Viel:2005qj}, but it is a sufficiently small
portion of the dark matter to be consistent with structure formation
constraints, provided another particle provides an additional and
dominant cold or warm component.

\subsection{Particle Colliders}

Remarkably, the $\Lambda$WDM and $\Lambda$CWDM scenarios may be
differentiated at particle colliders.  In contrast to weak-scale
gravitinos, light gravitinos interactions are stronger than
gravitational, as they are enhanced by their spin 1/2 longitudinal
Goldstino components.  These interactions depend on the
supersymmetry-breaking scale $F$ and are stronger for lighter
gravitinos.  For example, a Bino-like neutralino has decay width
\begin{equation}
\Gamma(\tilde{B} \to \gamma \gravitino) = 
\frac{\cos^2 \theta_W m_{\tilde{B}}^5}{16 \pi F^2} \ ,
\end{equation}
where $\theta_W$ is the weak mixing angle, corresponding to a decay
length
\begin{equation}
c \tau \simeq 22~\cm \left[ \frac{\mgravitino}{100~\ev} \right]^2
\left[ \frac{100~\gev}{m_{\tilde{B}}^2} \right]^5 \ .
\end{equation}

The two light gravitino scenarios therefore predict decay lengths that
are either shorter or longer than typical sizes of particle detectors.
In a Bino LSP GMSB scenario, supersymmetry events will be seen through
missing energy in the $\Lambda$WDM scenario, but will be characterized
by two high energy photons in the $\Lambda$CWDM scenario.  Such
diphoton events are the subject of ongoing searches at the
Tevatron~\cite{Aaltonen:2009tp} and will be spectacular at the LHC.
Similar results hold for the stau NLSP scenario and the decay
$\tilde{\tau} \to \tau \gravitino$: supersymmetry events will appear
with pairs of metastable heavy charged particles in the $\Lambda$WDM
scenario, but will be characterized by two high energy taus in the
$\Lambda$CWDM scenario.  The observation of these various event types,
when interpreted in the gravitino dark matter framework, may therefore
have strong cosmological implications, for example, indicating an era
of late entropy production in the $\Lambda$WDM case, or strongly
implying the existence of another, dominant, form of dark matter in
the $\Lambda$CWDM scenario.

\section{HIDDEN DARK MATTER}
\label{sec:hidden}

As noted in \secref{introduction}, despite great recent progress, all
solid evidence for dark matter is gravitational.  There is also strong
evidence against dark matter having strong or electromagnetic
interactions.  A logical possibility, then, is hidden dark matter,
that is, dark matter that has no SM gauge interactions.  Hidden dark
matter has been explored for decades and brings with it a great deal
of model building freedom, leading to a large and diverse class of
candidates~\cite{Kobsarev:1966,Blinnikov:1982eh,Foot:1991bp,%
Hodges:1993yb,Berezhiani:1995am}.  Unfortunately, this
freedom comes at a cost: by considering hidden dark matter, one
typically loses (1) connections to the central problems of particle
physics discussed in \secref{problems}, (2) the WIMP miracle, and (3)
predictivity, since most hidden dark matter candidates have no
non-gravitational signals, which are most likely required if we are to
identify dark matter.

Recently, however, some hidden dark matter models have been shown to
have some, or even all, of the three properties listed above, putting
them on more solid footing and providing extra structure and
motivation for this framework.  In this section, we first consider the
general possibility of hidden sectors, but then focus on this subset
of hidden dark matter candidates and explore their properties and
implications for detection.

\subsection{Thermal Freeze Out}

\subsubsection{CONSTRAINTS ON TEMPERATURE AND DEGREES OF FREEDOM}

The thermal history of hidden sectors may differ from that of the
visible sector.  However, the hidden sector's temperature, along with
its ``size,'' is constrained, to the extent that it affects the
cosmological history of the visible sector.

One of the leading constraints on hidden sectors is provided by
BBN. The success of BBN is highly sensitive to the expansion rate of
the Universe at time $\tBBN \sim 1 - 1000~\s$.  Light degrees of
freedom in a hidden sector change the expansion rate of the Universe
and thereby impact BBN, even if they have no SM interactions.  The
constraint is conventionally quoted as a bound on $\neff$, the
effective number of light neutrino species, and may be taken to be
$\neff = 3.24 \pm 1.2$ (95\%
CL)~\cite{Cyburt:2004yc,Fields:2006ga,Ciarcelluti:2008vs}.  This implies
\begin{equation}
g_*^h(\ThBBN)
\left(\frac{\ThBBN}{\TBBN} \right)^4 =\frac{7}{8}\, \cdot 2 \cdot
(\neff - 3) \le 2.52 \ (\text{95\% CL}) \ ,
\label{bound}
\end{equation}
where $\ThBBN$ and $\TBBN$ are the temperatures of the hidden and
visible sectors at time $\tBBN$.  This is a significant
constraint~\cite{Kolb:1985bf}; for example, \eqref{bound} excludes a
hidden sector that is an exact copy of the SM with $g_*^h(\ThBBN) =
10.75$, assuming it has the same temperature as the visible sector, so
$\ThBBN = \TBBN$.

As evident in \eqref{bound}, however, this statement is highly
sensitive to the hidden sector's temperature.  If the observable and
hidden sectors are not in thermal contact, the hidden sector may be
colder than the observable sector.  This would be the case if, for
example, the inflaton couplings to the observable and hidden sectors
are not identical, so that they reheat to different
temperatures~\cite{Hodges:1993yb,Berezhiani:1995am}.  Alternatively,
the observable and hidden sectors may initially have the same
temperature, either because they have the same inflaton couplings or
because they are in thermal contact, but may cool independently and
have different temperatures at later times.  For hidden sector
temperatures that are now half of the observable sector's temperature,
hundreds of degrees of freedom, equivalent to several copies of the SM
or MSSM, may be accommodated.

\subsubsection{THE WIMPLESS MIRACLE}
\label{sec:wimplessmiracle}

It is desirable for hidden dark matter to have naturally the correct
relic density, just as in the case of WIMPs.  One way to achieve this
goal would be to duplicate the couplings and mass scales of the
visible sector in the hidden sector so that the WIMP miracle is
satisfied in the hidden sector.  Given the discussion above, this is
certainly possible if the hidden sector has fewer light degrees of
freedom than the visible sector or is slightly colder.

At the same time, this possibility is both rather unmotivated and far
too rigid.  As discussed in \secref{wimpmiracle}, the thermal relic
density of a stable particle with mass $m_X$ annihilating through
interactions with couplings $g_X$ is
\begin{equation}
\Omega_X \sim \langle \sigma_A v \rangle^{-1} 
\sim \frac{m_X^2}{g_X^4} \ .
\label{omega}
\end{equation}
The WIMP miracle is the fact that, for $m_X \sim \mweak$ and $g_X \sim
\gweak \simeq 0.65$, $\Omega_X$ is roughly $\OmegaDM \approx 0.23$.

\Eqref{omega} makes clear, however, that the thermal relic density
fixes only one combination of the dark matter's mass and coupling, and
other combinations of $(m_X, g_X)$ can also give the correct
$\Omega_X$.  In the SM, $g_X \sim \gweak$ is the only choice
available, but in a general hidden sector, with its own matter content
and gauge forces, other values of $(m_X, g_X)$ may be realized.  Such
models generalize the WIMP miracle to the ``WIMPless miracle'': dark
matter that naturally has the correct relic density, but does not
necessarily have a weak-scale mass or weak
interactions~\cite{Feng:2008ya}.

\subsection{Candidates}
\label{sec:hiddencandidates}

\begin{figure}[tbp]
\begin{center}
\includegraphics[width=0.70\columnwidth]{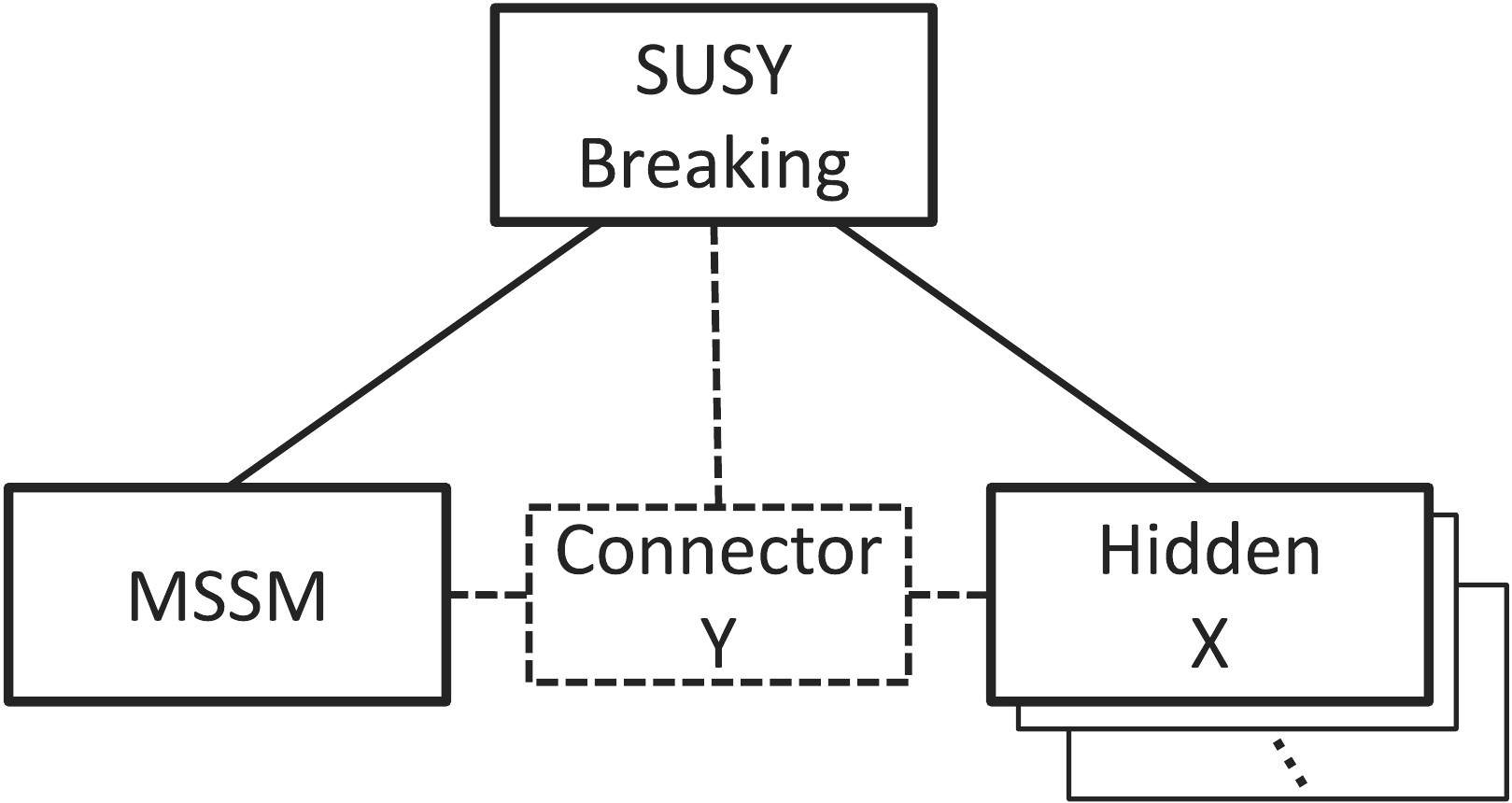}
\caption{Sectors of supersymmetric models.  Supersymmetry breaking is
mediated by gauge interactions to the MSSM and the hidden sector,
which contains the dark matter particle $X$.  An optional connector
sector contains fields $Y$, charged under both MSSM and hidden sector
gauge groups, which induce signals in direct and indirect searches and
at colliders.  There may also be other hidden sectors, leading to
multi-component dark matter.  {}From~\protect\citet{Feng:2008ya}.
}
\label{fig:sectors}
\end{center}
\end{figure}

The WIMPless miracle is naturally realized in particle physics
frameworks that have several other motivations. As an example,
consider the supersymmetric models with GMSB described previously in
\secref{gmsb}.  These models necessarily have several sectors, as
shown in \figref{sectors}.  The supersymmetry-breaking sector includes
the fields that break supersymmetry dynamically and the messenger
particles that mediate this breaking to other sectors through gauge
interactions.  The MSSM sector includes the fields of supersymmetric
extension of the SM.  In addition, supersymmetry breaking may be
mediated to one or more hidden sectors.  The hidden sectors are not
strictly necessary, but there is no reason to prevent them, and hidden
sectors are ubiquitous in such models originating in string theory.

\begin{figure}[tbp]
\includegraphics[width=0.70\columnwidth]{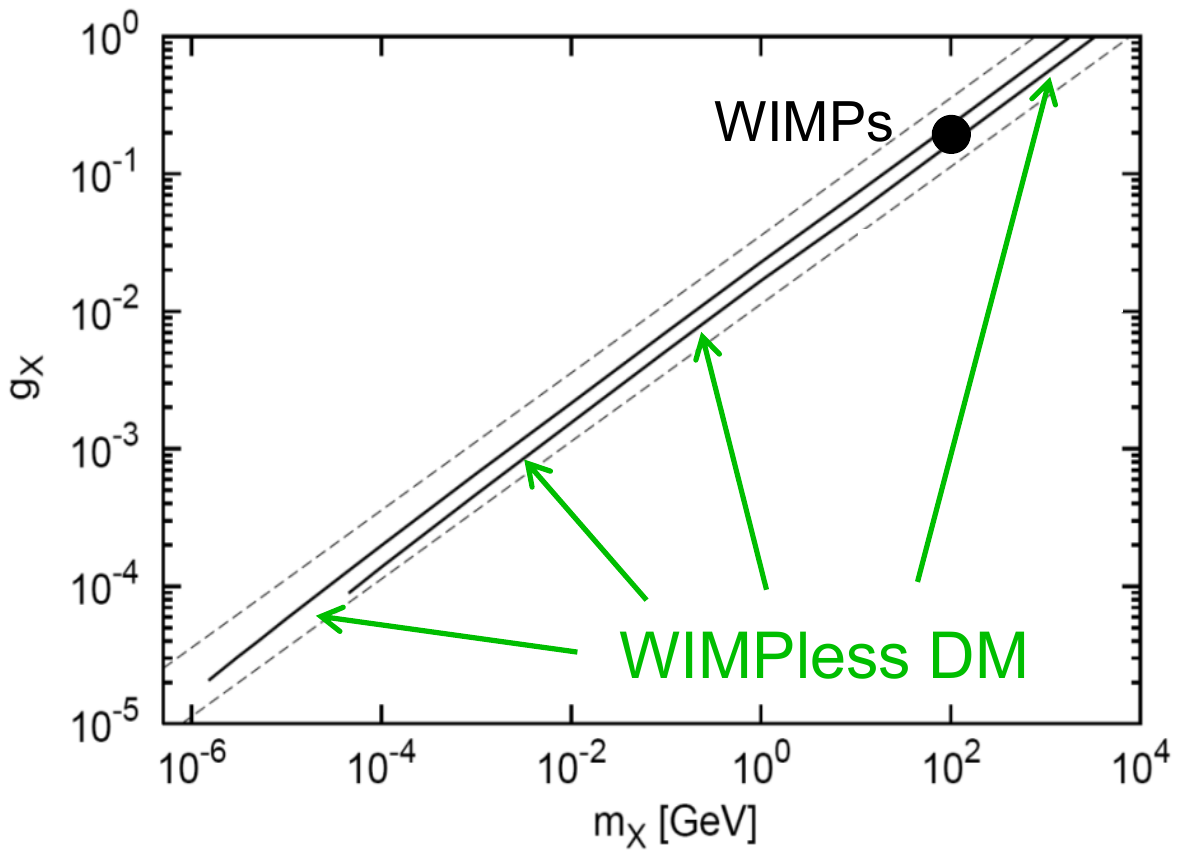}
\caption{Contours of $\Omega_X h^2 = 0.11$ in the $(m_X,g_X)$ plane
for hidden to observable reheating temperature ratios $\ThRH/\TRH=0.8$
(upper solid) and 0.3 (lower solid), where the hidden sector is a
1-generation flavor-free version of the MSSM.  Also plotted are lines
of $\mweak \equiv (m_X/g^2_X)g'^2 = 100~\gev$ (upper dashed) and 1 TeV
(lower dashed).  The WIMPless hidden models generalize the WIMP
miracle to a family of models with other dark matter masses and
couplings.  {}From~\citet{Feng:2008mu}.  }
\label{fig:wimpless_mxgx}
\end{figure}

As described in \secref{gmsb}, the essential feature of GMSB models is
that they elegantly address the new physics flavor problem by
introducing generation-independent squark and slepton masses of the
form
\begin{equation}
\label{mmass}
m \sim \frac{g^2}{16 \pi^2} \frac{F}{\mmess} \ .
\end{equation}
The generic feature is that superpartner masses are proportional to
gauge couplings squared times the ratio $F/\mmess$, which is a
property of the supersymmetry-breaking sector.  With analogous
couplings of the hidden sector fields to hidden messengers, the hidden
sector superpartner masses are
\begin{equation}
\label{mxmass}
m_X \sim \frac{g_X^2}{16 \pi^2} \frac{F}{\mmess} \ ,
\end{equation}
where $g_X$ is the relevant hidden sector gauge coupling.  As a
result,
\begin{equation}
\frac{m_X}{g_X^2} \sim \frac{m}{g^2} \sim
\frac{F}{16 \pi^2 \mmess} \ ;
\label{mxgx}
\end{equation}
that is, $m_X/g_X^2$ is determined solely by the
supersymmetry-breaking sector.  As this is exactly the combination of
parameters that determines the thermal relic density of \eqref{omega},
the hidden sector automatically includes a dark matter candidate that
has the desired thermal relic density, irrespective of its mass.  This
has been verified numerically for a concrete hidden sector model; the
results are shown in \figref{wimpless_mxgx}.  This property relies on
the relation $m_X \propto g_X^2$, which is common to other frameworks
for new physics that avoid flavor-changing problems, such as
anomaly-mediated supersymmetry breaking.  The ``coincidence'' required
for WIMPless dark matter may also be found in other settings; see,
\eg, \citet{Hooper:2008im}.

WIMPless and other hidden sector models also naturally open the
possibility of dark forces in the hidden sector.  In the WIMPless
scenarios just described, this possibility arises naturally if one
attempts to understand why the hidden sector particle is stable.  This
is an important question; after all, in these GMSB models, all SM
superpartners decay to the gravitino.  In the hidden sector, an
elegant way to stabilize the dark matter is through U(1) charge
conservation.  This possibility necessarily implies massless gauge
bosons in the hidden sector.  Alternatively, the hidden sector may
have light, but not massless, force carriers, as described in
\secref{indirect}.  In all of these cases, the dynamics of the hidden
sector may have many interesting astrophysical implications, some of
which are discussed in \secref{wimplessastrophysics}.

In summary, hidden sector dark matter models may in fact be motivated
by leading problems in particle physics, and may even have naturally
the correct relic density, through a generalization of the WIMP
miracle to the WIMPless miracle.  In fact, the third virtue discussed
in \secref{hidden}, predictivity, may also be restored, and we now
describe non-gravitational signals of hidden dark matter.

\subsection{Direct Detection}

The decoupling of the WIMP miracle from WIMPs has many possible
implications and observable consequences.  In the case that the dark
matter is truly hidden, it implies that there are no prospects for
direct or indirect detection.  Signals must be found in astrophysical
observations, as in the case of superWIMPs.  Alternatively, there may
be connector sectors containing particles that mediate interactions
between the SM and the hidden sector through non-gauge (Yukawa)
interactions (see \figref{sectors}).  Such connectors may generate
many signals with energies and rates typically unavailable to WIMPs.

\begin{figure}[tbp]
\includegraphics[width=0.70\columnwidth]{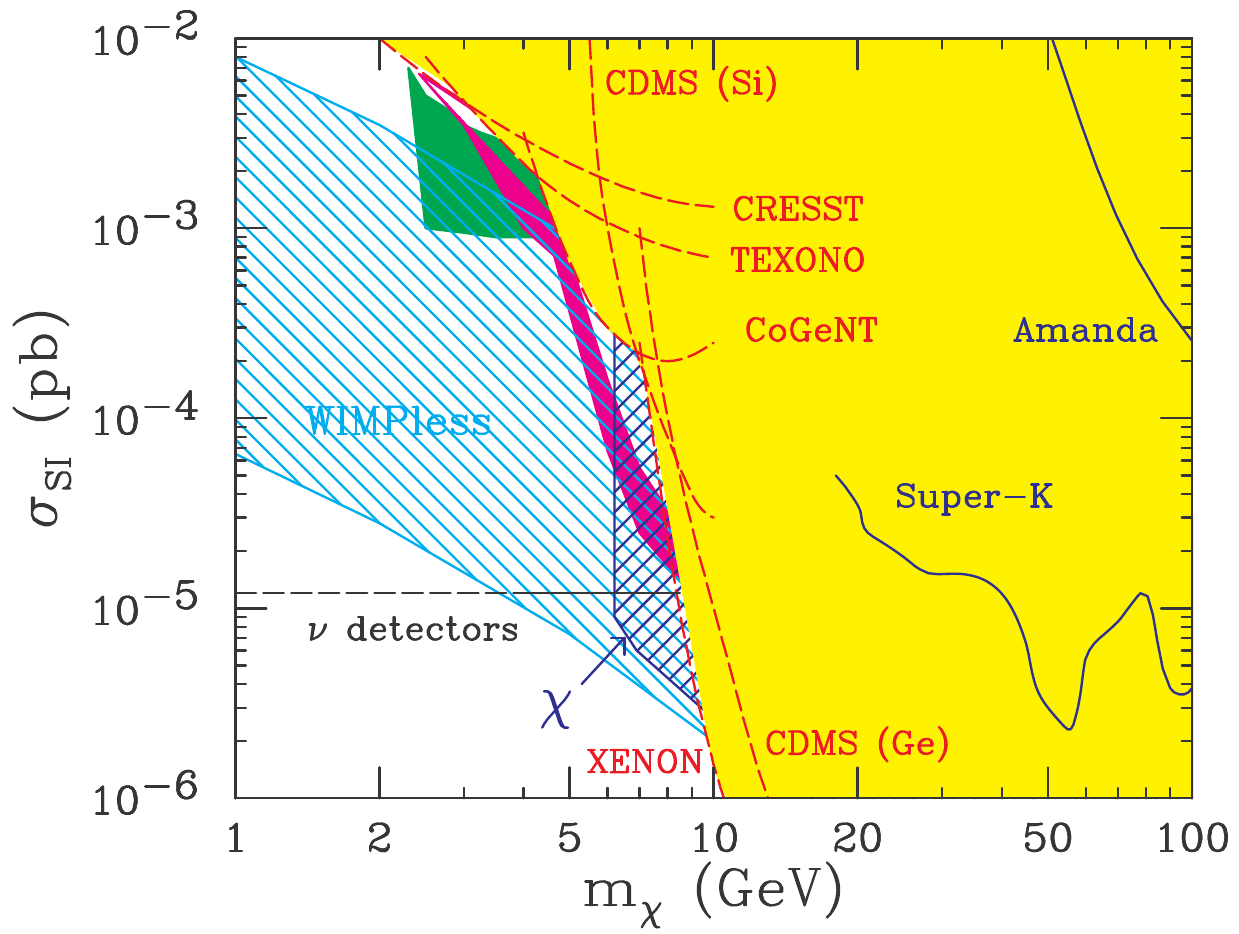}
\caption{Direct detection cross sections for spin-independent
$X$-nucleon scattering in the low dark matter mass $m_X$ region.  The
large hatched region is the predictions of WIMPless models with
connector mass $m_Y = 400~\gev$ and $0.3 <\lambda_b < 1.0$.  The small
hatched region is the prediction of neutralino models considered
in~\citet{Bottino:2003iu}.  The magenta shaded region is DAMA-favored
given channeling and no streams~\cite{Petriello:2008jj}, and the
medium green shaded region is DAMA-favored at 3$\sigma$ given streams
and no channeling~\cite{Gondolo:2005hh}.  The light yellow shaded
region is excluded by the direct detection experiments
CRESST~\cite{Angloher:2002in}, CDMS~\cite{Akerib:2005kh},
XENON10~\cite{Angle:2007uj}, TEXONO~\cite{Lin:2007ka}, and
CoGeNT~\cite{Aalseth:2008rx}, as indicated.  The blue contours are the
published Super-Kamiokande~\cite{Desai:2004pq} and
AMANDA~\cite{Braun:2009fr} exclusion limits, and the black line is a
projection of future neutrino detector sensitivity.
{}From~\citet{Kumar:2009ws}.
\label{fig:superkdirect}
}
\end{figure}

As an example, first consider direct detection.  As discussed in
\secref{direct}, the DAMA and CoGeNT signals may be explained without
violating other bounds if the dark matter mass and spin-independent
cross section are in the ranges $(m_X, \sigmaSI) \sim (1-10~\gev,
10^{-41} - 10^{-39}~\cm^2)$. Such masses are low for conventional
WIMPs, and the cross section is also very high.  In WIMPless models,
however, where the thermal relic density is achieved for a variety of
dark matter masses, such masses are perfectly natural.  Furthermore, a
WIMPless particle $X$ may couple to the SM through Yukawa interactions
\begin{equation}
{\cal L} = \lambda_f X \bar{Y}_L f_L
+ \lambda_f X \bar{Y}_R f_R \ ,
\label{connector}
\end{equation}
where $Y$ is a vector-like connector fermion, and $f$ is a SM fermion.
Taking $f$ to be the $b$ quark, and the $Y$ mass to be 400 GeV,
consistent with current bounds, these couplings generate
spin-independent scattering cross sections given in
\figref{superkdirect}.  We see that WIMPless dark matter may explain
the DAMA and CoGeNT results.  Other proposed hidden dark matter
explanations of DAMA and CoGeNT include those
of~\citet{Foot:2008nw,Kim:2009ke}.

\subsection{Indirect Detection}

WIMPless dark matter also provides new target signals for indirect
detection.  For WIMPs, annihilation cross sections determine both the
thermal relic density and indirect detection signals.  The thermal
relic density therefore constrains the rates of indirect detection
signals.  In the WIMPless case, however, this connection is weakened,
since the thermal relic density is governed by hidden sector
annihilation and gauge interactions, while the indirect detection
signals are governed by the interactions of \eqref{connector}.

This provides a wealth of new opportunities for indirect detection
experiments looking for photons, positrons, and other annihilation
products.  As an example, WIMPless dark matter may be detected through
its annihilation to neutrinos in the Sun by experiments such as
Super-Kamiokande.  Although such rates depend on the competing cross
sections for capture and annihilation, the Sun has almost certainly
reached its equilibrium state, and the annihilation rate is determined
by the scattering cross section~\cite{Desai:2004pq}.  The prospects
for Super-Kamiokande may therefore be compared to direct detection
rates~\cite{Desai:2004pq,Hooper:2008cf,Feng:2008qn,Kumar:2009ws}.  The
results are given in \figref{superkdirect}.  In the near future,
Super-Kamiokande and other neutrino detectors may be able to probe the
low mass regions corresponding to the DAMA and CoGeNT signals.

\subsection{Particle Colliders}

If hidden dark matter does not interact with SM particles, there are
no collider signals.  However, there may be connector particles.
Since these necessarily have SM interactions, they may be produced
with large cross sections at colliders, and since they necessarily
have hidden charge, their decays may be non-standard, leading to
unusual signatures.

As an example, consider the connector sector interactions specified in
\eqref{connector}.  In the hadronic version with $f=q$, the $Y$
necessarily have strong interactions, and so will be similar to 4th
generation quarks; for $f=\ell$, the $Y$ particles are similar to 4th
generation leptons.  The existence of 4th generation quarks and
leptons is constrained by direct collider searches and also by
precision electroweak measurements from LEP and the SLC, but is far
from excluded~\cite{Kribs:2007nz}.  4th generation quarks with masses
in the range $300~\gev \alt m_{T', B'} \alt 600~\gev$ are consistent
with all data.  Strongly-interacting connector particles with mass in
this range will be produced in large numbers at the LHC, and also at
the Tevatron.  At the same time, such connector particles are unlike
standard 4th generation quarks, which dominantly decay through
charged-current modes, such as $t' \to W b$ and $b' \to W
t$. Connector particles will decay through $T' \to t X$, and $B'\to
bX$, similar to squarks, except that squarks typically decay through
cascades, which produce leptons. Hidden sector dark matter therefore
provides new motivation for relatively simple signatures with missing
energy carried away by the hidden dark matter~\cite{Alwall:2010jc}.
Hidden sectors and connectors may also impact the properties of
SM particles, with consequences for colliders running below the energy
frontier~\cite{McKeen:2009rm}.

Hidden sector gauge forces may also have other observable effects.
For example, hidden photons may mix with SM photons through kinetic
mixing terms $F^{\mu\nu} F^h_{\mu\nu}$, leading to fractionally
charged particles~\cite{Holdom:1985ag} that are subject to a wide
variety of probes~\cite{Davidson:2000hf,Perl:2004qc}.

\subsection{Astrophysical Signals}
\label{sec:wimplessastrophysics}

As explained in \secref{hiddencandidates}, an elegant way to stabilize
a hidden sector dark matter candidate is through hidden charge
conservation, much like the electron is stabilized by U(1) charge
conservation in the SM.  This requires that the dark matter have
hidden charge and interact through hidden photons or other gauge
bosons.  More generally, hidden dark matter may interact through
massive, but light, force carriers.  In both cases, the hidden dark
matter has significant self-interactions.  Velocity-independent
(``hard sphere'') self-interactions have been extensively studied, for
example, in the strongly self-interacting framework
of~\citet{Spergel:1999mh}.  Recent work on hidden dark matter has
motivated new interest in velocity-dependent cross sections, namely,
the classic Coulomb scattering cross section
\begin{equation}
\frac{d\sigma}{d\Omega} = 
\frac{\alpha_X^2}{m^2_X v^4\sin^4 \frac{\theta}{2} } 
\end{equation}
and its generalization to Yukawa scattering for massive gauge bosons.

Charged hidden dark matter has many astrophysical implications: (1)
bound state formation and Sommerfeld-enhanced annihilation after
chemical freeze out may reduce its relic
density~\cite{Dent:2009bv,Zavala:2009mi}; (2) similar effects greatly
enhance dark matter annihilation in protohalos at redshifts of $z \sim
30$~\cite{Kamionkowski:2008gj}; (3) Compton scattering off hidden
photons delays kinetic decoupling, suppressing small scale
structure~\cite{Feng:2009mn}; and (4) Rutherford scattering makes such
dark matter collisional~\cite{Ackerman:2008gi,Feng:2009mn}.

\begin{figure}[tbp]
\includegraphics[width=0.70\columnwidth]{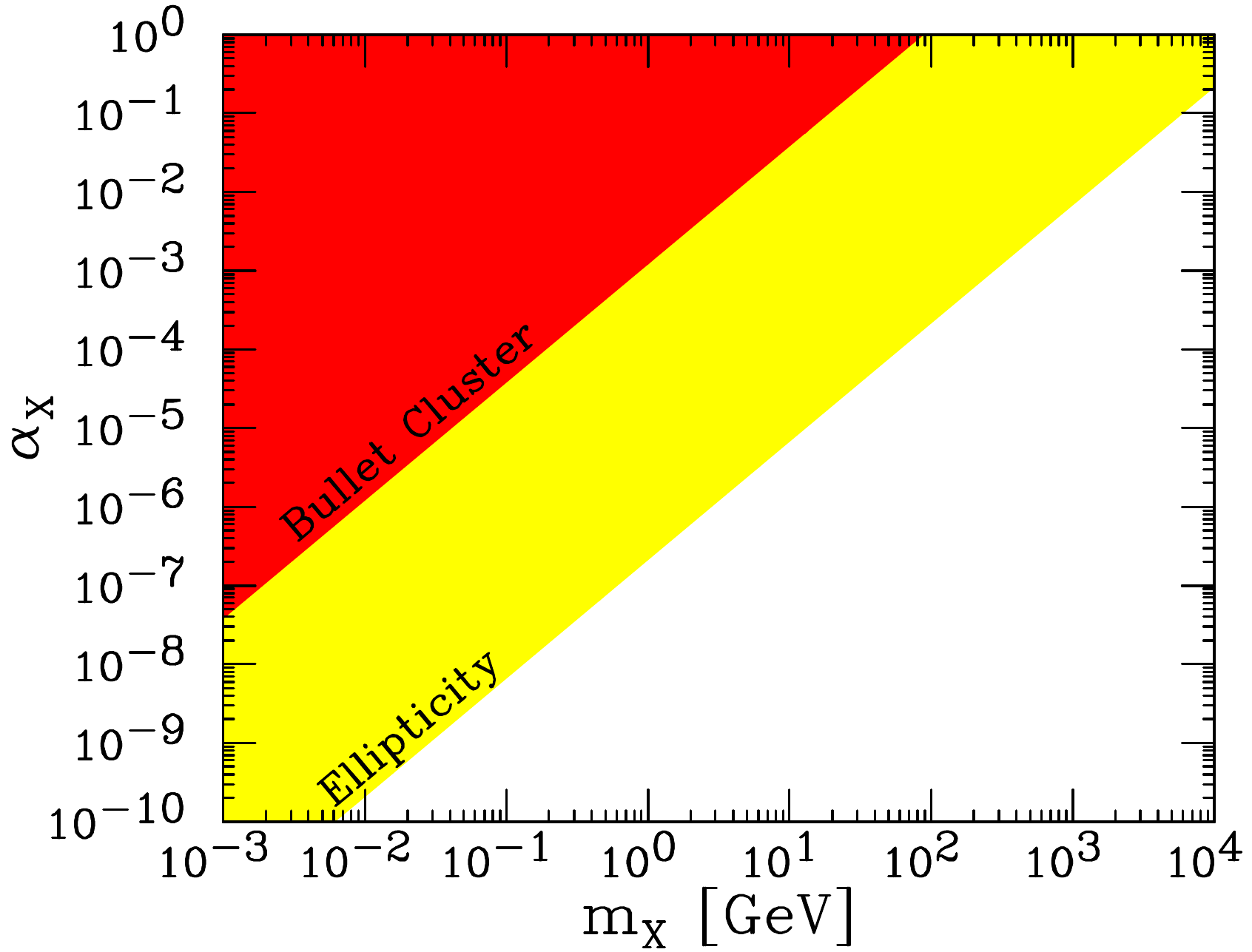}
\caption{Allowed regions in $(m_X, \alpha_X)$ plane, where $m_X$ is
the mass of dark matter charged under hidden electromagnetism with
fine-structure constant $\alpha_X$.  The shaded regions are excluded
by constraints from Bullet Cluster bounds on self-interactions (dark
red) and the observed ellipticity of galactic dark matter halos (light
yellow).  {}From~\citet{Feng:2009mn}.
\label{fig:alphamx}}
\end{figure}

The last possibility, of collisional dark matter, may lead to a number
of observable effects.  A well-known probe of dark matter
self-interactions is provided by the Bullet Cluster, a rare system
where a subcluster is seen to be moving through a larger cluster with
relative velocity $\sim 4500~\km/\s$~\cite{Clowe:2006eq}. Through
observations in the optical and X-ray and strong and weak
gravitational lensing observations, it is clear that dark matter
tracks the behavior of stars, which are collisionless, rather than the
gas.  These observations have allowed stringent bounds to be placed on
the dark matter self-interaction strength.  For velocity-independent
cross sections, the Bullet Cluster observations imply
$\sigma_{\text{DM}} /m_X \alt 3000~\gev^{-3}$ ($\sigma_{\text{DM}}
/m_X \alt 0.7~\cm^2/\g$).  These are the most direct constraints on
the self-interaction of dark matter.  They have been adapted to the
case of velocity-dependent cross sections, leading to the bounds on
dark matter mass $m_X$ and coupling strength $\alpha_X$ shown in
\figref{alphamx}.

Self-interactions also allow dark matter particles to transfer energy.
Self-interactions that are strong enough to create ${\cal O}(1)$
changes in the velocities of dark matter particles will isotropize the
velocity dispersion and create spherical halos. These expectations are
borne out by simulations of self-interacting dark matter in the hard
sphere limit~\cite{Spergel:1999mh,Dave:2000ar,Yoshida:2000bx}.  The
existence of elliptical dark matter halos may therefore also constrain
self-interactions~\cite{MiraldaEscude:2000qt}.  The average rate for
dark matter particles to change velocities by ${\cal O}(1)$ factors is
\begin{equation}
\Gamma_k=\int d^3v_1 d^3v_2 f(v_1) f(v_2) 
\left(n_X \vrel \sigma_T \right)
\left(\vrel^2 / v_0^2\right) ,    
\end{equation}
where $f(v) = e^{-v^2/v^2_0} / (v_0\sqrt\pi)^3$ is the dark matter's
assumed (Maxwellian) velocity distribution, $n_X$ is its number
density, $\vrel=|\vec{v}_1-\vec{v}_2|$, and $\sigma_T = \int
d\Omega_\ast (d\sigma/d\Omega_\ast)(1- \cos \theta_\ast)$ is the
energy transfer cross section, where $\theta_\ast$ is the scattering
angle in the center-of-mass frame.  X-ray observations have
established the ellipticity of the dark matter halo of the elliptical
galaxy NGC 720~\cite{Buote:2002wd,Humphrey:2006rv}, and requiring
$\Gamma_k^{-1} > 10^{10}~\text{Gyr}$ for this system also constrains
self-interacting hidden dark matter, as shown in \figref{alphamx}.
Note that these constraints are much stronger than those from the
Bullet Cluster: for elliptical halos with dark matter velocities
$v\sim 10^{-3}$, the cross section is greatly enhanced relative to the
Bullet Cluster with its larger velocities.

The possibility that dark matter is stabilized by hidden charge
conservation also motivates other astrophysical signals, such as time
delays of light passing through dark matter; see, \eg,
\citet{Gardner:2010qh}.

\section{STERILE NEUTRINOS}
\label{sec:sterileneutrinos}

The evidence for neutrino mass described in \secref{neutrinoproblem}
requires new physics beyond the SM.  This problem may be resolved by
adding right-handed neutrinos $\nu^{\alpha} \equiv \nu_R^{\alpha}$, so
that neutrinos may get mass through the same mechanism that generates
masses for the quarks and charged leptons.  For the mass terms to be
allowed under the symmetries of the SM, the right-handed neutrinos
must have no SM gauge interactions.  One may therefore also add a
gauge-invariant term to the Lagrangian involving only two right-handed
neutrinos --- the so-called Majorana mass term.

The SM is therefore extended to include $N$ sterile neutrinos by
adding the terms
\begin{equation}
{\cal L}_{\nu_R} = 
\bar{\nu}^{\alpha} i \slashed{D} \nu^{\alpha} 
- \left( \lambda_{i\beta}^{\nu} \bar{L}^i \nu^\beta \tilde{\phi} 
+ \text{h.c.} \right)
- \frac{1}{2} M_{\alpha\beta} \bar{\nu}^{\alpha} \nu^{\beta} \ ,
\end{equation}
where $\lambda_{i\beta}^{\nu}$ are the neutrino Yukawa couplings,
$M_{\alpha\beta}$ is the Majorana mass matrix, and $\alpha , \beta =
1, \ldots, N$, where $N \ge 2$ so that at least two neutrino states
are massive.  When electroweak symmetry is broken, the Higgs field
gets a vacuum expectation value.  The neutrino mass eigenstates are
then determined by diagonalizing the complete $(3+N) \times (3+N)$
neutrino mass matrix
\begin{equation}
m_{\nu} = \left( \begin{array}{cc}
0 & \lambda_{i \beta} \langle \phi \rangle \\
\lambda^*_{\alpha j} \langle \phi \rangle & M_{\alpha \beta} \end{array}
\right) .
\end{equation}
Mass eigenstates that are dominantly linear combinations of
left-handed neutrinos are called ``active'' neutrinos, and those that
are dominated by right-handed neutrino components are called
``sterile'' neutrinos.

An elegant idea for explaining the very small neutrino masses is the
see-saw mechanism. In this framework, one assumes $\lambda^{\nu} \sim
{\cal O}(1)$ and $M \gg \langle \phi \rangle$.  There are then $N$
large neutrino masses $\sim M$, and three small neutrino masses $\sim
\lambda^{\nu\, 2} / M$.  For $M \sim 10^{14}~\gev$, near the grand
unification scale, one gets the desired light neutrino masses.  In
this case, sterile neutrinos are beyond the range of experiments and
are not dark matter candidates.

On the other hand, given the range of fermion masses illustrated in
\figref{sm_particles}, there is clearly a range of Yukawa couplings in
the SM and there is no strong reason to assume $\lambda^{\nu} \sim
{\cal O}(1)$.  In general, one may, then, have light sterile
neutrinos, which may be dark matter candidates. We denote this
neutrino $\nu_s$, with mass $m_s$ and mixing angle $\theta$ defined by
\begin{equation}
\nu_s = \cos \theta \, \nu_R + \sin \theta \, \nu_L \ ,
\end{equation}
where $\nu_R$ ($\nu_L$) is a linear combination of right-handed
(left-handed) gauge eigenstates.

\subsection{Production Mechanisms}
\label{sec:productionnu}

Sterile neutrinos may be produced in a number of ways.  The relic
density depends on the sterile neutrino mass and mixing angle, and all
of the mechanisms require very small masses and mixing angles for
sterile neutrinos to be viable dark matter candidates.  These
parameters are not independently motivated by other theoretical
arguments, and so sterile neutrinos do not naturally have the right
relic density.  At the same time, sterile neutrinos with these masses
and mixings may explain some observations, described in
\secsref{nuindirect}{nuastrophysical}, which may be taken as
observational evidence for these parameters.

\begin{figure}[tbp]
\includegraphics[width=0.70\columnwidth]{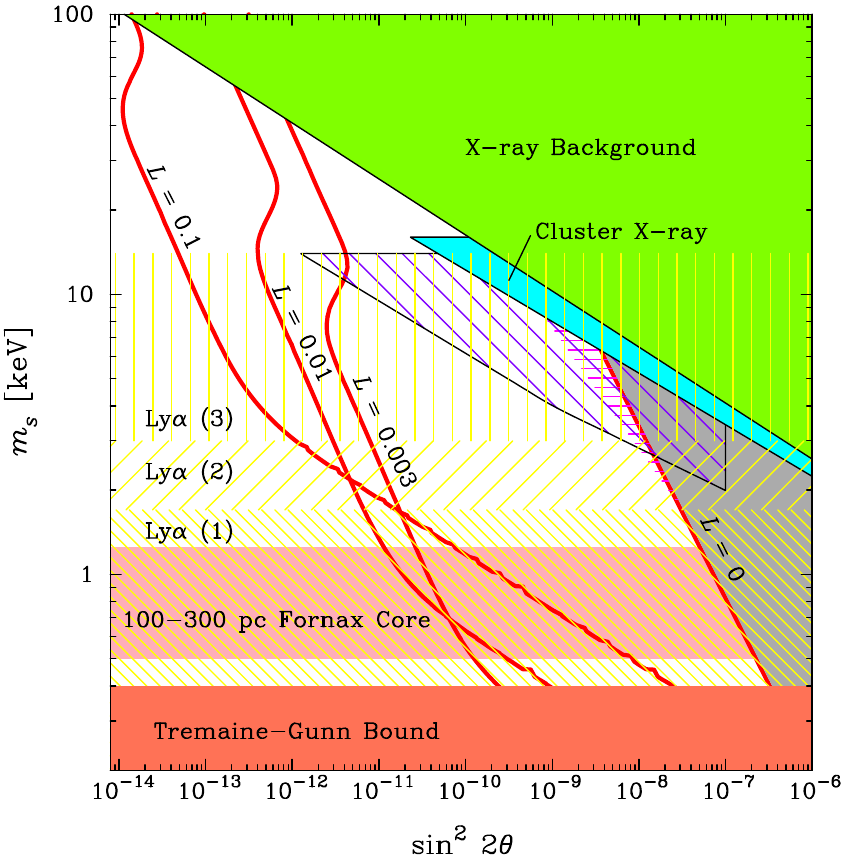}
\caption{Regions of the sterile neutrino dark matter $(\sin^2 2\theta,
m_s)$ parameter plane. On the $L=0$ ($L=0.003, 0.01, 0.1$) contours,
sterile neutrinos produced by oscillations (resonance production) have
relic densities consistent with being all of the dark matter.  The
grey region to the right of the $L=0$ contour is excluded by
overclosure. The regions denoted ``X-ray Background'' and ``Cluster
X-ray'' are excluded by null searches for X-rays from $\nu_s \to
\gamma \nu_a$; the diagonal wide-hatched region is the projected reach
of future X-ray searches.  The ``Fornax'' and horizontal hatched
regions are favored to explain a core in the Fornax dwarf galaxy and
pulsar kicks, respectively.  The Tremaine-Gunn phase space density
bound and a variety of Lyman-$\alpha$ forest constraints on small
scale structure place lower bounds on $m_s$ as shown.
{}From~\citet{Abazajian:2006yn}. 
\label{fig:nulimits}}
\end{figure}

Sterile neutrinos may be produced by oscillations at temperatures $T
\sim 100~\mev$~\cite{Dodelson:1993je}.  Sterile neutrinos were never
in thermal equilibrium, but the resulting distribution is near
thermal.  The results of numerical studies may be approximated by
\begin{equation}
\Omega_{\nu_s} \approx 0.2 \, \frac{\sin^2 2\theta}{10^{-8}} 
\left[ \frac{m_s}{3~\kev} \right]^{1.8} \ ;
\end{equation}
this is shown in \figref{nulimits} as the $L=0$ contour.  This
production mechanism is always present, and the region to the right of
this contour is excluded by overclosure.

The sterile neutrino relic abundance may be enhanced if the Universe
has a non-zero lepton number asymmetry
\begin{equation}
L = \frac{ n_{\nu_a} - n_{\bar{\nu}_a}} { n_{\gamma} } \ .
\end{equation}
For values of $L \agt 10^{-3}$ that are nevertheless small enough to
be well within current limits from BBN, this can allow sterile
neutrinos to be all of the dark matter for smaller mixing angles and
masses, as shown in \figref{nulimits}~\cite{Shi:1998km}. For further
work on the relic density in both the $L=0$ and $L > 0$ scenarios,
see~\citet{Asaka:2006nq,Laine:2008pg}.

Sterile neutrinos may also be produced at higher temperatures, for
example, in the decay of heavy particles.  One example follows from
introducing a singlet Higgs boson field $S$, which couples to
right-handed neutrinos through $ - \frac{1}{2} \lambda^S_{\alpha\beta}
S \bar{\nu}^{\alpha} \nu^{\beta}$.  When $S$ gets a vacuum expectation
value, this term becomes a Majorana mass term.  At temperatures of $T
\sim m_S$, this term also produces sterile neutrinos through the
decays $S \to \bar{\nu}_s \nu_s$, and this may be a significant source
of colder sterile
neutrinos~\cite{Shaposhnikov:2006xi,Kusenko:2006rh,Petraki:2007gq}.

\subsection{Indirect Detection}
\label{sec:nuindirect}

The dominant decay of sterile neutrinos is through $\nu_s \to \nu_L
\bar{\nu}_L \nu_L$.  However, sterile neutrinos may also decay through
a loop-level process to a photon and an active neutrino with branching
ratio $27 \alpha / (8 \pi) \approx 1/128$.  The radiative decay width
is~\cite{Pal:1981rm}
\begin{equation}
\Gamma (\nu_s \to \gamma \nu_a) 
= \frac{9 \alpha}{2048 \pi^4} G_F^2 \sin^2 2\theta \, m_s^5
\simeq \frac{1}{1.5\times 10^{32}~\s} \frac{\sin^2 2 \theta}{10^{-10}} 
\left[ \frac{m_s}{\kev} \right]^5 \ .
\end{equation}
For the allowed sterile neutrino parameters, the sterile neutrino's
lifetime is much longer than the age of the Universe, as required for
it to be dark matter.

At the same time, even if a small fraction of sterile neutrinos decay,
this may be observed.  Because the radiative decay is two-body, the
signal is a mono-energetic flux of X-rays with energy $E_{\gamma}
\simeq m_s/2$. Such signals may be seen by the XMM-Newton, Chandra
X-ray, and Suzaku observatories.  Null results exclude the upper right
shaded region of \figref{nulimits}, and these constraints have been
updated in later analyses (see, \eg,
\citet{Laine:2008pg,Abazajian:2009hx}).  There is also reported
evidence of a signal in Chandra data, consistent with $(m_s, \sin^2 2
\theta) \sim (5~\kev, 3 \times 10^{-9})$, in the heart of the allowed
parameter
space~\cite{Loewenstein:2009cm,Boyarsky:2010ci,Kusenko:2010sq}.
Future observations from the International X-ray Observatory may
extend sensitivities to the entire range of parameters plotted in
\figref{nulimits}~\cite{Abazajian:2009hx}.

\subsection{Astrophysical Signals}
\label{sec:nuastrophysical}

Sterile neutrinos are the classic warm dark matter candidate.  Their
warmth depends on the production mechanism, however.  For the three
production mechanisms discussed in \secref{productionnu}, the sterile
neutrino free-streaming length is roughly
\begin{equation}
\lambda_{\text{FS}} \approx R \, \frac{\kev}{m_s} \
,
\end{equation}
where $R = 0.9$, $0.6$, and $0.2$ Mpc for production from
oscillations, $L$-enhanced production (depending on $L$), and
production through Higgs decay, respectively~\cite{Kusenko:2009up}.
This implies that bounds on small scale structure, for example,
through Lyman-$\alpha$ constraints, depend on the production
mechanism.  For production by oscillations, current bounds require
$m_s \agt 10~\kev$, effectively excluding this production mechanism as
a source for all of dark matter (see \figref{nulimits}).  For colder
production mechanisms, however, the Lyman-$\alpha$ bounds on $m_s$ are
weaker, and these mechanisms may viably produce all of the dark
matter.  We see, though, that for much of sterile neutrino parameter
space, power on small scales is reduced, providing an observable
difference from vanilla cold dark matter.  For further discussion of
small scale structure constraints on sterile neutrinos,
see~\citet{Abazajian:2006yn,Boyarsky:2008xj,Abazajian:2009hx,%
Primack:2009jr}.

In addition to its impact on small structure and the X-ray spectrum,
sterile neutrinos may have other astrophysical effects, for example,
on the velocity distribution of pulsars and on the formation of the
first stars.  For reviews, see~\citet{Boyarsky:2009ix,Kusenko:2009up}.

\section{AXIONS}
\label{sec:axions}

Axions are motivated by the strong CP problem described in
\secref{strongcpproblem}~\cite{Peccei:1977hh,Wilczek:1977pj,%
Weinberg:1977ma}.  The axion solution follows from introducing a new
pseudoscalar field $a$ with coupling
\begin{equation}
{\cal L}_a = - \frac{g_3^2}{32 \pi^2} \frac{a}{f_a}
\epsilon^{\mu\nu\rho\sigma} G_{\mu\nu}^{\alpha}
G_{\rho\sigma}^{\alpha} \ ,
\label{agg}
\end{equation}
where $f_a$ is a new mass scale, the axion decay constant.  This term
makes the coefficient of $\epsilon^{\mu\nu\rho\sigma}
G_{\mu\nu}^{\alpha} G_{\rho\sigma}^{\alpha}$ dynamical.  The vacuum
energy depends on this coefficient, and it relaxes to a minimum where
the EDM of the neutron is very small and consistent with current
bounds.

As we will see, the allowed parameters for axions imply that they are
extremely light and weakly-interacting, providing yet another
qualitatively different dark matter candidate well-motivated by
particle physics~\cite{Preskill:1982cy,Abbott:1982af,Dine:1982ah}.
For a general review of axions, see, \eg, \citet{Asztalos:2006kz}.

\subsection{Production Mechanisms}

The axion's mass and interactions are determined by $f_a$ up to
model-dependent constants that are typically ${\cal O}(1)$.  The
axion's mass is
\begin{equation}
m_a = \frac{\sqrt{m_u m_d}}{m_u + m_d} \, m_{\pi} f_{\pi} \frac{1}{f_a}
\approx 6~\mu\ev \, \left( \frac{10^{12}~\gev}{f_a} \right) ,
\end{equation}
where $m_u \simeq 4~\mev$, $m_d \simeq 8~\mev$, and $m_{\pi} \simeq
135~\mev$ are the up quark, down quark, and pion masses, and $f_{\pi}
\simeq 93~\mev$ is the pion decay constant.  

Axions interact with gluons, through the term of \eqref{agg}, and also
with fermions.  At loop-level, they also interact with photons through
the coupling
\begin{equation}
{\cal L}_{a\gamma \gamma} 
= - g_{\gamma} \frac{\alpha}{\pi} \frac{a}{f_a} \vec{E} \cdot \vec{B}
\equiv - g_{a\gamma\gamma} \, a \, \vec{E} \cdot \vec{B} \ ,
\label{agammagamma}
\end{equation}
where $\alpha$ is the fine-structure constant and $g_\gamma$ is a
model-dependent parameter.  For two well-known possibilities, the
KSVZ~\cite{Kim:1979if,Shifman:1979if} and
DFSZ~\cite{Dine:1981rt,Zhitnitsky:1980tq} axions, $g_{\gamma}$ is
$-0.97$ and 0.36, respectively.  

The axion's mass is bounded by several independent constraints.  The
coupling of \eqref{agammagamma} implies that axions decay with
lifetime
\begin{equation}
\tau (a \to \gamma \gamma)  = \frac{64 \pi}{g_{a\gamma\gamma}^2 m_a^3} 
\simeq \frac{8.8 \times 10^{23}~\s}{g_\gamma^2} 
\left( \frac{\ev}{m_a} \right)^5 \ .
\label{alifetime}
\end{equation}
For axions to live longer than the age of the Universe, $m_a \alt
20~\ev$.  Other astrophysical bounds are even more stringent.  In
particular, axions may be produced in astrophysical bodies and then
escape, leading to a new source of energy loss.  Constraints from the
longevity of red giants and the observed length of the neutrino pulse
from Supernova 1987a, along with other astrophysical constraints,
require $f_a \agt 10^9~\gev$, implying $m_a \alt
10~\text{meV}$~\cite{Raffelt:2006cw}.

There are several possible production mechanisms for axion dark
matter.  {\em A priori} the most straightforward is thermal
production, as in the case of light gravitinos and sterile neutrinos.
Unfortunately, axions produced thermally would have a relic density of
$\Omega_a^{\text{th}} \sim 0.22 \, (m_a/80~\ev)$ and be hot dark
matter.  In addition, \eqref{alifetime} implies that axions with mass
$\sim 80~\ev$ have lifetimes shorter than the age of the Universe, and
so this mechanism cannot produce axions that are the bulk of dark
matter.

There are, however, several non-thermal production mechanisms linked
to the rich cosmological history of the axion field.  As the Universe
cools to a temperature $T \sim f_a$, the axion field takes values that
vary from place to place. This is known as the Peccei-Quinn (PQ) phase
transition.  Although the value of the axion field now is fixed to
minimize the vacuum energy and solve the strong CP problem, at
temperatures $T \agt \gev$, other effects dominate the vacuum energy,
and all values of the axion field are equally possible.

If inflation occurs after the PQ phase transition, then our observable
Universe lies in a patch with a single value of the axion field.  At
$T \sim \gev$, the axion field then relaxes to its minimum.  This
``vacuum realignment'' generates a relic density~\cite{Bae:2008ue}
\begin{equation}
\Omega_a \simeq 0.4 \, \theta_i^2 \left( \frac{f_a}{10^{12}~\gev}
\right)^{1.18} ,
\label{vacuumrealignment}
\end{equation}
where $\theta_i$ is the initial vacuum misalignment angle, assuming
the relic density is not diluted by late entropy production.  The
requirement $\Omega_a \le \OmegaDM$ implies $f_a \alt 10^{12}~\gev \,
\theta_i^{-2}$.  

On the other hand, inflation may occur before the PQ phase transition.
This has two effects.  Our observable Universe then consists of a
mixture of many patches with different $\theta_i$, and the relic
density from vacuum re-alignment is that given in
\eqref{vacuumrealignment}, but with an effective $\theta_i \sim {\cal
O}(1)$. In addition, since many different patches are observable, the
boundaries between patches, topological defects such as domain walls
and axionic strings generated during the PQ phase transition, may have
observable effects.  Production from domain wall decay is expected to
be sub-dominant~\cite{Chang:1998tb} to vacuum realignment, but the
relic density of axions radiated by axionic strings may be of the same
order or even an order of magnitude
larger~\cite{Hagmann:2000ja,Battye:1994au,Yamaguchi:1998gx}.

To summarize, if inflation occurs after the PQ transition, the allowed
window of axion parameter space is roughly
\begin{eqnarray}
10^{12}~\gev \ \theta_i^{-2} \agt &f_a& \agt 10^9~\gev
\nonumber \\
6~\mu\ev \ \theta_i^2 \ \alt &m_a& \alt 6~\text{meV} \ ,
\label{axionconstraints}
\end{eqnarray}
where $\theta_i$ is an arbitrary constant less than 1.  If inflation
occurs before, then \eqref{axionconstraints} applies with $\theta_i
\sim 1$, and axion string production may imply a slightly stronger
upper bound on $f_a$.

The lower bound on $m_a$ arises from requiring that axions don't
overclose the Universe.  When this bound is saturated, axions may be
all of the dark matter, and so this is the preferred target region for
dark matter searches. In contrast to WIMPs and superWIMPs, axions do
not naturally have the correct relic density: there is a range of
possible $m_a$, and there is no reason {\em a priori} to be in the
allowed window or near it's lower boundary.

Note, however, that, if inflation occurred after the PQ transition,
the lower bound on $m_a$ depends sensitively on $\theta_i$. In this
case, if $\theta_i \ll 1$, axions may be all of the dark matter even
for smaller $m_a$ and larger $f_a$.  This latter possibility has some
theoretical attractions, as it implies $f_a$ near $\mgut \simeq
10^{16}~\gev$ may be allowed and provides an avenue for anthropic
selection effects to favor axion densities near the observed
value~\cite{Linde:1987bx,Tegmark:2005dy}.  As we will see, however, if
axions are even lighter and more weakly-coupled axions than naively
expected, they will be beyond detection for the foreseeable future.

\subsection{Direct Detection}

\begin{figure}[tbp]
\includegraphics[width=0.70\columnwidth]{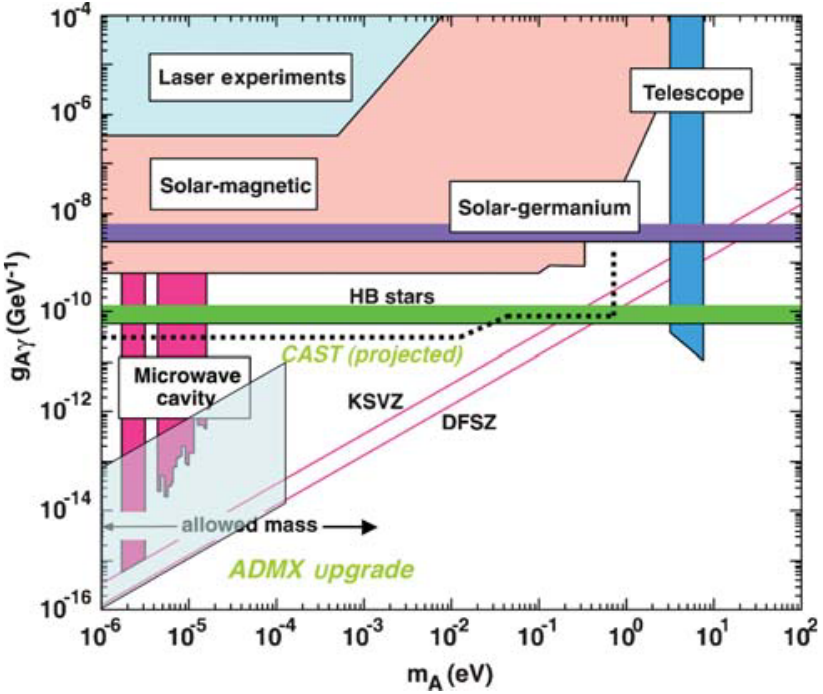}
\caption{Current constraints (shaded) and theoretical predictions for
KSVZ and DFSZ axions in the $(m_a, g_{A\gamma} = g_{a\gamma\gamma})$
plane.  ``Allowed mass'' denotes roughly the window of axion masses
allowed by \eqref{axionconstraints}; see text for details.
{}From~\citet{Asztalos:2006kz}.
\label{fig:axionlimits}
}
\end{figure}

Axions may be detected directly by looking for their scattering with
SM particles in the laboratory.  Current and projected limits from
direct detection axion experiments are shown in \figref{axionlimits},
along with the theoretical predictions for KSVZ and DFSZ axions.

For cosmological axions, given all of the caveats discussed above, the
favored region of axion parameter space, in which axions may be all of
the dark matter, may be taken to be $1~\mu\ev \alt m_a \alt
100~\mu\ev$. In this region, the leading experimental results are from
the Axion Dark Matter Experiment (ADMX).  ADMX searches for
cosmological axions by looking for the resonantly-enhanced conversion
of dark matter axions to photons through scattering off a background
magnetic field, the Primakoff process $a \gamma^* \to
\gamma$~\cite{Sikivie:1983ip}.  This is a scanning experiment --- one
must run at a given frequency to be sensitive to axions of a given
mass.  Once the desired sensitivity in coupling $g_{a \gamma \gamma}$
has been reached, one then moves to another frequency.  Theoretical
expectations therefore play a role in setting the search strategy.  To
date, ADMX has probed down to the level of KSVZ predictions for axion
masses of a few $\mu\ev$.  In the coming decade, ADMX will continue
running, along with another experiment, New CARRACK, also based on the
Primakoff process.  ADMX expects to extend its sensitivity to DFSZ
predictions for $\mu\ev < m_a < 100~\mu\ev$.

In addition to searches for cosmological axions, there are also
searches for axions produced in the core of the Sun.  These and other
experiments are reviewed in~\citet{Asztalos:2006kz}.

\section{CONCLUSIONS}

Current observations support a remarkably simple model of the Universe
consisting of baryons, dark matter, and dark energy, supplemented by
initial conditions determined by an early epoch of inflation.  If
scientific progress is characterized by periods of confusion, which
are resolved by neat and tidy models, which are then launched back
into confusion by further data, the current era is most definitely of
the neat and tidy sort.

Dark matter may be the area that launches us back into confusion with
further data.  The microscopic properties of dark matter are as much
of a mystery now as they were in the 1930's. In the next few years,
however, searches for dark matter through a variety of means discussed
here will discover or exclude many of the most promising candidates.

At its core, the dark matter problem is highly interdisciplinary.
Rather than attempt a summary of this review, we close with some
optimistic, but plausible, scenarios for the future in which
experiments from both particle physics and astrophysics are required
to identify dark matter.  Consider the following examples:

\begin{itemize}
\setlength{\itemsep}{1pt}\setlength{\parskip}{0pt}\setlength{\parsep}{0pt}
\item {\em Scenario 1}: Direct detection experiments see a dark
matter signal in spin-independent scattering.  This result is
confirmed by the LHC, which sees a missing energy signal that is
followed up by precision measurements pointing to a 800 GeV
Kaluza-Klein gauge boson.  Further LHC studies constrain the
Kaluza-Klein particle's predicted thermal relic density to be
identical at the percent level with $\OmegaDM$, establishing a new
standard cosmology in which the dark matter is composed entirely of
Kaluza-Klein dark matter, cosmology is standard back to 1 ns after the
Big Bang, and the Universe has extra dimensions.  Direct and indirect
detection rates are then used to constrain halo profiles and
substructure, ushering in a new era of dark matter astronomy.

\item {\em Scenario 2}: The LHC discovers heavy, charged particles
that are apparently stable.  Together the LHC and International Linear
Collider determine that the new particles are staus, predicted by
supersymmetry.  Detailed follow-up studies show that, if these staus
are absolutely stable, their thermal relic density is larger than the
total mass of the Universe! This paradox is resolved by further
studies that show that staus decay on time scales of a month to
gravitinos. Careful studies of the decays determine that the amount of
gravitinos in the Universe is exactly that required to be dark matter,
providing strong quantitative evidence that dark matter is entirely in
the form of gravitinos, and providing empirical support for
supergravity and string theory.

\item {\em Scenario 3}: An X-ray experiment discovers a line
signal.  Assuming this results from decaying sterile neutrinos, the
photon energy determines the neutrino's mass $m = 2 E_{\gamma}$, the
intensity determines the neutrino mixing angle ($I \propto
\sin^2\theta$), and the image morphology determines the dark matter's
spatial distribution.  {}From the mass and radial distribution,
theorists determine the free-streaming length.  This favors production
from Higgs decays over production by oscillations, leading to
predictions of non-standard Higgs phenomenology, which are then
confirmed at the LHC.  Additional information on neutrino parameters
from the LHC strengthens the hypothesis of sterile neutrino dark
matter, and the energy distribution of the narrow spectral line is
then used to study the expansion history of the Universe and dark
energy.

\end{itemize}

These scenarios are, of course, highly speculative and idealized, but
they illustrate that, even in ideal scenarios that we have studied and
understand, close interactions between many subfields will be
required.  At the same time, if any of the ideas discussed here is
correct, there are promising prospects for the combination of
detection methods in particle physics and astrophysics to identify
dark matter in the not-so-distant future.

\section*{ACKNOWLEDGMENTS}

I am grateful to my colleagues and collaborators at UC Irvine and
elsewhere for their many valuable insights.  I also thank the
participants of the Keck Institute for Space Studies program
``Shedding Light on the Nature of Dark Matter'' for stimulating
interactions, and Alex Kusenko, David Sanford, and Hai-Bo Yu for
comments on this manuscript.  This work was supported in part by NSF
Grant No.~PHY--0653656.

\providecommand{\href}[2]{#2}\begingroup\raggedright\endgroup

%\bibliography{araarefs}

\end{document}